\documentclass[aps,amsmath,amssymb,superscriptaddress,floatfix,twocolumn]{revtex4-2}
\usepackage{amsmath}
\usepackage{changepage}
\usepackage{epsfig}
\usepackage{epstopdf}
\usepackage[colorlinks=true,
            linkcolor=blue,
            citecolor=blue,
            urlcolor=blue]{hyperref}
\usepackage{lipsum}
\usepackage{booktabs}
\usepackage{hhline}
\usepackage{multirow}
\usepackage{braket}
\usepackage{multirow}
\usepackage{makecell}
\usepackage{amsmath}
\usepackage{mathtools}
\usepackage{multirow}
\usepackage[table]{xcolor}
\definecolor{paritygray}{RGB}{240,240,240}
\usepackage{tikz}
\usetikzlibrary{calc}

\begin{document}
\title{ Engineering \(XY\)  and \(ZZ\) interactions between two remote superconducting qubits via a multimode waveguide}
\author{Da-Wei Wang}
\affiliation{School of Integrated Circuits, Tsinghua University, Beijing 100084, China}
\author{Yu-xi Liu}
\thanks{yuxiliu@mail.tsinghua.edu.cn}
\affiliation{School of Integrated Circuits, Tsinghua University, Beijing 100084, China}
\affiliation{Frontier Science Center for Quantum Information, Beijing 100084, China}

\begin{abstract}
Waveguides provide a flexible platform for long-range interactions between superconducting qubits. Here, we study how to engineer the effective \(XY\) and \(ZZ\) interactions between two qubits via qubit frequency assignments and multimode interference determined by the mode configuration and standing-wave parities of a finite multimode waveguide. For a single-mode waveguide, we find that  \(XY\) and \(ZZ\) interactions cannot be simultaneously suppressed to zero and the residual qubit-qubit coupling always exists. For example, the \(XY\)  interaction may be suppressed to a negligible level  when two qubits and  the  waveguide mode satisfy the two-photon resonant condition, but the  \(ZZ\) interaction cannot be switched off.  For multimode configurations within the central dispersive window, the engineered interaction types are very different for odd and even numbers of modes. Even-mode configurations can readily produce strong \(XY\) interactions, whereas odd-mode configurations can strongly suppress the \(XY\)  interaction while retaining a finite \(ZZ\) interaction. Full-Hamiltonian simulations of minimal two- and three-mode configurations confirm \(XY\)-dominated SWAP dynamics and \(ZZ\)-dominated CZ-type dynamics, respectively. These results establish a frequency assignment criterion of qubits and waveguide modes for selectively engineering long-range qubit-qubit interaction types induced by the waveguide with finite multimodes.
\end{abstract}

\maketitle
\section{Introduction}

Distributed quantum information processing based on superconducting quantum circuits requires coherent interactions between spatially separated qubits~\cite{arute2019,kjaergaard2020,krantz2019,Gu2017,monroe2014, PhysRevX.14.041030}.  Circuit and waveguide quantum electrodynamics provide natural routes to create qubit-qubit interactions through virtual-photon exchanges of photon modes inside either transmission-line resonators or waveguides~\cite{blais2004,blais2021,wallraff2004,majer2007,PhysRevX.9.011021,brehm2021,sillanpaa2007}. Unlike single-mode resonators, transmission-line waveguides of finite length support discrete multimode standing-waves with a finite free spectral range (FSR)~\cite{malekakhlagh2016,egger2013,solgun2022,he2025,7gr4-nk6m}. Thus, they are intermediate between single-mode resonators and continuous-mode waveguide. Multimode waveguides have been exploited for coherent communication, state transfer, long-range entanglement generation, and
remote logic operations~\cite{xx24-r7q6,kurpiers2018,axline2018,campagne2018,zhong2019,leung2019,magnard2020,chang2020,grebel2024,song2025,8x68-cx1w,
PRXQuantum.4.010314,yuan2026}. When two qubits are coupled to a multi-mode waveguide, several discrete waveguide modes are simultaneously involved  to dispersively induce the effective qubit-qubit interaction and produce structured interference among several photon pathways. These modes, dispersively coupled to the qubits, form   a dispersive spectral window.

Photon-mediated qubit-qubit interactions depend on the large detuning between the qubits and photon modes, where the single-photon resonance energy exchanges are suppressed. The detunings together with the spatially dependent standing-wave coupling determine the magnitudes and phases of qubit-qubit interaction strengths~\cite{majer2007,solgun2022}. Consequently, changes in the qubit frequencies  relative to the standing-wave spectrum can qualitatively reorganize the interference among photon pathways and result in different types of qubit-qubit interactions, e.g., either \(XY\) or \(ZZ\) interaction~\cite{solgun2022,filipp2011,deng2025}.  Thus, similar to those of avoiding spectral collisions, reducing unwanted interactions,  and selecting suitable operating points for two-qubit gates~\cite{PhysRevResearch.4.023079,Klimov2024,PhysRevLett.117.170502,yr9d-7z8k,mundada2019,sung2021},  the frequency assignment plays a very important role in engineering photon mediated qubit-qubit interactions.

A central question is whether frequency assignment alone can be used to selectively produce different types of effective qubit--qubit interactions. This is nontrivial because the \(XY\) interaction usually arises from second-order virtual single-photon processes~\cite{majer2007,blais2021,PhysRevA.108.043703,solgun2022}, whereas the \(ZZ\) interaction generally originates from higher-order virtual processes involving the nonlinearities of the qubits and two-excitation states~\cite{PhysRevLett.91.167005,xwny-vkft,mckay2016,mundada2019,PhysRevLett.125.240503}.
Consequently, suppressing one interaction channel while retaining the other is difficult in fixed-coupling architectures.
Conventional approaches address this problem using tunable coupling elements, which provide additional control over the interaction strengths and phases ~\cite{chen2014,yan2018,foxen2020,sung2021,yuan2026}.
Although powerful, such active approaches require additional circuit elements and control resources, and may introduce extra decoherence channels.

An alternative route is to exploit interference among multiple coupling pathways. Experiments on multipath couplers have demonstrated strong suppression of static \(ZZ\) interactions while maintaining a finite \(XY\) interaction, thereby separating interaction channels that are normally linked to fixed-coupling circuits~\cite{PhysRevX.13.031035,PhysRevLett.127.130501,PhysRevLett.125.200504,PhysRevX.14.041050,PhysRevApplied.16.024037}. More recently, multimode waveguide coupling between remote superconducting qubits was used to realize high-contrast \(XY\) and \(ZZ\) interactions, including nearly \(ZZ\)-free \(XY\) operation and remote entangling gates~\cite{10.1063/5.0280529}. Complementary theoretical works, which relate effective \(XY\) and \(ZZ\) interactions in multimode circuit-QED systems to the pole structure of the electromagnetic environment, establish multimode interference as a general mechanism for engineering qubit--qubit interactions~\cite{solgun2022,2h4m-mg2p}. However, how the qubit and photon-mode frequencies affect the balance between \(XY\) and \(ZZ\) interactions remains an open question in multimode transmission-line waveguides and multipath couplers.

We here study the coupling of two transmon qubits mediated by a finite multimode transmission-line waveguide and show how qubit-frequency assignment can be used to select the effective interaction types for a given waveguide-mode configuration. For a single-mode waveguide, the suppression of the second-order \(XY\) interaction coincides with the two-photon resonance between two qubits and the waveguide mode,  and thus the \(XY\) and \(ZZ\) interactions are not simultaneously switchable.  For multimode configurations, we identify a mode criterion: the strong suppression of the \(XY\) interaction becomes accessible when there exists a waveguide working mode with a frequency between the frequencies of the two qubits. For symmetric mode distributions within the dispersive spectral window, this criterion is different for odd and even number modes. Even-mode configurations contain only symmetric mode pairs which cannot make either one of \(XY\) and \(ZZ\) interactions negligibly small, whereas odd-mode configurations contain an additional unpaired central mode that can make the \(XY\) interaction negligibly small while the \(ZZ\) interaction remains finite. Full-Hamiltonian simulations of the minimal two- and three-mode configurations confirm the corresponding \(XY\)-dominated SWAP and \(ZZ\)-dominated CZ-type dynamics. These results establish a qubit-frequency-assignment criterion, determined jointly by the mode configuration and standing-wave parity, for selecting long-range \(XY\)- and \(ZZ\)-dominated interactions in finite multimode waveguides.

The remainder of this paper is organized as follows. In Sec.~\ref{sec:model}, we introduce the multimode circuit model and the effective \(XY\) and \(ZZ\) interactions.
 In Sec.~\ref{sec:mode_allocation}, we analyze how the photon-mode configuration and qubit-frequency assignment determine the interaction structure and establish the corresponding frequency-allocation criterion.
 In Sec.~\ref{sec:gate_dynamics}, we verify the resulting \(XY\)- and \(ZZ\)-dominated gates through full Hamiltonian simulations. In Sec.~\ref{sec:discussion}, we discuss the experimental feasibility. Finally, we summarize  the main results in Sec.~\ref{sec:conclusion}.

\section{Model and Effective Interactions}

\label{sec:model}
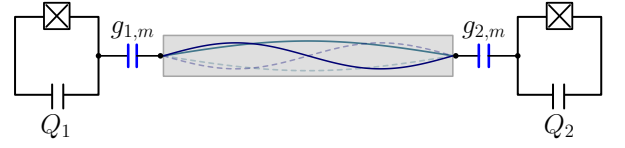
\begin{figure}[htbp]
    \centering
    \resizebox{0.9\columnwidth}{!}{%
   \begin{tikzpicture}[
        scale=0.85,
        transform shape,
        line cap=round,
        line join=round
    ]
    % ============================================================
    % Parameters
    % ============================================================
    \def\yc{1.50}
    % ============================================================
    % Central waveguide
    % ============================================================
    % 背景灰色波导
    \fill[gray!25]
        (4.10,0.95) rectangle (11.90,2.05);
    \draw[gray!70, line width=1pt]
        (4.10,0.95) rectangle (11.90,2.05);
    % ------------------------------------------------------------
    % 驻波模式 (standing-wave Modes in Waveguide)
    % 波导范围: x从 4.10 到 11.90, 长度 L = 7.80, 中心 y = 1.50
    % ------------------------------------------------------------

    % 基模 n = 1
    \draw[cyan!45!black, line width=1.1pt, smooth, domain=4.10:11.90, samples=100]
        plot (\x, {\yc + 0.4*sin((\x - 4.10)/7.80 * 180)});
    \draw[cyan!45!black, line width=1.1pt, dashed, opacity=0.35, smooth, domain=4.10:11.90, samples=100]
        plot (\x, {\yc - 0.4*sin((\x - 4.10)/7.80 * 180)});

    % 二阶模 n = 2
    \draw[blue!45!black, line width=1.0pt, smooth, domain=4.10:11.90, samples=100]
        plot (\x, {\yc + 0.35*sin((\x - 4.10)/7.80 * 360)});
    \draw[blue!45!black, line width=1.0pt, dashed, opacity=0.35, smooth, domain=4.10:11.90, samples=100]
        plot (\x, {\yc - 0.35*sin((\x - 4.10)/7.80 * 360)});

    % ============================================================
    % LEFT SIDE
    % ============================================================

    % left loop (方角矩形)
    \draw[line width=1.3pt]
        (0.10,0.45) rectangle (2.35,2.55);

    % upper white element (SQUID / 交叉元件)
    \fill[white]
        (0.88,2.22) rectangle (1.57,2.93);

    \draw[line width=1.2pt]
        (0.88,2.22) rectangle (1.57,2.93);

    \draw[line width=1.2pt]
        (0.94,2.28) -- (1.51,2.87);

    \draw[line width=1.2pt]
        (0.94,2.87) -- (1.51,2.28);

    % lower element: 电容（两根平行竖线）
    \fill[white] (1.1,0.08) rectangle (1.40,0.73); % 擦除背景线
    \draw[line width=1.3pt] (1.1,0.08) -- (1.1,0.73); % 左极板
    \draw[line width=1.3pt] (1.4,0.08) -- (1.4,0.73); % 右极板

    % label Q1
    \node[font=\huge] at (1.20,-0.4) {$Q_1$};

    % connection node
    \fill (2.35,\yc) circle (0.075);

    % wire to coupling capacitor
    \draw[line width=1.1pt]
        (2.35,\yc) -- (3.15,\yc);

    % coupling capacitor (耦合电容：两根平行竖线)
    \draw[blue,line width=1.7pt]
        (3.15,1.20) -- (3.15,1.80);

    \draw[blue,line width=1.7pt]
        (3.38,1.20) -- (3.38,1.80);

    % label g_{1,m}
    \node[font=\huge] at (3.27,2.18) {$g_{1,m}$};

    % wire to waveguide
    \draw[line width=1.1pt]
        (3.38,\yc) -- (4.02,\yc);

    \fill (4.02,\yc) circle (0.075);

    % ============================================================
    % RIGHT SIDE
    % ============================================================

    % right loop (方角矩形)
    \draw[line width=1.3pt]
        (13.65,0.45) rectangle (15.90,2.55);

    % upper white element (SQUID / 交叉元件)
    \fill[white]
        (14.43,2.22) rectangle (15.12,2.93);

    \draw[line width=1.2pt]
        (14.43,2.22) rectangle (15.12,2.93);

    \draw[line width=1.2pt]
        (14.49,2.28) -- (15.06,2.87);

    \draw[line width=1.2pt]
        (14.49,2.87) -- (15.06,2.28);

    % lower element: 电容（两根平行竖线）
    \fill[white] (14.6,0.08) rectangle (14.9,0.73); % 擦除背景线
    \draw[line width=1.3pt] (14.6,0.08) -- (14.6,0.73); % 左极板
    \draw[line width=1.3pt] (14.9 ,0.08) -- (14.9 ,0.73); % 右极板

    % label Q2
    \node[font=\huge] at (14.77,-0.4) {$Q_2$};

    % connection node
    \fill (13.65,\yc) circle (0.075);

    % waveguide side node
    \fill (11.98,\yc) circle (0.075);

    % wire from waveguide
    \draw[line width=1.1pt]
        (11.98,\yc) -- (12.62,\yc);

    % coupling capacitor (耦合电容：两根平行竖线)
    \draw[blue,line width=1.7pt]
        (12.60,1.20) -- (12.60,1.80);

    \draw[blue,line width=1.7pt]
        (12.87,1.20) -- (12.87,1.80);

    % label g_{2,m}
    \node[font=\huge] at (12.74,2.18) {$g_{2,m}$};

    % wire to right loop
    \draw[line width=1.1pt]
        (12.85,\yc) -- (13.65,\yc);

    \end{tikzpicture}%
    }
\caption{Two spatially separated transmons are coupled to a finite length transmission-line waveguide, that supports several standing-wave modes.
}
\label{model}
\end{figure}

As schematically shown in Fig.~\ref{model},   we consider two spatially separated superconducting qubits, e.g., transmons, coupled to a finite-length microwave  transmission-line waveguide that supports several standing-wave modes, each with frequency \(\omega_{r,m}\). Each transmon qubit is modeled as an anharmonic oscillator. The total Hamiltonian within the rotating wave approximation is written as
\begin{equation}
\hat H=\hat H_0+\hat V ,
\label{eq:total_hamiltonian}
\end{equation}
with
\begin{subequations}
\label{eq:Hamiltonian_parts}
\begin{align}
\hat H_0
&=
\sum_{\mu=1,2}
\left[
\omega_\mu \hat b_\mu^\dagger \hat b_\mu
+
\frac{\alpha_\mu}{2}
\hat b_\mu^{\dagger 2}\hat b_\mu^2
\right]
+
\sum_m \omega_{r,m} \hat a_m^\dagger \hat a_m ,
\label{eq:H0}
\\
\hat V
&=
\sum_{\mu=1,2}\sum_m
g_{\mu m}
\left(
\hat b_\mu^\dagger \hat a_m
+
\hat a_m^\dagger \hat b_\mu
\right).
\label{eq:interaction_hamiltonian}
\end{align}
\end{subequations}
Here, \(\hat b_\mu\) is the annihilation operator of the \(\mu\)th  transmon with  its anharmonicity \(\alpha_\mu<0\). \(\hat a_m\) is the  microwave photon annihilation operator of the \(m\)th waveguide mode with the frequency \(\omega_{r,m}\). We assume that all frequencies of coupled modes form a spectrum window and use \(\omega_{r,0}\) as the reference frequency at the center of the spectrum window, with an approximately uniform free spectral range \(\Delta_{\rm FSR}\).
The coupling strength between the \(\mu\)th transmon qubit and the \(m\)th mode  is denoted by \(g_{\mu m}\). We take
\begin{equation}
g_{\mu m}
=g_\mu u_m(x_\mu)
\sqrt{\frac{\omega_{r,m}}{\omega_{r,0}}}.
\label{eq:general_mode_dependent_coupling}
\end{equation}
where the factor \(\sqrt{\omega_{r,m}/\omega_{r,0}}\) accounts for the frequency dependence of the vacuum fluctuation amplitude within the finite bandwidth considered here~\cite{PhysRevLett.101.080502,Zhong2021}. $u_m(x_\mu)$ is the mode function of the photons. $g_\mu$ is the dipole-like transition element of the \(\mu\)th transmon qubit.

For a waveguide of the length \(L\), the standing-wave mode function may be written as \(u_m(x)\simeq C_m\sin(n_m\pi x/L)\), where \(n_m\) is the longitudinal-mode index and \(C_m\) is the mode-normalization amplitude. Within the selected spectral window, the weak mode dependence of \(C_m\) is neglected and absorbed into  \(g_\mu\). For two transmon qubits located at $x_1$ and $x_2$, respectively, they have relation \(x_2=L-x_1\), the mode functions satisfy \(u_m(x_2)=(-1)^{n_m+1}u_m(x_1)\). We therefore define the relative standing-wave parity factor \(s_m=\mathrm{sgn}[u_m(x_1)u_m(x_2)]\). Since adjacent modes satisfy \(n_{m+1}=n_m+1\),  the adjacent modes have different relative parities, i.e., \(s_{m+1}=-s_m\)~\cite{PhysRevLett.79.5242,Vogell2017}.

Hereafter, we assume that the qubits are not resonant with any waveguide mode. That is, we focus on the dispersive interaction between the qubits and waveguide  modes, satisfying the condition
$
|\Delta_{\mu m}|
=
|\omega_\mu-\omega_{r,m}|
\gg
|g_{\mu m}|
$,
where the waveguide modes are only virtually populated and can be perturbatively eliminated.
To second order of the qubit--mode coupling~\cite{Schrieffer1966}, virtual photons mediate an effective   \(XY\)  interaction between two transmon qubits. We can obtain  the resulting effective Hamiltonian as
\begin{equation}
\hat H_{\rm eff}^{(2)}
=
\sum_{\mu=1,2}
\tilde\omega_\mu
\hat b_\mu^\dagger \hat b_\mu+
\frac{\alpha_\mu}{2}
\hat b_\mu^{\dagger 2}\hat b_\mu^2
+
J_{\rm eff}
\left(
\hat b_1^\dagger \hat b_2
+
\hat b_2^\dagger \hat b_1
\right),
\label{eq:second_order_eff_hamiltonian}
\end{equation}
where \(\tilde\omega_\mu\) are the Lamb-shifted transmon frequencies.
The effective \(XY\)  interaction strength \(J_{\rm eff}\)  is a sum of mode-resolved amplitudes \(J_m\),
\begin{subequations}
\label{eq:Jeff_and_Jm}
\begin{align}
J_{\rm eff}
&=
\sum_m J_m ,
\label{eq:Jeff}
\\
J_m
&=
\frac{g_{1m}g_{2m}}{2}
\left(
\frac{1}{\Delta_{1m}}
+
\frac{1}{\Delta_{2m}}
\right)\equiv \frac{g_{1m}g_{2m}}{2} A_m
\label{eq:Jm}
\end{align}
\end{subequations}
with $A_m=1/\Delta_{1m}+1/\Delta_{2m}$. Thus, \(J_{\rm eff}\) is controlled by the linear interference of virtual single-photon exchange pathways.
Projecting Eq.~\eqref{eq:second_order_eff_hamiltonian}
onto the two-level subspace of each transmon gives
the effective \(XY\) interaction in terms of the
Pauli operators, with
\(\hat b_\mu\rightarrow\sigma_-^{(\mu)}\) and
\(\hat b_\mu^\dagger\hat b_\mu
\rightarrow (I+\sigma_z^{(\mu)})/2\).
The transmon anharmonicities do not contribute to this projection, but they play an important role in higher-order virtual processes that induce an effective    $ZZ$ interaction  proportional to $\sigma_z^{(1)}\sigma_z^{(2)}$.

The $ZZ$ interaction strength can be obtained
by two methods. The first one is to numerically
diagonalize the full Hamiltonian in Eq.~\eqref{eq:total_hamiltonian}
and extract the $ZZ$ interaction from the
eigenenergies as~\cite{PhysRevLett.127.130501}
\begin{equation}
\zeta
=
E_{\widetilde{11}}
-
E_{\widetilde{10}}
-
E_{\widetilde{01}}
+
E_{\widetilde{00}},
\label{eq:zeta_definition}
\end{equation}
Here, $E_{\widetilde{pq}}$ with $p,q=0,1$ denote the
eigenenergies of the eigenstates
$|\widetilde{pq}\rangle$, corresponding to the coupled transmon-waveguide system described by the full Hamiltonian in Eq.~\eqref{eq:total_hamiltonian},  which are the dressed states formed by the waveguide modes and transmon qubits.
For example, the dressed state
$|\widetilde{11}\rangle$ can be expressed as
\begin{equation}
\begin{aligned}
|\widetilde{11}\rangle={}&
c_{11}|11;\mathbf{0}\rangle
+c_{20}|20;\mathbf{0}\rangle
+c_{02}|02;\mathbf{0}\rangle\\
&+\sum_m\bigl(
c_{10,m}|10;1_m\rangle
+c_{01,m}|01;1_m\rangle
+c_{2m}|00;2_m\rangle
\bigr)\\
&+\sum_{m<n}c_{mn}|00;1_m1_n\rangle.
\end{aligned}
\label{eq:dressed_11_expansion}
\end{equation}
Here, $\mathbf{0}$ denotes the vacuum of all
waveguide modes. The state $|11;\mathbf{0}\rangle$
is the bare state with both transmons in their
first excited states, while
$|20;\mathbf{0}\rangle$ and
$|02;\mathbf{0}\rangle$ denote the doubly
excited states of the first and second
transmons, respectively.
The states $|10;1_m\rangle$ and
$|01;1_m\rangle$ contain one waveguide photon
in the $m$th mode, whereas
$|00;2_m\rangle$ and
$|00;1_m1_n\rangle$ describe two photons
in the same and different waveguide modes,
respectively.
The expansion coefficients are determined
by the full Hamiltonian and represent
the probability amplitudes of the
corresponding bare-state components.
In the limit of vanishing qubit--waveguide
couplings, $c_{11}\rightarrow1$ and
all other coefficients vanish, so that
$|\widetilde{11}\rangle$ reduces to
$|11;\mathbf{0}\rangle$.
The other dressed computational states
are defined analogously by their
corresponding bare states in this limit.

The second method is to calculate the $ZZ$
interaction analytically using perturbation
theory. The leading
$ZZ$ interaction arises at fourth order
from virtual processes involving the
transmon anharmonicities and multimode
waveguide excitations
~\cite{yuan2026,deng2025}.
In particular, the energy correction to
$|11;\mathbf{0}\rangle$ involves virtual
transitions through intermediate states
appearing in Eq.~\eqref{eq:dressed_11_expansion},
such as $|20;\mathbf{0}\rangle$,
$|02;\mathbf{0}\rangle$,
$|00;2_m\rangle$, and
$|00;1_m1_n\rangle$ ($m<n$).
Combining the fourth-order energy corrections
to the computational states according to
Eq.~\eqref{eq:zeta_definition}, we obtain
\begin{equation}
\zeta^{(4)}
=
-\frac{2C_2^2}{\delta+\alpha_1}
+\frac{2C_1^2}{\delta-\alpha_2}
+2(C_1+C_2)K,
\label{eq:zz_fourth_main}
\end{equation}
where $\delta=\omega_1-\omega_2$,
$C_\mu=\sum_m g_{1m}g_{2m}/\Delta_{\mu m}$,
and $K=\sum_m g_{1m}g_{2m}/
(\Delta_{1m}\Delta_{2m})$.
The analytical and numerical results show good agreement under the perturbative conditions, as shown in  the Appendix~\ref{app:conditional_interaction}.

%\begin{figure}[t]
%\centering
%\includegraphics[width=0.45\textwidth]{Figure1.png}
%\caption{Virtual transition contributing to the \(ZZ\) interaction \(\zeta\) from \(|11;0_m\rangle\), including transmon nonlinear states and same- and cross-mode two-photon states.}
%\label{fig:zz_virtual_channels}
%\end{figure}

Combining the second-order $XY$ interaction
and the leading-order $ZZ$ interaction,
the effective Hamiltonian in the
computational subspace can be written as
\begin{equation}
\begin{aligned}
\hat H_{\rm eff}
={}&
\sum_{\mu=1,2}
\frac{\widetilde{\omega}_\mu}{2}
\sigma_z^{(\mu)}
+
J_{\rm eff}
\left(
\sigma_+^{(1)}\sigma_-^{(2)}
+
\sigma_-^{(1)}\sigma_+^{(2)}
\right)
+
\frac{\zeta}{4}
\sigma_z^{(1)}\sigma_z^{(2)},
\end{aligned}
\label{eq:effective_qubit_hamiltonian}
\end{equation}
where $J_{\rm eff}$ and $\zeta$ denote the
effective $XY$ and $ZZ$ interaction
strengths, respectively.
 In contrast to the \(XY\)  interaction, which arises from second-order virtual single-photon processes and is governed by the linear superposition in Eq.~\eqref{eq:Jeff_and_Jm}, the \(ZZ\)  interaction originates from fourth-order virtual processes involving multiple excitation processes. Consequently, the \(XY\)  interaction suppression condition \(\sum_m J_m \simeq 0\) does not generally imply \(\zeta \simeq 0\).
The ratio \(J_{\rm eff}/\zeta\) determines the dominant two-qubit dynamics. For conditional-phase operations via \(ZZ\)  interaction, one seeks
$|J_{\rm eff}|/|\zeta|\ll1$, so that population exchange is suppressed and the dynamics are dominated by conditional phase accumulation. Conversely,  SWAP or iSWAP requires $|J_{\rm eff}|/|\zeta|\gg 1$, so that coherent excitation swap dominates over unwanted conditional phase accumulation. Below, we show how qubit-frequency assignment affects the balance between these interaction types for different numbers and configurations of interacting photon modes in the waveguide with the qubits.

\section{Qubit-Frequency Assignment and Effective Interactions}
\label{sec:mode_allocation}
In this section, we investigate how the assignment of qubit frequencies relative to a finite multimode waveguide spectrum affects the effective \(XY\) and \(ZZ\) interactions. For a given mode configuration, changing the qubit frequencies modifies the qubit--mode detunings and hence the signs and relative weights of the mode-resolved virtual-photon pathways. Together with the standing-wave parity structure, this frequency assignment determines whether two effective interactions reinforce or suppress one another. We first analyze the single- and two-mode cases and then generalize the resulting frequency-allocation method to symmetric multimode configurations.

\subsection{Single-Mode Case}

We first consider the case $\Delta_\text{FSR}\gg g_{\mu m}$,  in which the two transmon qubits are dispersively coupled to, e.g., a single central mode of transmission-line waveguide with the frequency \(\omega_{r,0}\), and effect of other modes on the qubits is negligibly small.  In this case, the effective exchange \(XY\)  interaction is
\begin{equation}
J_{\rm eff}=
s\frac{|g_{1m}||g_{2m}|}{2}
\left(
\frac{1}{\Delta_1}
+
\frac{1}{\Delta_2}
\right),
\label{eq:J_single_parity}
\end{equation}
where \(\Delta_\mu=\omega_\mu-\omega_{r,0}\), and
\(s=\operatorname{sgn}(g_{1m}g_{2m})=\pm1\) is the relative standing-wave parity factor.
Equation~\eqref{eq:J_single_parity} already contains the basic interference mechanism of exchange cancellation. Since the product \(s\) of parities  only changes the global sign of \(J_{\rm eff}\), it cannot change the strength of the \(XY\)  interaction.

Let us now consider different frequency assignments. (i) For the case that $\Delta_1$ and $\Delta_2 $ have the same sign, i.e., $\omega_1,\,\omega_2 >\omega_{r,0}$ or $\omega_1,\,\omega_2 < \omega_{r,0}$, $|J_{\rm eff}|$ is always nonzero. Away from higher-order resonances, the \(XY\) interaction is second order in the qubit--waveguide coupling, whereas the
\(ZZ\) interaction arises from higher-order virtual processes. Consequently, the \(XY\) interaction generally dominates in this case. (ii) For the case that the signs of $\Delta_1$ and $\Delta_2 $ are different but \(\Delta_1\neq-\Delta_2\),  then the \(XY\)  and  \(ZZ\) interactions coexist.  However, the strength of the \(XY\)  interaction can be zero when the condition \(\Delta_1=-\Delta_2\), called the  \(XY\) interaction suppression condition,   or equivalently \(\omega_1+\omega_2=2\omega_{r,0}\), is satisfied. Thus, for the coupling of two qubits to a single waveguide mode, the suppression of  the \(XY\)  interaction requires the photon mode frequency to be chosen  between the frequencies of the two qubits.  Physically, this corresponds to destructive interference between two virtual-photon pathways with opposite detunings.

 Interestingly, the frequency condition \(\omega_1+\omega_2=2\omega_{r,0}\) results in the degeneracy of states $|11,0_{r,0}\rangle$ and $|00,2_{r,0}\rangle$ for the two qubits and two photons.
However, this energy degeneracy does not by itself imply resonant hybridization with the two-photon state. In fact, the leading effective coupling to the same-mode two-photon state is proportional to \(1/\Delta_1+1/\Delta_2\) and therefore vanishes under the \(XY\) interaction suppression condition.
 Thus, the single-mode  \(XY\) interaction suppression condition coincides with the two-photon resonant condition. However, this does not mean  that the \(ZZ\) interaction  is derived from this condition.  As shown in Appendix~\ref{app:conditional_interaction},  the finite \(ZZ\) interaction actually originates from other higher-order virtual processes that are not based on the condition \(\omega_1+\omega_2=2\omega_{r,0}\) of the \(XY\) interaction suppression.  That is, once the condition \(\omega_1+\omega_2=2\omega_{r,0}\) is satisfied, the \(XY\) interaction is negligibly small, and \(ZZ\)  interaction dominates.

\begin{table}[t]
	\centering
	\caption{Detuning conditions and the corresponding effective interactions for two qubits coupled to a single waveguide mode.}
	\begin{ruledtabular}
		\begin{tabular}{c|c|c}
			Detuning condition & \(XY\) interaction & \(ZZ\) interaction \\
			\hline
			\(\Delta_1\Delta_2>0\)
			& \(J_{\rm eff}\neq0\)
			& $ \zeta \neq 0 $ \\
			\hline
			\(\Delta_1\Delta_2<0,\ \Delta_1+\Delta_2\neq0\)
			& \(J_{\rm eff}\neq0\)
			&$ \zeta \neq 0 $ \\
			\hline
			\(\Delta_1+\Delta_2=0\)
			& \(J_{\rm eff}=0\)
			& $ \zeta \neq 0 $
		\end{tabular}
	\end{ruledtabular}
	\label{table1}
\end{table}

\begin{figure}[b]
  \centering
  \includegraphics[width=0.5\textwidth]{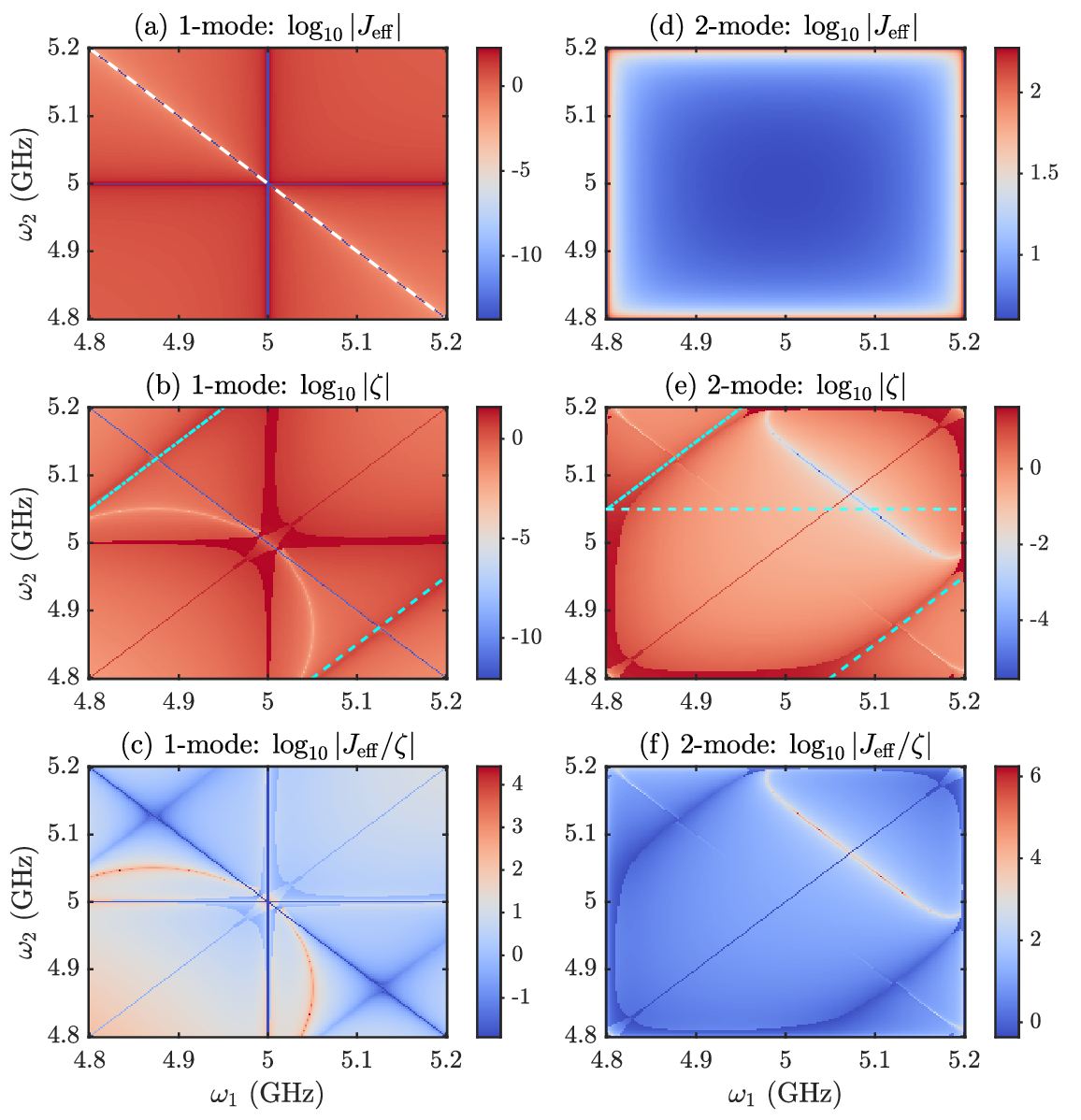}
\caption{
Comparison of the effective \(XY\) and \(ZZ\) interactions mediated by
 single waveguide mode [(a)--(c)] and two waveguide modes [(d)--(f)].
The logarithmic interaction strengths
\(\log_{10}|J_{\rm eff}|\) are shown in (a) and (d), while
\(\log_{10}|\zeta|\) are shown in (b) and (e).
The corresponding ratios
\(\log_{10}|J_{\rm eff}/\zeta|\) are shown in (c) and (f)
in the \((\omega_1,\omega_2)\) parameter plane.
For the single-mode case, the mode frequency is
\(\omega_r=\omega_{r,0}\), whereas for the two-mode case the mode
frequencies are
\(\omega_{r,1}=\omega_{r,0}-\Delta_{\rm FSR}/2\) and
\(\omega_{r,2}=\omega_{r,0}+\Delta_{\rm FSR}/2\).
The parameters are
\(\Delta_{\rm FSR}=0.4~\mathrm{GHz}\),
\(g_\mu/2\pi=0.02~\mathrm{GHz}\), and
\(\alpha_\mu/2\pi=-0.25~\mathrm{GHz}\).
}
  \label{J}
\end{figure}

\begin{table*}[t]
\centering
\caption{
Detuning and parity structure of two waveguide
modes for different qubit-frequency configurations.
Without loss of generality, we assume
$\omega_{r,1}<\omega_{r,2}$ and $\omega_1<\omega_2$,
with alternating standing-wave parities $s_2=-s_1$.
$XY$ cancellation is possible when the two
mode-resolved contributions have opposite signs
and satisfy $J_1=-J_2$. Here, $\Delta_{\mu m}=\omega_{\mu}-\omega_{r,m}$ with $\mu=1,\,2$.
}
\label{table2}

\renewcommand{\arraystretch}{1.2}
\setlength{\tabcolsep}{4pt}
\small

\begin{ruledtabular}
\begin{tabular}{ccccccc}
Frequency configuration
& Mode frequnecy
& Detuning
& $A_m$
& Parity $s_m$
& $\operatorname{sgn}(J_m)$
& XY cancellation
\\
\hline

% I. Central frequency window
\multirow{2}{*}{
$\omega_{r,1}<\omega_1<\omega_2<\omega_{r,2}$
}
& $\omega_{r,1}$
& $\Delta_{\mu1}>0$
& $A_1>0$
& $s_1$
& $+s_1$
& \multirow{2}{*}{Forbidden}
\\
& $\omega_{r,2}$
& $\Delta_{\mu2}<0$
& $A_2<0$
& $-s_1$
& $+s_1$
&
\\
\hline

% II. Both qubits below both modes
\multirow{2}{*}{
$\omega_1<\omega_2<\omega_{r,1}<\omega_{r,2}$
}
& $\omega_{r,1}$
& $\Delta_{\mu1}<0$
& $A_1<0$
& \cellcolor{paritygray}$s_1$
& \cellcolor{paritygray}$-s_1$
& \multirow{2}{*}{Possible}
\\
& $\omega_{r,2}$
& $\Delta_{\mu2}<0$
& $A_2<0$
& \cellcolor{paritygray}$-s_1$
& \cellcolor{paritygray}$+s_1$
&
\\
\hline

% III. First mode between the two qubits
\multirow{2}{*}{
$\omega_1<\omega_{r,1}<\omega_2<\omega_{r,2}$
}
& $\omega_{r,1}$
& $\Delta_{11}<0,\ \Delta_{21}>0$
& $A_1\gtrless0$
& \cellcolor{paritygray}$s_1$
& \cellcolor{paritygray}$\pm s_1$
& \multirow{2}{*}{Possible if $A_1<0$}
\\
& $\omega_{r,2}$
& $\Delta_{\mu2}<0$
& $A_2<0$
& \cellcolor{paritygray}$-s_1$
& \cellcolor{paritygray}$+s_1$
&
\\
\hline

% IV. Second mode between the two qubits
\multirow{2}{*}{
$\omega_{r,1}<\omega_1<\omega_{r,2}<\omega_2$
}
& $\omega_{r,1}$
& $\Delta_{\mu1}>0$
& $A_1>0$
& \cellcolor{paritygray}$s_1$
& \cellcolor{paritygray}$+s_1$
& \multirow{2}{*}{Possible if $A_2>0$}
\\
& $\omega_{r,2}$
& $\Delta_{12}<0,\ \Delta_{22}>0$
& $A_2\gtrless0$
& \cellcolor{paritygray}$-s_1$
& \cellcolor{paritygray}$\mp s_1$
&
\\
\hline

% V. Both qubits above both modes
\multirow{2}{*}{
$\omega_{r,1}<\omega_{r,2}<\omega_1<\omega_2$
}
& $\omega_{r,1}$
& $\Delta_{\mu1}>0$
& $A_1>0$
& \cellcolor{paritygray}$s_1$
& \cellcolor{paritygray}$+s_1$
& \multirow{2}{*}{Possible}
\\
& $\omega_{r,2}$
& $\Delta_{\mu2}>0$
& $A_2>0$
& \cellcolor{paritygray}$-s_1$
& \cellcolor{paritygray}$-s_1$
&
\\
\hline

% VI. Both modes between the two qubits
\multirow{2}{*}{
$\omega_1<\omega_{r,1}<\omega_{r,2}<\omega_2$
}
& $\omega_{r,1}$
& $\Delta_{11}<0,\ \Delta_{21}>0$
& $A_1\gtrless0$
& \cellcolor{paritygray}$s_1$
& \cellcolor{paritygray}$\pm s_1$
& \multirow{2}{*}{Possible if $A_1A_2>0$}
\\
& $\omega_{r,2}$
& $\Delta_{12}<0,\ \Delta_{22}>0$
& $A_2\gtrless0$
& \cellcolor{paritygray}$-s_1$
& \cellcolor{paritygray}$\mp s_1$
&
\\

\end{tabular}
\end{ruledtabular}
\end{table*}

This behavior is illustrated in Figs.~\ref{J}(a) to (c). The suppression of \(J_{\rm eff}\) in Fig.~\ref{J}(a) occurs along the line \(\omega_1+\omega_2=2\omega_{r,0}\), which coincides with the enhancement of \(\zeta\) in Fig.~\ref{J}(b). In addition to this two-photon degeneracy, Fig.~\ref{J}(b) also shows enhanced \(ZZ\)  interaction near the transmon nonlinear resonances
\begin{subequations}
\begin{align}
|11,0_{r,0}\rangle &\leftrightarrow |20,0_{r,0}\rangle,
&
\omega_2-\omega_1 &\simeq \alpha_1 ,
\label{eq:20_resonance}
\\
|11,0_{r,0}\rangle &\leftrightarrow |02,0_{r,0}\rangle,
&
\omega_1-\omega_2 &\simeq \alpha_2 ,
\label{eq:02_resonance}
\end{align}
\end{subequations}
which are referred to as a leakage condition. These features actually originate from the frequency crowding~\cite{Wilhelm} with virtual transitions through the noncomputational states \(|20\rangle\) and \(|02\rangle\), see Appendix~\ref{app:conditional_interaction} for details. The additional curved structures in \(\zeta\) are attributed to higher-order virtual photon processes and therefore do not necessarily follow simple resonant condition  \(\omega_1+\omega_2=2\omega_{r,0}\)  in the \((\omega_1,\omega_2)\) plane.  Accordingly, the small ratio of \(|J_{\rm eff}/\zeta|\) in Fig.~\ref{J}(c)  near this line is associated with the information leakage to the higher energy levels of qubits rather than to the two-photon state of the photon mode.

Table~\ref{table1} summarizes the effective \(XY\) and \(ZZ\) interactions for different qubit-frequency assignments. Three features are evident. First, a finite \(ZZ\) interaction generally remains because it originates from higher-order virtual processes. Second, the \(XY\) interaction can be strongly suppressed when \(\omega_1+\omega_2=2\omega_{r,0}\). Third, suppression of the \(XY\) interaction does not generally eliminate the \(ZZ\) interaction, thus the residual qubit-qubit coupling always exists.

\subsection{Two-Mode Case}

We now consider that two qubits are simultaneously coupled to two neighboring
waveguide modes with frequencies \(\omega_{r,1}\) and \(\omega_{r,2}\),
where \(\omega_{r,1}<\omega_{r,2}\). In this case, each mode-resolved  \(XY\) interaction strength is
\begin{equation}
J_m=
s_m
\frac{|g_{1m}||g_{2m}|}{2}
A_m,
\label{eq:Jm_general}
\end{equation}
and thus, the total strength of the \(XY\) interaction  is  $J_{\rm eff}=J_1+J_2 $.
At first sight, the presence of two virtual-photon pathways seems
sufficient to allow destructive interference.
However, for a self-consistent description of two waveguide modes, the qubit
frequencies must be assigned between frequencies of two chosen waveguide modes to avoid interaction with other photon modes, i.e.,
\begin{equation}
\omega_{r,1}<\omega_1,\omega_2<\omega_{r,2}.
\label{eq:two_mode_window}
\end{equation}
If either qubit frequency lies outside this interval, then an adjacent
waveguide mode omitted from the two-mode truncation can become
spectrally comparable or closer to such qubit  frequency, thus the two-mode description is no longer guaranteed to
remain self-consistent.

Within the frequency window in Eq.~\eqref{eq:two_mode_window}, the lower waveguide mode lies below both qubits and the upper waveguide mode lies above both giving \(A_1>0\) and \(A_2<0\).   At the same time, neighboring waveguide modes have different parities  \(s_1=-s_2\). Consequently, the two mode-resolved \(XY\) interaction strengths \(J_1\) and \(J_2\) have the same sign and interfere constructively rather than destructively as summarized  in  the first row of  Table~\ref{table2}.
The corresponding numerical results in Figs.~\ref{J}(d)--\ref{J}(f) confirm our analytical analysis.
The suppression region of the \(XY\) interaction cannot be found in Fig.~\ref{J}(d) for experimentally valid parameters, while the \(ZZ\) interaction remains finite in
Fig.~\ref{J}(e).
Accordingly, the ratio \(|J_{\rm eff}/\zeta|\) in Fig.~\ref{J}(f)
remains comparatively large over most of the valid two-mode frequency
window. Thus, the configuration of two waveguide modes with the condition in Eq.~(\ref{eq:two_mode_window}) predominantly induces
\(XY\)-dominated interaction and is naturally suited to
SWAP-type dynamics,  the \(ZZ\) coupling can be made negligibly small in this case.  This behavior differs from conventional two-path interference in
engineered couplers, where the relative signs or phases of different
coupling pathways can be independently designed to suppress the
\(XY\) interaction~\cite{PhysRevLett.127.130501,PhysRevLett.125.200504,
PhysRevApplied.16.024037}.
In the present transmission-line waveguide, by contrast, the relative
signs of the couplings between the qubits and transmission-line waveguide modes  are fixed by the standing-wave
parity structure once the two neighboring waveguide modes are selected.

The above discussions show that the \(XY\) interaction cannot be switched off under the condition in Eq.~(\ref{eq:two_mode_window}). However,
if two microwave photon modes are not confined to the same waveguide, i.e., with each mode belonging to a different waveguide,  then the qubit frequencies are not limited by the condition in Eq.~(\ref{eq:two_mode_window}) and the  \(XY\) interaction can be switched off by choosing the expected detunings between the qubits and the waveguide modes. As summarized in Table~\ref{table2}, the factors \(A_m\) can have different signs by assigning different configurations of the qubit frequencies, allowing the two mode-resolved \(XY\) interaction strengths  to
acquire opposite signs and resulting in
\(
J_{\rm eff}=J_1+J_2=0
\)
when all of the parameters are properly chosen. Thus,  the \(XY\) interaction cancellation is mathematically allowed within an approximation of
two waveguide modes once the qubit frequencies are not limited by the condition in Eq.~(\ref{eq:two_mode_window}). Such cancellation points, however, do not necessarily correspond to a physically self-consistent approximation of two waveguide modes. If
the qubit frequencies do not satisfy the condition in Eq.~(\ref{eq:two_mode_window}), additional neighboring waveguide modes may become
spectrally relevant, and the validity of the truncation for two waveguide modes
must therefore be reexamined. A detailed discussion of this outside-window
case is provided in Appendix~\ref{app:boundary_effect}. We would also like to mention that if two qubits are coupled to two harmonic oscillators (e.g., central modes of two independent waveguides), the assignment of the qubit frequencies is free of the condition in  Eq.~(\ref{eq:two_mode_window}) and no parity needs to be considered for these harmonic oscillators. In this case, the interference between the two mode-resolved $XY$ interactions can be controlled by the factors $A_m$, whose signs depend on the detunings between the qubits and the oscillator modes. As indicated by the shaded regions in Table~\ref{table2}, the two contributions can acquire opposite signs, so that allowing the effective $XY$ interaction to be cancelled when the parameters are properly chosen such that $J_1=-J_2$.

\subsection{Intermediate-Mode Criterion in Symmetric Multimode Configurations}

\begin{figure}[h]
  \centering
  \includegraphics[width=0.44\textwidth]{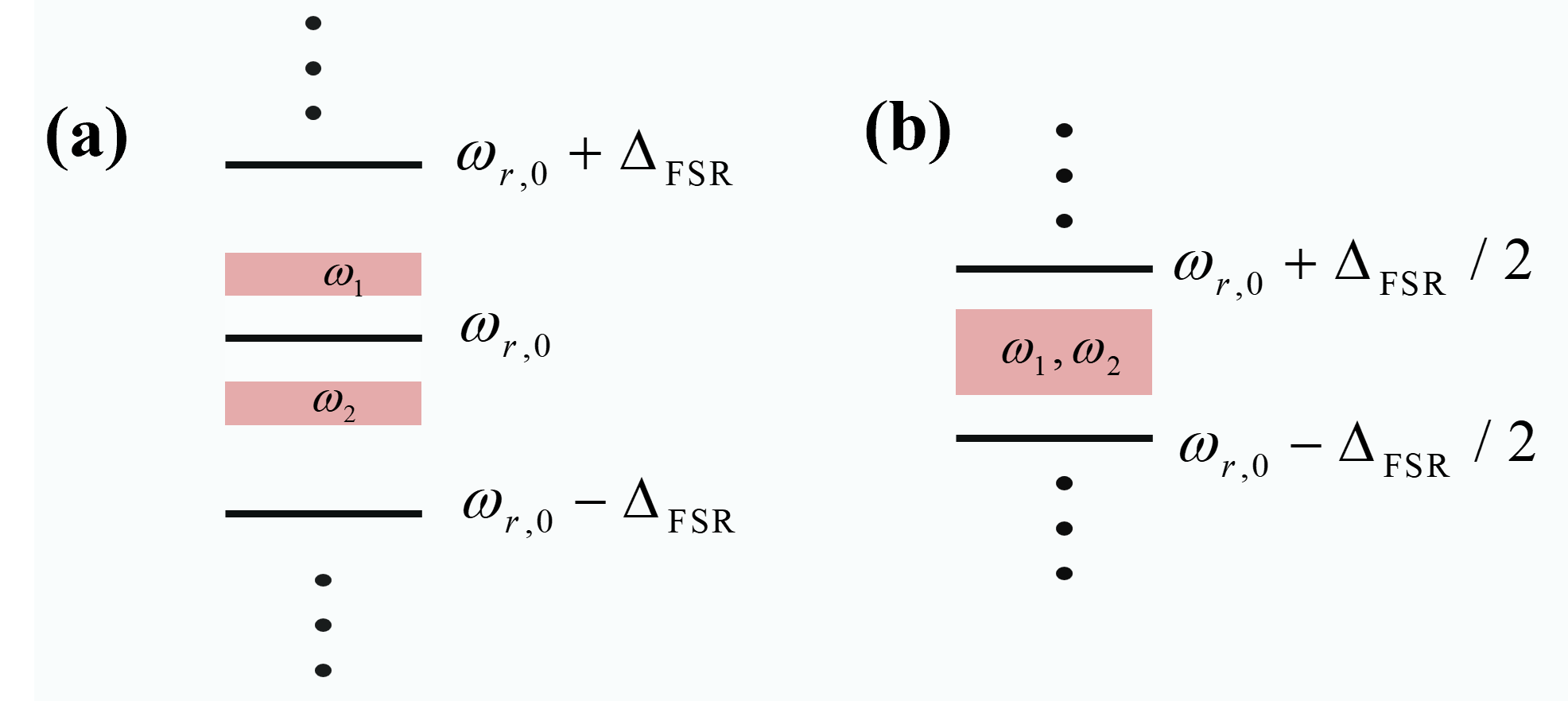}
\caption{
 Central qubit-frequency allocation windows for even- and odd-mode configurations. Black horizontal lines denote the operational waveguide modes, and the shaded regions denote the central windows considered in the finite-mode analysis.
}
  \label{s}
\end{figure}

Let us finally study a general case that two qubits are dispersively coupled to $N$ microwave transmission-line waveguide modes, arranged around a reference frequency \(\omega_{r,0}\), with a free spectral range \(\Delta_{\rm FSR}\). Thus, all frequencies of these coupled modes form a truncated spectral window, which is hereafter referred to  as  central finite-mode window. Below, we consider two special cases that the number of coupled waveguide modes is either odd or even.

For an odd number \(N=2M+1\) of coupled waveguide modes, the chosen spectrum of waveguide modes contains a central mode, in which its frequency can be chosen as a reference frequency \(\omega_{r,0}\),  as shown in Fig.~\ref{s}(a). Thus, all waveguide mode frequencies are given as \(\omega_{r,\pm m}=\omega_{r,0}\pm m\Delta_{\rm FSR}\), where \(m=0,1,\ldots, M\).  In this case, we assume that the qubit frequencies are taken around this central mode frequency, i.e.,
\begin{equation}
|\omega_\mu-\omega_{r,0}|<\frac{\Delta_\text{FSR}}{2}\quad
\mu=1,\, 2.
\label{eq:odd_mode_window}
\end{equation}
Within this condition, the frequency of the central mode among coupled waveguide modes is the closest one to the qubit frequencies. The frequencies of other coupled waveguide modes relative to qubits symmetrically distribute two sides of central frequency as shown in Fig.~\ref{s}(a). It is clear that the waveguide modes with the frequencies \(\omega_{r, m}\) and \(\omega_{r, -m}\)  are symmetric mode pair with $\omega_{r,0}$.

For an even number \(N=2M\) of coupled waveguide modes, we cannot take the reference frequency \(\omega_{r,0}\) as any one of coupled waveguide modes. Instead, all coupled waveguide modes are located symmetrically around \(\omega_{r,0}\) with frequencies \(\omega_{r,\pm m}=\omega_{r,0}\pm\left(m-1/2\right)\Delta_{\rm FSR}\), where \(m=1,\ldots,M\), as shown in Fig.~\ref{s}(b). In this case,  we assume that  both qubit frequencies satisfy the condition
\begin{equation}
\omega_{r,0}-\frac{\Delta_{\rm FSR}}{2}
<
\omega_\mu
<
\omega_{r,0}+\frac{\Delta_{\rm FSR}}{2},
\quad
\mu=1,2 .
\label{eq:even_mode_window}
\end{equation}
Also, the waveguide modes with the frequencies \(\omega_{r, m}\) and \(\omega_{r, -m}\) form a symmetric mode pair.

Equations~(\ref{eq:odd_mode_window}) and (\ref{eq:even_mode_window})  define the physically relevant regimes in which the corresponding odd- and even-mode descriptions remain self-consistent. Outside these conditions, the number of coupled waveguide modes should be changed according to the qubit frequencies. For example, as the qubit frequencies approach the boundaries of the truncated spectrum, modes omitted from the finite $N$ mode description may become spectrally relevant to the qubits, thus additional transmission-line modes must be incorporated.

\begin{figure}[t]
  \centering
  \includegraphics[width=0.5\textwidth]{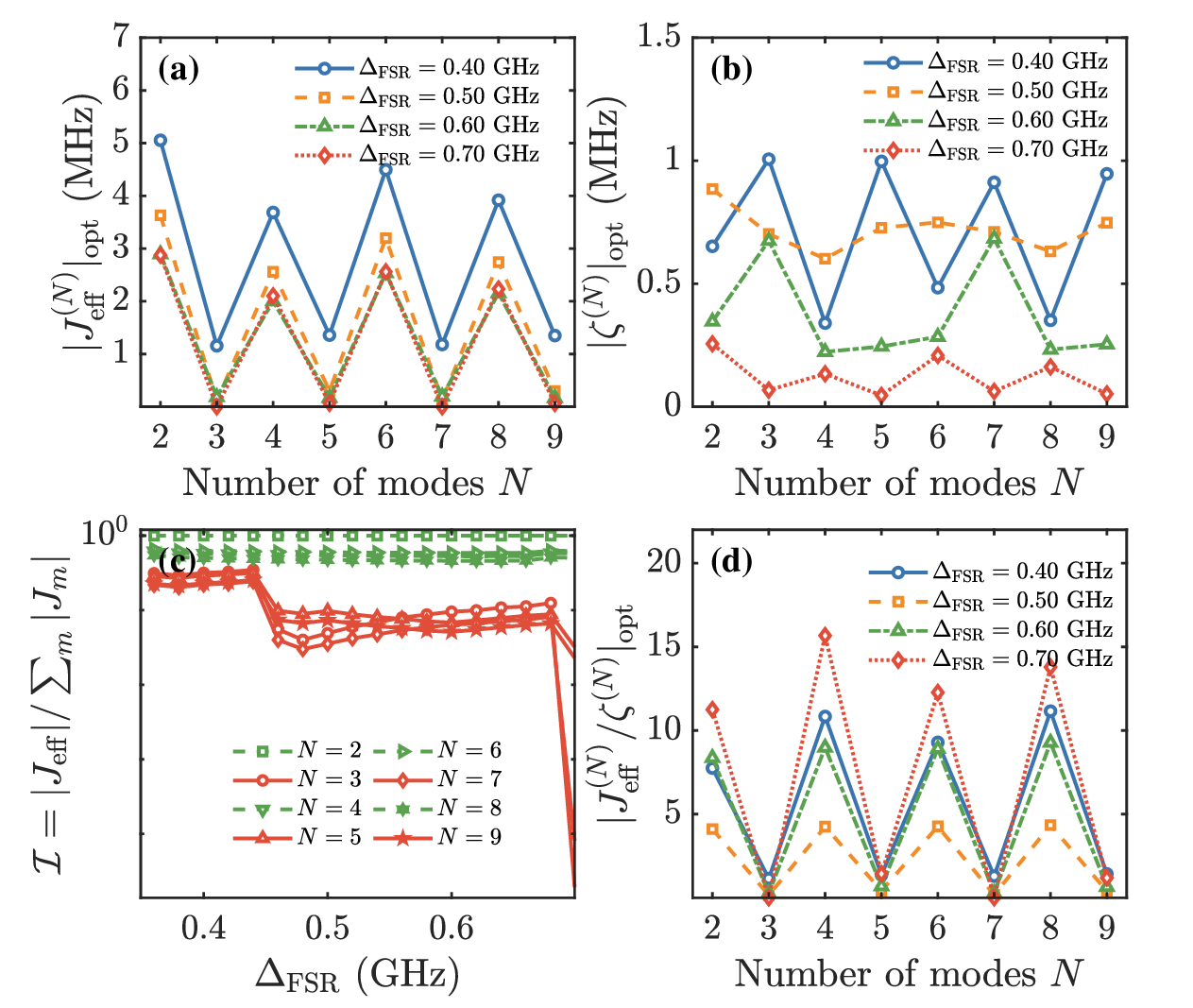}
\caption{
Optimized interaction characteristics in finite multimode transmission-lines.
(a) Optimized \(XY\)  interaction
\(|J_{\rm eff}^{(N)}|_{\rm opt}\).
(b) \(ZZ\)  interaction strength \(|\zeta^{(N)}|_{\rm opt}\) evaluated at
the $XY$  interaction suppressed operating points.
(c)   \(XY\) interaction cancellation factor
\(\mathcal{I}=|J_{\rm eff}|/\sum_m |J_m|\).
(d) Optimized interaction ratio
\(|J_{\rm eff}^{(N)}/\zeta^{(N)}|_{\rm opt}\).
The parameters are the same as in Fig.~\ref{J}.
}
  \label{fig:exchange_suppression_scaling}
\end{figure}

The coupling induced by single- and two waveguide modes shows that the \(XY\) interaction suppression is governed by both the arrangement of qubit frequencies relative to the spectrum of the waveguide modes and the number of the waveguide modes. We further find that a key condition to suppressing the \(XY\) interaction is the presence of a waveguide mode between the two qubit frequencies. We refer to this as the intermediate-mode criterion.  This criterion manifests itself as an odd-even distinction when the qubit frequencies satisfy the condition either in Eq.~(\ref{eq:odd_mode_window}) or in Eq.~(\ref{eq:even_mode_window}).    Within these spectral configurations, there is a central mode that can lie between the two qubits for an odd number of the waveguide modes, whereas two qubit frequencies can only be between two central modes for an even number of the waveguide modes. The odd-even behavior discussed below is therefore a consequence of the frequency allocation of these waveguide modes, rather than a universal property determined solely by the number of waveguide modes.
The  odd-even distinction is directly reflected in the mode-resolved \(XY\)  interaction contributions \(J_m\) defined in Eq.~\eqref{eq:Jeff_and_Jm}. The sign of each \(J_m\)  is determined by both the qubit--mode detunings and the standing-wave parity factor.

 For an even number of coupled waveguide modes with mode assignment as shown in Eq.~(\ref{eq:even_mode_window}), we assume that $\omega_{r,-m} < \omega_1, \,\omega_2 <\omega_{r,m}$. Thus, $A_m$ and $A_{-m}$  have different signs. At the same time, $s_m$ and $s_{-m}$  have different parities. Thus, \(J_{-m}\) and \(J_{m}\) have the same sign and the corresponding \(XY\) interaction strengths induced by the modes $m$ and $-m$ tend to interfere constructively rather than destructively.  We know that $J_{-1}$ and $J_{1}$ are leading terms in the summation for the total $XY$ coupling strength
 \begin{equation}
 J_{\rm eff}^{(2M)}=\sum_{m=1}^{M}(J_{m}+J_{-m})
 \end{equation}
 induced by the $2M$ modes, thus the $XY$ interaction strength is always nonzero. The $ZZ$ interaction strength can be made negligibly small but is nonzero for this even mode configuration. We note that the enhancement of the $XY$ coupling strength  induced by  the symmetric modes $m$ and $-m$ is consistent with the results of a recent study on superconducting qubits coupled remotely via a waveguide~\cite{10.1063/5.0280529},  in which the contributions of two adjacent waveguide  modes to the \(XY\) interaction share the same effective sign, thereby maintaining substantial \(XY\) coupling strength. In the experiments~\cite{10.1063/5.0280529}, an approximately \(ZZ\)-free operating point was obtained through a separate cancellation among higher-excitation contributions. Our analysis on the \(XY\) interaction induced by multimode is more general than the two-mode case in the region of the XY interaction suppression

For an odd number of coupled waveguide modes, the effective \(XY\) interaction induced by these modes can be written as
\begin{equation}
J_{\rm eff}^{(2M+1)}
=
J_0
+
\sum_{m=1}^{M}
\left(
J_{-m}+J_{+m}
\right).
\label{eq:odd_exchange_sum}
\end{equation}
The \(XY\) interaction strength \(J_0\) of the central mode contribution is unpaired, it therefore provides an additional virtual exchange pathway that can compensate the   $XY$ coupling from the $M$ symmetric mode pairs. By appropriately choosing the qubit frequencies relative to the central mode, destructive interference can yield
\(J_{\rm eff}^{(2M+1)}\simeq0\). Importantly, the \(ZZ\)  interaction strength \(\zeta\) is not from the same linear interference condition as for the \(XY\) interaction strength and therefore is nonzero under the condition that the \(XY\) interaction strength is zero. Thus, odd-mode configurations can result in \(|J_{\rm eff}|\ll|\zeta|\), enabling predominantly \(ZZ\) interaction with strong suppression of the $XY$ interaction.
%
%\begin{table}[t]
%    \centering
%      \textcolor{blue}{
%    \caption{Odd--even interference structure within the central qubit-frequency window (\(N=1\) excluded).}
%    \label{tab:odd_even_summary}
%    \setlength{\tabcolsep}{2pt}
%    \renewcommand{\arraystretch}{1.1}
%    \begin{ruledtabular}
%        \begin{tabular}{l| l| l}
%            & Even \(N=2M\) & Odd \(N=2M+1\) \\
%            \hline
%            Mode struct. & Symm. pairs & Pairs + central mode \\     \hline
%            Interf. struct. & constructive & destructive \\     \hline
%            Exch. suppr. & Disfavored  & Possible \\
%        \end{tabular}
%    \end{ruledtabular}
%    }
%\end{table}

The odd-even distinction is shown in Fig.~\ref{fig:exchange_suppression_scaling} by numerically optimized \(XY\) and \(ZZ\) interaction strengths. For each mode configuration, the qubit frequencies are optimized within the dispersive regime to minimize the   \(XY\)  interaction, with the exclusion of single photon, transmon anharmonicity assisted, and two photon resonances.  Figure~\ref{fig:exchange_suppression_scaling}(a) shows that the optimized odd-mode configuration achieves stronger suppression of the \(XY\) interaction than that for even-mode configuration, while retaining a finite \(ZZ\)  interaction, as shown in Fig.~\ref{fig:exchange_suppression_scaling}(b). To distinguish genuine destructive interference from a uniform weakening of the individual exchange channels, we introduce the \(XY\) interaction cancellation factor
\begin{equation}
\mathcal I
=
\frac{|J_{\rm eff}|}{\sum_m |J_m|}.
\label{eq:cancellation_factor}
\end{equation}
A value of $\mathcal I$ approaching unity indicates that the mode-resolved exchange interactions add constructively, whereas \(\mathcal I\ll1\) means strong cancellation in their signed sum. Figure~\ref{fig:exchange_suppression_scaling}(c) shows that the odd-mode configurations exhibit substantially smaller values of \(\mathcal I\) than those for even-mode configurations. The suppression of
\(J_{\rm eff}\) therefore arises not merely from weaker individual exchange channels, but also from destructive interference of these individual exchange channels.
Consistently, the optimized ratio \(|J_{\rm eff}^{(N)}/\zeta^{(N)}|\) displays an odd--even oscillation in Fig.~\ref{fig:exchange_suppression_scaling}(d). Odd-mode configurations can be in the \(ZZ\)-dominated regime \(|J_{\rm eff}|\ll|\zeta|\), whereas even-mode configurations remain \(XY\)  interaction dominated.

These results establish the intermediate-mode criterion as the relevant structural condition for the suppression of the \(XY\) interaction in the finite coupled waveguide modes with different configurations considered here. Within the assumption that qubit-frequencies are around that of the central mode, this criterion manifests itself as an odd-even distinction: odd mode configurations can be used to strongly suppress  the effective \(XY\) interaction without simultaneously eliminating the \(ZZ\) interaction. Even mode configurations cannot be used to suppress the effective \(XY\) interaction, but can make the \(ZZ\) interaction negligibly small.

\begin{figure}[t]
 \centering
 \includegraphics[width=0.5\textwidth]{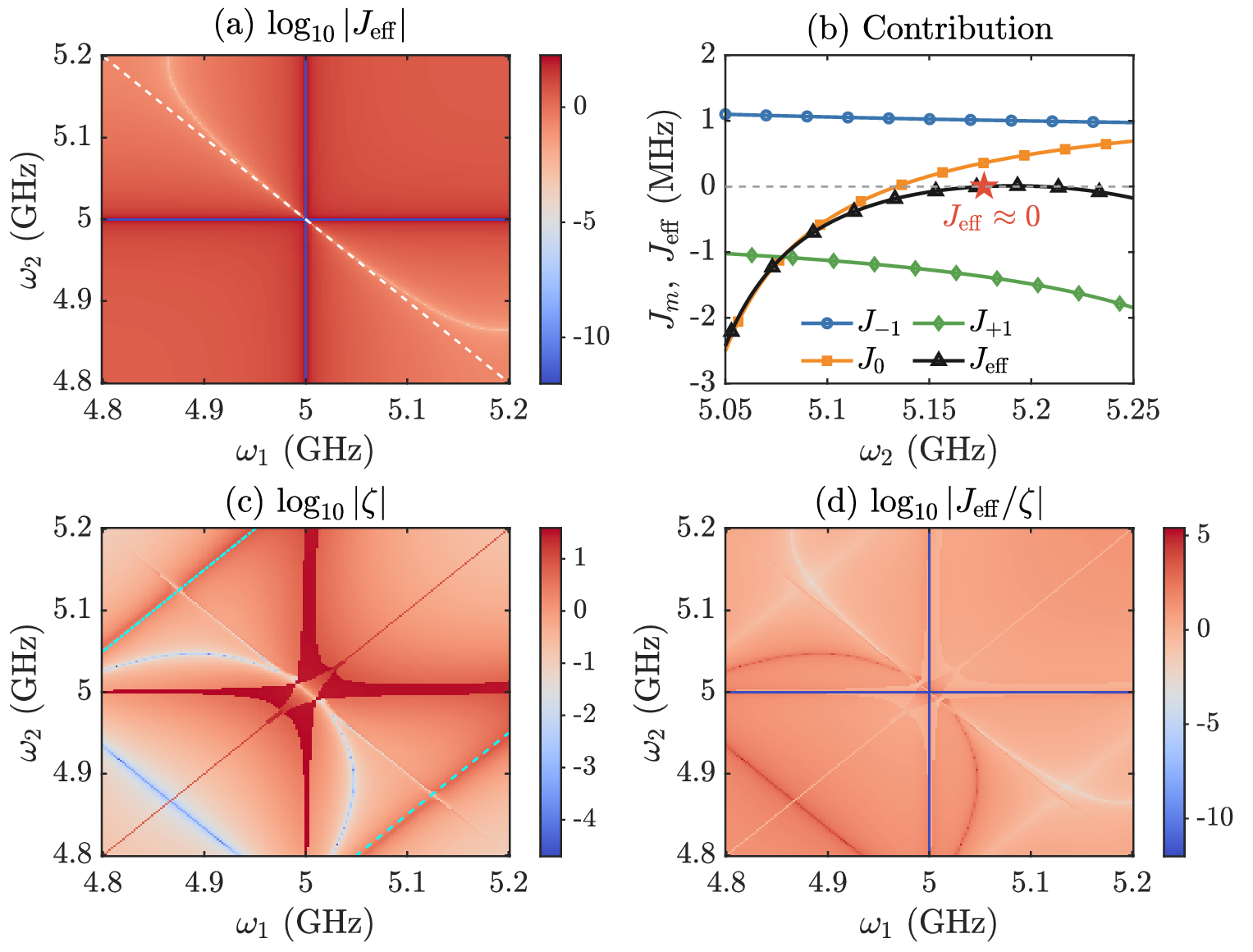}
  \caption{
(a) Effective \(XY\)  interaction \(\log_{10}|J_{\rm eff}|\). The dashed line
indicates the single-mode two-photon resonance condition
\(\omega_1+\omega_2=2\omega_{r,0}\).
(b) Mode-resolved exchange contributions as functions of \(\omega_2\) for
fixed \(\omega_1/2\pi=4.866~{\rm GHz}\).
(c) \(ZZ\)  interaction \(\log_{10}|\zeta|\).
(d) Interaction ratio \(\log_{10}|J_{\rm eff}/\zeta|\).
The parameters are \(\omega_{r,0}/2\pi=5.0~{\rm GHz}\),
\(\Delta_{\rm FSR}=0.4~{\rm GHz}\), \(g_\mu/2\pi=0.015~{\rm GHz}\), and
\(\alpha_\mu/2\pi=-0.25~{\rm GHz}\).
  }
   \label{fig:mode_resolved_J}
\end{figure}

Having established this general intermediate-mode criterion, we now consider the three coupled waveguide modes as an example for the minimal odd number of multimode realization. In this case, the central mode lies between the two qubits and provides the unpaired pathway to construct destructive interference. The resulting strengths for  \(J_{\rm eff}\)  and $ZZ$ interactions are shown in Fig.~\ref{fig:mode_resolved_J}. Figure~\ref{fig:mode_resolved_J}(a) shows a region in which the effective \(XY\) interaction is strongly suppressed. Importantly, this region does not coincide with the single-mode two-photon resonance condition \(\omega_1+\omega_2=2\omega_{r,0}\), showing that the suppression does not originate from resonance-enhanced hybridization. Instead, the mode-resolved decomposition in Fig.~\ref{fig:mode_resolved_J}(b) demonstrates that the small value of \(J_{\rm eff}\) results from the destructive interference in the sum of the $XY$ interaction contributions $J_{-1}, \;J_0$ and $J_1$ induced by three coupled waveguide modes.

Figure~\ref{fig:mode_resolved_J}(c) shows that the \(ZZ\) interaction remains finite in the region of  the $XY$ interaction suppression. This behavior is expected because \(\zeta\) originates from higher-order virtual processes and is not governed by the same linear interference condition as \(J_{\rm eff}\). Consequently, the ratio \(|J_{\rm eff}/\zeta|\) in Fig.~\ref{fig:mode_resolved_J}(d) exhibits very low value in the region of $XY$ interaction suppression. This region is a candidate for which the \(XY\) interaction is strongly suppressed while the \(ZZ\) interaction remains finite.

We emphasize that the odd--even distinction identified above is not a universal mode-number rule, but a consequence of the spectral and parity
structure within the corresponding central finite-mode window. Outside this window, even-mode configurations may also exhibit destructive exchange interference at the level of a truncated model. The essential issue, however, is whether such a truncation remains physically self-consistent. Once a qubit is moved outside the spectral window spanned by the retained modes, additional neighboring modes can become spectrally comparable and may no longer be neglected.
In this sense, the three-mode configuration provides the minimal central-window configuration in which the $XY$ interaction cancellation can be realized while retaining
a natural and self-consistent finite-mode description. The case corresponding to outside spectral window of two modes is analyzed in Appendix~\ref{app:boundary_effect}.

\section{Gate Dynamics}
\label{sec:gate_dynamics}
The frequency-allocation analysis of previous sections identifies two representative interaction regimes within the central finite-mode window. The configuration for even number of waveguide modes retains a sizable \(XY\) interaction and therefore favors SWAP gate operation, whereas the odd configuration can strongly suppress the \(XY\) interaction while preserving a finite \(ZZ\) interaction for controlled phase (CZ) gate operation. We now verify these predictions by simulating representative operating points using the full finite-mode Hamiltonian.

\begin{figure}[t]
 \centering
 \includegraphics[width=0.48\textwidth]{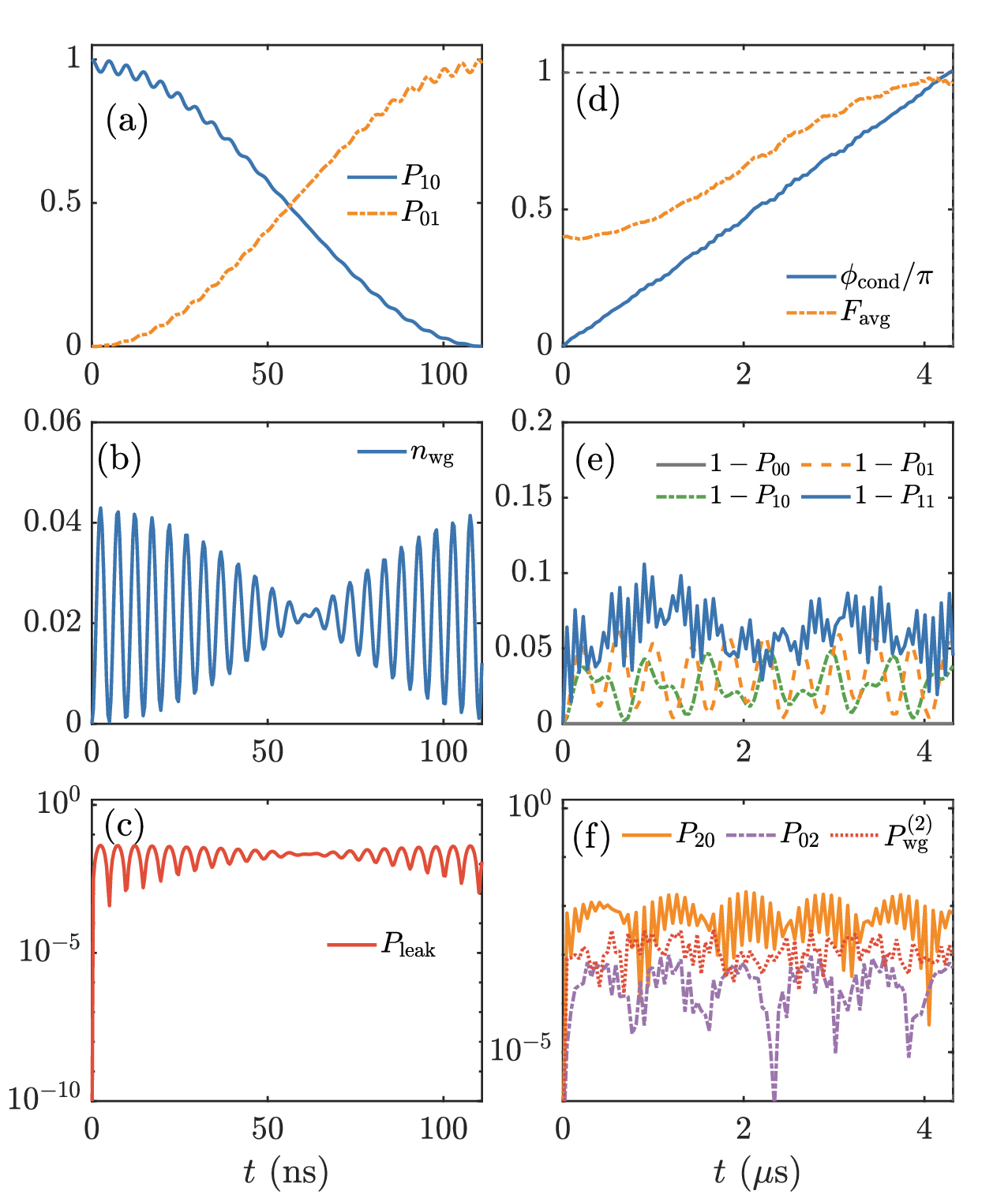}
\caption{
Comparison between \(XY\)-dominated and \(ZZ\)-dominated dynamics in finite multimode transmission-lines. The common parameters are \(\Delta_{\rm FSR}=0.4~{\rm GHz}\), \(g_\mu/2\pi=0.015~{\rm GHz}\), and \(\alpha_\mu/2\pi=-0.25~{\rm GHz}\). The left column, panels (a)--(c), shows the two-mode configuration with \(\omega_{r,1}/2\pi=5.0~{\rm GHz}\) and \(\omega_{r,2}/2\pi=5.4~{\rm GHz}\), at the selected qubit operating point \(\omega_1/2\pi=\omega_2/2\pi=5.2~{\rm GHz}\). The right column, panels (d)--(f), shows the three-mode configuration with \(\omega_{r,m}/2\pi=4.6,\,5.0,\,5.4~{\rm GHz}\), at \(\omega_1/2\pi=5.176~{\rm GHz}\) and \(\omega_2/2\pi=4.866~{\rm GHz}\). (a) Coherent population transfer between the computational states \(|10\rangle\) and \(|01\rangle\).  (b) Total waveguide occupation \(n_{\rm wg}\)  for an initial \(|10\rangle\) state. (c) The leakage out of the single-excitation subspace for two qubits \(P_{\rm leak}=1-P_{10}-P_{01}\). (d) Conditional phase accumulation \(\phi_{\rm cond}\) and the corresponding average gate fidelity \(F_{\rm avg}\). (e) Deviation from the computational basis states during the conditional-phase evolution, shown as \(1-P_{ij}\). (f) Regarding the non-computational evolution channel for the initial state $|11\rangle$, the process involves the population of the Transmon doubly excited states $P_{20}$ and $P_{02}$, as well as the collective two-photon state $P_{\rm wg}^{(2)}$ in the waveguide.
}
\label{fig:gate_dynamics}
\end{figure}

To ensure that the selected working points of qubit frequencies are in the dispersive regime, we require the dressed computational states \( |\widetilde{ij}\rangle \), adiabatically connected to the bare states \( |ij\rangle \), to satisfy
$ \min_{ij}|\langle ij|\widetilde{ij}\rangle|^2
>O_{\rm min}$ and $|\zeta|>\zeta_{\rm min}$
with, e.g.,
\(O_{\rm min}=0.95\)
and
\(\zeta_{\rm min}=0.1~{\rm MHz}\).
These conditions ensure that (i) the computational subspace remains well defined with sufficiently weak hybridization~\cite{larssen2026}, and (ii) the effective interaction is strong enough to guarantee a practical gate speed.
Figure~\ref{fig:gate_dynamics} compares representative gate dynamics for
even- and odd-mode configurations. Panels (a)-(c) show a two-mode system, corresponding to  $XY$  interaction dominated SWAP gate,
whereas panels (d)-(f) show a three-mode system, corresponding to $ZZ$ dominated CZ gate.

For the two-mode configuration, the  frequencies of two waveguide modes are chosen as \(\omega_{r,1}/2\pi=5.0~{\rm GHz}\) and \(\omega_{r,2}/2\pi=5.4~{\rm GHz}\), and the qubit frequencies are chosen as  \(\omega_1/2\pi=\omega_2/2\pi=5.2~{\rm GHz}\). All these parameters are experimentally accessible. The qubits are therefore dispersively detuned from both waveguide modes, suppressing real photon excitations, while the dispersive couplings between the waveguide modes and qubits induce constructive interference for two mode-resolved $XY$ interaction strengths. Consequently, the effective interaction satisfies \(|J_{\rm eff}|\gg|\zeta|\), making the system in the \(XY\)-dominated regime. As shown in Fig.~\ref{fig:gate_dynamics}(a), an excitation initially prepared in \(|10\rangle\) is coherently transferred to \(|01\rangle\), reaching nearly complete population inversion at the optimal SWAP time. The population within the single-excitation computational space remains close to unity throughout the time evolution process. This indicates that the dynamics is well described by an effective \(XY\) interaction. Figure~\ref{fig:gate_dynamics}(b) shows the corresponding occupation of the transmission-line modes. The waveguide modes are only transiently populated and remain far below unity, consistent with a virtual-photon-mediated exchange process. The leakage out of the single-excitation subspace for two qubits, \(P_{\rm leak}=1-P_{10}-P_{01}\), is shown in Fig.~\ref{fig:gate_dynamics}(c). It is less than \textcolor{blue}{ 0.04} and remains small compared to the single-excitation population, confirming that the two-mode configuration naturally supports the SWAP-type gate operation arising  from  \(XY\) interaction induced by the two waveguide modes.

The three-mode configuration displays qualitatively different behavior. In this case, we assume: (i) the work points of the qubit frequencies are chosen from the \(XY\) interaction suppressed region identified in Fig.~\ref{fig:mode_resolved_J}; (ii) the couplings of the qubits to the waveguide modes are in the dispersive regime; (iii)  the dressed computational states retain large overlaps with the corresponding bare states; and (iv) the \(ZZ\) interaction is sufficiently strong for CZ gate operation.

At the selected operating point, \(\omega_1/2\pi=5.176~{\rm GHz}\) and \(\omega_2/2\pi=4.866~{\rm GHz}\), the   \(XY\) interaction is strongly suppressed, \(|J_{\rm eff}|/2\pi\sim10^{-4}~{\rm MHz}\), while the \(ZZ\) interaction remains finite, \(|\zeta|/2\pi\sim0.12~{\rm MHz}\). The resulting ratio \(|J_{\rm eff}/\zeta|\sim10^{-3}\) places the system deep in the \(ZZ\)-dominated regime, \(|J_{\rm eff}|\ll|\zeta|\).
To verify that this selected operating point does not result in a near-resonant enhancement for the $XY$ interaction through the fourth-order perturbation, we further decompose  the fourth-order perturbation into different two-photon processes. As shown in Appendix~\ref{app:conditional_interaction}, the selected operating point avoids the near-resonant enhancement region associated with  \(|20\rangle\), \(|02\rangle\) of the qubit states and two-photon processes of the same waveguide mode and cross waveguide modes.

As shown in Fig.~\ref{fig:gate_dynamics}(d), the conditional phase accumulates nearly linearly and reaches \(\phi_{\rm cond}\simeq\pi\) at the optimal gate time, where the average gate fidelity also reaches its maximum.
The average gate fidelity is evaluated from the computational-subspace propagator after removing the global and single-qubit dynamical phases~\cite{PhysRevLett.117.170502},
\begin{equation}
F_{\rm avg}(t)
=
\frac{
\operatorname{Tr}
\left[
U_{\rm corr}^{\dagger}(t)U_{\rm corr}(t)
\right]
+
\left|
\operatorname{Tr}
\left[
U_{\rm CZ}^{\dagger}U_{\rm corr}(t)
\right]
\right|^2
}
{20},
\label{eq:Favg_main}
\end{equation}
where \(U_{\rm CZ}=\operatorname{diag}(1,1,1,-1)\). Since \(U_{\rm corr}\) is obtained by projecting the full propagator onto the computational subspace without renormalization, this definition includes both coherent gate errors and leakage. It is worth noting that our aim here is not to optimize a fully calibrated experimental CZ gate by using the full Hamiltonian simulation. We only want to verify that the working point selected by the intermediate-mode criterion indeed produces  \(ZZ\) dominated interaction with strongly suppressed $XY$ interaction.

The computational-state preservation in Fig.~\ref{fig:gate_dynamics}(e) shows that the evolution is largely confined to the computational subspace, although residual coherent oscillations are visible in the bare-state populations. The largest deviation occurs for the \(|11\rangle\) state, reflecting its stronger virtual coupling to nearby two-excitation states. Since the simulation uses bare states as the input and readout basis, the part of this oscillation stems from the finite dressing of the computational states. Therefore, the gate performance may be further improved by using a dressed basis, or equivalently by calibrating the gate with respect to the adiabatically connected dressed computational states~\cite{PhysRevApplied.19.054050,PhysRevA.87.022309,PhysRevA.90.022307,yuan2026}.
Figure~\ref{fig:gate_dynamics}(f) shows the leakage to the dominant noncomputational states during the three-mode evolution. The leakage mainly involves the transmon two-excitation states and the two-photon states of the waveguide modes. Among these leakage channels, the population of the noncomputational state \(|20\rangle\) makes a large contribution, consistent with the analysis in Fig.~\ref{fig:zz_channels} of  Appendix~\ref{app:conditional_interaction}, where the \(|20\rangle\)-mediated leakage is the dominant component near the selected working point. Although these populations remain small compared with the computational-state population,   they provide a direct indication of leakage-induced infidelity in the bare-state gate dynamics.

The above two examples illustrate that the $XY$ and $ZZ$  couplings between two remote qubits can be engineered by different configurations of the waveguide modes for realizing different logic gates through dynamical evolutions.  Even-mode configurations naturally favor \(XY\)-dominated dynamics and  are therefore suitable for SWAP-type gate operations. Odd-mode configurations, by contrast, can strongly suppress the \(XY\) interaction while retaining a finite $ZZ$  interaction, thereby enabling CZ gate operations. Furthermore, the dynamical simulations presented here should be regarded as a verification of the coherent interaction mechanism rather than as a gate optimization protocol. A quantitative evaluation of experimental gate performance would also require consideration of superconducting-qubit relaxation, pure dephasing, and photon loss in the transmission-line modes, which are beyond the scope of the present work. Nevertheless, some qualitative conclusions can still be drawn. Qubit relaxation will inevitably reduce the gate fidelity. Moreover, although the dispersive mechanism suppresses the occupation of real photons, the finite photon population in the transmission-line modes means that photon loss can further reduce the fidelity~\cite{PhysRevLett.123.210501,deng2025,7gr4-nk6m}.

\section{Discussion}
\label{sec:discussion}

The results obtained here show that the frequency assignment is a possible way to adjust the effective interactions between two qubits, mediated by a multimode transmission-line waveguide.
An important issue for superconducting circuit implementations is whether the interference mechanism identified in finite-mode models remains valid when additional waveguide modes are included. In a realistic transmission-line waveguide, modes far detuned from the qubits generally contribute weakly because their virtual-photon amplitudes are largely suppressed~\cite{blais2004,blais2021}. The effective qubit-qubit interactions can therefore be determined mainly determined by a finite spectral window around the qubit frequencies, rather than by all waveguide modes equally. The scaling results in Fig.~\ref{fig:exchange_suppression_scaling} further show that the same interference tendency persists when additional neighboring modes are included. Nevertheless, the odd-even distinction identified here applies to the symmetric standing-wave configurations within the central frequency window and should not be interpreted as a universal classification of  multimode waveguides. In practice, the relevant mode frequencies and position-dependent couplings must therefore be taken into account when applying the criterion of the frequency assignment.

For superconducting quantum processors, this  interference mechanism could be combined with tunable qubit frequencies~\cite{yan2018,PhysRevApplied.16.054020} to achieve different types of interactions. For example, by adjusting the qubit frequencies relative to the waveguide modes, the effective interactions could be changed between $XY$-dominated and $ZZ$-dominated regimes, providing a possible route to interaction adjustment without dedicated tunable couplers~\cite{chen2014,sung2021}. However, the present results are obtained for ideal coherent dynamics, whereas practical gate operations are also affected by waveguide loss, qubit relaxation, and residual interactions. Future work may investigate open-system gate fidelities, robustness against mode-frequency disorder and position-dependent couplings~\cite{ls5b-279m},
extensions to multi-qubit interaction engineering and crosstalk suppression, and the crossover from finite standing-wave spectra to continuum waveguide regimes~\cite{PRXQuantum.6.010342,rieck2026,kv28-r6w4}.

%---------------------------------------------------------------------------
\section{Conclusion}
\label{sec:conclusion}

We have investigated the effective $XY$ and $ZZ$ interactions between two transmon qubits induced by a multimode transmission-line waveguide. The $XY$ interaction is determined by the interference sum of mode-resolved virtual-photon exchange contributions with signs, whereas the higher-order $ZZ$ interaction is not governed by the same interference condition. For a single-mode waveguide, the suppression of the $XY$ interaction coincides with a  two-photon resonance, and thus the $XY$ interaction cannot be suppressed in off two-photon resonant case.
For symmetric multimode configurations within the central dispersive window, the frequency assignment, together with the standing-wave parity structure, determine the interference
between different exchange pathways. The resulting odd--even distinction allows strong suppression of the $XY$ interaction in odd-mode configurations while retaining a finite $ZZ$ interaction.
Full-Hamiltonian simulations of minimal two- and three-mode configurations confirm the corresponding $XY$-dominated SWAP and $ZZ$-dominated CZ-type logical gates. These results establish a frequency assignment criterion, determined jointly by the mode configuration and standing-wave parity, for selecting long-range interaction types in finite multimode waveguides. Our study here may be helpful for the quantum network connected by waveguides  and the quantum computation based waveguide couplers.

\section{acknowledgments}
D.W. is supported by the funding from China Postdoctoral Science Foundation (2025M773397).	
Y.X is supported by the National Natural Science Foundation of China under Grant No. 92365209 and Beijing Key Laboratory of Quantum AI.

\section{Appendix}
\appendix
%---------------------------------------------------------------------------
\section{Leading-Order Channel Interpretation of the \(ZZ\) Interaction}
\label{app:conditional_interaction}

In this Appendix, we analyze the virtual processes contributing
to the effective \(ZZ\) interaction using perturbation theory.
We first derive the second-order exchange interaction and then
calculate the fourth-order energy corrections associated with
different virtual transitions. In particular, we consider
processes involving doubly excited transmon states and
same- and cross-mode two-photon states of the waveguide.

The total Hamiltonian is
\begin{equation}
H=H_0+V,
\end{equation}
where
\begin{equation}
H_0=
\sum_{\mu=1,2}
\left[
\omega_\mu b_\mu^\dagger b_\mu
+
\frac{\alpha_\mu}{2}
b_\mu^{\dagger2}b_\mu^2
\right]
+
\sum_m\omega_{r,m}a_m^\dagger a_m,
\end{equation}
and
\begin{equation}
V=
\sum_{\mu,m}
g_{\mu m}
\left(
b_\mu^\dagger a_m+
a_m^\dagger b_\mu
\right).
\end{equation}
Here, \(g_{\mu m}\) is taken to be real.
We define the detuning
\(\Delta_{\mu m}=\omega_\mu-\omega_{r,m}\)
and assume the dispersive regime
\(|g_{\mu m}|\ll|\Delta_{\mu m}|\).
The relevant higher-excitation energy differences
are also assumed to be sufficiently large for
nondegenerate perturbation theory.

%---------------------------------------------------------------------------
\subsection{Second-order exchange interaction}

The Schrieffer--Wolff (SW) transformation~\cite{Schrieffer1966}
is defined by
\begin{equation}
H_{\rm eff}=e^{-S}He^S,
\end{equation}
where \(S=-S^\dagger\).
To eliminate the qubit--waveguide coupling to first order,
the generator satisfies
\(
V+[H_0,S]=0.
\)
For the lowest transmon transition, the corresponding
part of the generator is
\begin{equation}
S_{01}
=
-\sum_{\mu,m}
\frac{g_{\mu m}}{\Delta_{\mu m}}
\left(
|1\rangle_\mu\langle0|a_m
-
a_m^\dagger|0\rangle_\mu\langle1|
\right).
\end{equation}
For transitions involving higher transmon levels,
the energy denominators depend on the corresponding
transition frequencies. In particular, the
\(1\leftrightarrow2\) transition involves
\(\Delta_{\mu m}+\alpha_\mu\).

Using the Baker--Campbell--Hausdorff
expansion~\cite{Sakurai2017}, we obtain
\begin{equation}
H_{\rm eff}
=
H+[H,S]
+\frac12[[H,S],S]+\cdots.
\end{equation}
With the first-order condition, this becomes
\begin{equation}
H_{\rm eff}
=
H_0+\frac12[V,S]+O(g^3).
\end{equation}
Projecting the second-order contribution onto
the zero-photon single-excitation subspace gives
the effective exchange interaction
\begin{equation}
H_{\rm ex}
=
J_{\rm eff}
\left(
b_1^\dagger b_2+b_2^\dagger b_1
\right),
\end{equation}
where the operators are restricted to the
single-excitation subspace and
\begin{equation}
J_{\rm eff}
=
\sum_m
\frac{g_{1m}g_{2m}}{2}
\left(
\frac1{\Delta_{1m}}
+
\frac1{\Delta_{2m}}
\right).
\label{eq:Jeff_appendix}
\end{equation}
This interaction originates from second-order
virtual single-photon exchange between
\(|10;0\rangle\) and \(|01;0\rangle\).

%---------------------------------------------------------------------------
\subsection{Fourth-order energy corrections}

To calculate the effective $ZZ$ interaction
analytically, we employ nondegenerate
 perturbation theory.
For a nondegenerate bare state $|i\rangle$,
the corresponding energy can be expanded as
\begin{equation}
E_i
=
E_i^{(0)}
+
E_i^{(2)}
+
E_i^{(4)}
+\cdots.
\label{eq:app_energy_expansion}
\end{equation}
Here, the odd-order corrections vanish
for the present rotating-wave Hamiltonian,
since each interaction changes the
waveguide photon number by one.

The second-order energy correction is
\begin{equation}
E_i^{(2)}
=
\sum_{k\neq i}
\frac{|V_{ik}|^2}
{E_i^{(0)}-E_k^{(0)}},
\label{eq:app_second_order_energy}
\end{equation}
where $V_{ik}=\langle i|V|k\rangle$.
For the computational bare states,
the second-order corrections are
$E_{10}^{(2)}=\sum_m g_{1m}^2/\Delta_{1m}$,
$E_{01}^{(2)}=\sum_m g_{2m}^2/\Delta_{2m}$,
and
$E_{11}^{(2)}=\sum_m
(g_{1m}^2/\Delta_{1m}
+g_{2m}^2/\Delta_{2m})$.
Since the vacuum state $|00;\mathbf{0}\rangle$
is decoupled from the other states,
$E_{00}^{(2)}=0$.
Consequently, the second-order contribution
to the $ZZ$ interaction vanishes,
\begin{equation}
\begin{aligned}
\zeta^{(2)}
&=
E_{11}^{(2)}
-
E_{10}^{(2)}
-
E_{01}^{(2)}
+
E_{00}^{(2)}
=0.
\end{aligned}
\label{eq:app_zz_second_order}
\end{equation}
Thus, the second-order energy corrections
do not generate a conditional energy shift.

The effective $ZZ$ interaction between
superconducting qubits has been extensively
studied in coupled-qubit and multimode-bus
architectures~\cite{DiCarlo2009,Zhao2023}.
Here, the leading nonzero contribution
arises at fourth order.
For a nondegenerate state $|i\rangle$,
the fourth-order energy correction
is~\cite{PhysRevB.87.024510,PhysRevApplied.16.024037}
\begin{equation}
\begin{aligned}
E_i^{(4)}
={}&
\sum_{\substack{k,l,n\\k,l,n\neq i}}
\frac{
V_{ik}V_{kl}V_{ln}V_{ni}
}{
(E_i^{(0)}-E_k^{(0)})
(E_i^{(0)}-E_l^{(0)})
(E_i^{(0)}-E_n^{(0)})
}
\\
&-
\sum_{\substack{k,l\\k,l\neq i}}
\frac{
|V_{ik}|^2|V_{il}|^2
}{
(E_i^{(0)}-E_k^{(0)})^2
(E_i^{(0)}-E_l^{(0)})
}.
\end{aligned}
\label{eq:app_fourth_order_energy}
\end{equation}
The first term describes four-step virtual
transitions through intermediate states,
whereas the second term accounts for
the lower-order energy corrections.

The fourth-order $ZZ$ interaction is
obtained by combining the energy corrections
to the computational states,
\begin{equation}
\begin{aligned}
\zeta^{(4)}
&=
E_{11}^{(4)}
-
E_{10}^{(4)}
-
E_{01}^{(4)}
+
E_{00}^{(4)}
\\
&=
E_{11}^{(4)}
-
E_{10}^{(4)}
-
E_{01}^{(4)},
\end{aligned}
\label{eq:app_zz_fourth_definition}
\end{equation}
where $E_{00}^{(4)}=0$ because
$|00;\mathbf{0}\rangle$ is decoupled
from the other states.

For the bare state $|11;\mathbf{0}\rangle$,
the virtual transitions in the first term
of Eq.~\eqref{eq:app_fourth_order_energy}
can be classified according to the
two-excitation intermediate states
\begin{equation}
|20;\mathbf{0}\rangle,\quad
|02;\mathbf{0}\rangle,\quad
|00;2_m\rangle,\quad
|00;1_m1_n\rangle
\quad (m<n).
\label{eq:app_virtual_states}
\end{equation}
The first two states correspond to
doubly excited transmon states,
whereas the latter two describe
two photons occupying the same or
different waveguide modes.
In the following, we evaluate the
corresponding fourth-order contributions
separately.

%---------------------------------------------------------------------------
\subsection{Virtual transitions through doubly excited transmon states}

We first consider the virtual transitions
involving \(|20;0\rangle\) and \(|02;0\rangle\).
For the \(|20;0\rangle\) state, a representative
fourth-order process is
\begin{equation}
|11;0\rangle
\rightarrow
|10;1_m\rangle
\rightarrow
|20;0\rangle
\rightarrow
|10;1_n\rangle
\rightarrow
|11;0\rangle.
\label{eq:app_path_20}
\end{equation}
The corresponding matrix elements are
\(\langle10;1_m|V|11;0\rangle=g_{2m}\) and
\(\langle20;0|V|10;1_m\rangle=\sqrt{2}g_{1m}\).
The remaining two matrix elements follow from
Hermitian conjugation.
The relevant energy differences are
\(E_{11}^{(0)}-E_{10;1_m}^{(0)}=\Delta_{2m}\) and
\(E_{11}^{(0)}-E_{20}^{(0)}
=\omega_2-\omega_1-\alpha_1\).
Substituting these expressions into
Eq.~\eqref{eq:app_fourth_order_energy},
we obtain
\begin{equation}
\begin{aligned}
E_{20}^{(4)}
&=
\sum_{m,n}
\frac{
2g_{1m}g_{2m}g_{1n}g_{2n}
}{
\Delta_{2m}\Delta_{2n}
(\omega_2-\omega_1-\alpha_1)
}
\\
&=
\frac{2}{\omega_2-\omega_1-\alpha_1}
\left(
\sum_m\frac{g_{1m}g_{2m}}{\Delta_{2m}}
\right)^2.
\end{aligned}
\label{eq:zeta20}
\end{equation}
This contribution depends on the energy
difference between \(|11;0\rangle\) and
\(|20;0\rangle\), which vanishes at the
bare resonance condition \(\omega_2-\omega_1\simeq\alpha_1\).

Similarly, the virtual process involving
\(|02;0\rangle\) can be written as
\begin{equation}
E_{02}^{(4)}
=
\frac{2}{\omega_1-\omega_2-\alpha_2}
\left(
\sum_m\frac{g_{1m}g_{2m}}{\Delta_{1m}}
\right)^2.
\label{eq:zeta02}
\end{equation}
This contribution is associated with
the bare resonance condition
\(\omega_1-\omega_2\simeq\alpha_2\).
Both contributions contain coherent sums
over the waveguide modes. Their strengths
therefore depend on the qubit--mode detunings
and the standing-wave parity structure.
The perturbative expressions are valid
only away from the corresponding resonances.

%---------------------------------------------------------------------------
\subsection{Two-photon virtual transitions}
\label{subsec:two_photon_virtual_transitions}

We next consider fourth-order virtual processes involving
two photons in the waveguide modes. These processes can
be divided into same-mode and cross-mode two-photon transitions.

For the same-mode case, the two photons occupy
the same waveguide mode, corresponding to the state
\(|00;2_m\rangle\). A representative virtual process is
\begin{equation}
|11;0\rangle
\rightarrow
|01;1_m\rangle
\rightarrow
|00;2_m\rangle
\rightarrow
|10;1_m\rangle
\rightarrow
|11;0\rangle.
\label{eq:app_same_path}
\end{equation}
The intermediate single-photon states are
\(|01;1_m\rangle\) and \(|10;1_m\rangle\),
with energy differences
\(E_{11}^{(0)}-E_{01;1_m}^{(0)}=\Delta_{1m}\) and
\(E_{11}^{(0)}-E_{10;1_m}^{(0)}=\Delta_{2m}\),
respectively. The energy difference involving
the two-photon state is
\(E_{11}^{(0)}-E_{00;2_m}^{(0)}
=\Delta_{1m}+\Delta_{2m}\).
The matrix elements involving the second photon
contain a factor of \(\sqrt{2}\) due to
bosonic enhancement.

Summing over the intermediate single-photon states,
the fourth-order contribution is
\begin{equation}
E_{\rm same}^{(4)}
=
\sum_m
\frac{2g_{1m}^2g_{2m}^2}
{\Delta_{1m}+\Delta_{2m}}
\left(
\frac{1}{\Delta_{1m}}
+
\frac{1}{\Delta_{2m}}
\right)^2.
\label{eq:zeta_same}
\end{equation}
The corresponding bare two-photon degeneracy condition is
\(\omega_1+\omega_2\simeq2\omega_{r,m}\).
However, the energy correction also depends on
the interference between the virtual pathways,
and a small energy denominator does not
necessarily lead to a large contribution.

For the cross-mode case, the two photons occupy
different waveguide modes, corresponding to
\(|00;1_m1_n\rangle\) with \(m<n\).
These states are connected to \(|11;0\rangle\)
through the intermediate single-photon states
\(|01;1_m\rangle\), \(|10;1_m\rangle\),
\(|01;1_n\rangle\), and \(|10;1_n\rangle\).
The corresponding fourth-order contribution is
\begin{equation}
E_{\rm cross}^{(4)}
=
\sum_{m<n}
\frac{
|\Lambda_{mn}^{\rm cross}|^2
}{
\omega_1+\omega_2-\omega_{r,m}-\omega_{r,n}
},
\label{eq:zeta_cross}
\end{equation}
where
\(\Lambda_{mn}^{\rm cross}
=g_{1m}g_{2n}(1/\Delta_{1m}+1/\Delta_{2n})
+g_{1n}g_{2m}(1/\Delta_{1n}+1/\Delta_{2m})\).
Unlike the same-mode case, no bosonic enhancement
factor arises because the photons occupy different
modes. The corresponding bare two-photon degeneracy
condition is
\(\omega_1+\omega_2\simeq\omega_{r,m}+\omega_{r,n}\).

Both contributions depend on the waveguide-mode
frequencies and coupling strengths, as well as
the interference between different virtual pathways.

%---------------------------------------------------------------------------
\subsection{Remaining fourth-order energy corrections}
To calculate these contributions, we define
\begin{equation}
B_\mu
=
\sum_m\frac{g_{\mu m}^2}{\Delta_{\mu m}},
\quad
A_\mu
=
\sum_m\frac{g_{\mu m}^2}{\Delta_{\mu m}^2},
\quad
C_\mu
=
\sum_m
\frac{g_{1m}g_{2m}}{\Delta_{\mu m}}
\label{eq:app_AB}
\end{equation}
For the state \(|11;0\rangle\), the
second term of the fourth-order
energy correction is
\begin{equation}
\begin{aligned}
E_{11,\rm counter}^{(4)}
={}&
-
\sum_{k,l\neq11}
\frac{
|V_{11,k}|^2|V_{11,l}|^2
}{
(E_{11}^{(0)}-E_k^{(0)})^2
(E_{11}^{(0)}-E_l^{(0)})
}
\\
={}&
-(A_1+A_2)(B_1+B_2).
\end{aligned}
\label{eq:app_counter_11}
\end{equation}
For the state \(|10;0\rangle\), the
fourth-order correction includes
virtual transitions through \(|01;0\rangle\),
\begin{equation}
|10;0\rangle
\rightarrow
|00;1_m\rangle
\rightarrow
|01;0\rangle
\rightarrow
|00;1_n\rangle
\rightarrow
|10;0\rangle.
\end{equation}
The corresponding path contribution is
\begin{equation}
E_{10 }^{(4)}
=
\frac{C_1^2}{\omega_1-\omega_2}.
\end{equation}
Including the fourth-order counterterm gives
\begin{equation}
E_{10}^{(4)}
=
\frac{C_1^2}{\omega_1-\omega_2}
-A_1B_1.
\label{eq:app_E10_fourth}
\end{equation}
Similarly, for the state \(|01;0\rangle\),
\begin{equation}
E_{01}^{(4)}
=
\frac{C_2^2}{\omega_2-\omega_1}
-A_2B_2.
\label{eq:app_E01_fourth}
\end{equation}
Since \(|00;0\rangle\) is the exact vacuum
eigenstate, \(E_{00}^{(4)}=0\).

These corrections are required when calculating
the conditional energy shift, even though they
do not correspond to additional
two-excitation intermediate states.

%---------------------------------------------------------------------------
\subsection{Fourth-order \(ZZ\) interaction}

Collecting the four classes of virtual
transitions and the counterterm, the
fourth-order energy correction of
\(|11;0\rangle\) is
\begin{equation}
\begin{aligned}
E_{11}^{(4)}
={}&
E_{20}^{(4)}
+
E_{02}^{(4)}
+
E_{\rm same}^{(4)}
+
E_{\rm cross}^{(4)}
+
E_{11,\rm counter}^{(4)}.
\end{aligned}
\label{eq:app_E11_total}
\end{equation}

Substituting the above results into  Eq.~\eqref{eq:app_zz_fourth_definition} gives
\begin{equation}
\begin{aligned}
\zeta^{(4)}
={}&
E_{20}^{(4)}
+
E_{02}^{(4)}
+
E_{\rm same}^{(4)}
+
E_{\rm cross}^{(4)}
-
A_1B_2-\\
&A_2B_1
+
\frac{C_2^2-C_1^2}{\omega_1-\omega_2}.
\end{aligned}
\label{eq:app_zz_all_channels}
\end{equation}

Equation~\eqref{eq:app_zz_all_channels}
contains the contributions from the
four classes of virtual transitions,
together with the remaining fourth-order
energy corrections.

Using the relations
\(\Delta_{1m}-\Delta_{2m}=\omega_1-\omega_2\)
and
\(\Delta_{1m}+\Delta_{2n}
=\Delta_{1n}+\Delta_{2m}\),
the same- and cross-mode contributions satisfy
\begin{equation}
\begin{aligned}
&E_{\rm same}^{(4)}
+E_{\rm cross}^{(4)}
-A_1B_2-A_2B_1=
\frac{C_2^2-C_1^2}{\omega_1-\omega_2}.
\end{aligned}
\label{eq:app_photon_identity}
\end{equation}
Substituting this identity into
Eq.~\eqref{eq:app_zz_all_channels}
gives
\begin{equation}
\begin{aligned}
\zeta^{(4)}
={}&
2C_2^2
\left(
\frac1{\omega_2-\omega_1-\alpha_1}
-
\frac1{\omega_2-\omega_1}
\right)
\\
&+
2C_1^2
\left(
\frac1{\omega_1-\omega_2-\alpha_2}
-
\frac1{\omega_1-\omega_2}
\right).
\end{aligned}
\label{eq:app_zz_simplified}
\end{equation}
This expression shows that the net
fourth-order \(ZZ\) interaction depends
on the transmon anharmonicities,
qubit frequencies, and the multimode
waveguide couplings and detunings.

In contrast, the second-order exchange
interaction is
\begin{equation}
J_{\rm eff}=\frac{C_1+C_2}{2}.
\end{equation}
Consequently, the exchange-cancellation
condition \(J_{\rm eff}=0\) does not
generally imply \(\zeta=0\).
The two interactions depend on different
combinations of the mode-resolved
virtual-photon contributions.
The preceding nondegenerate expressions
are applicable away from the corresponding
bare energy degeneracies.
Near such conditions, the relevant states
must be treated using degenerate perturbation
theory or direct diagonalization of
the full Hamiltonian.

%---------------------------------------------------------------------------
\begin{figure}[t]
\centering
\includegraphics[width=0.48\textwidth]{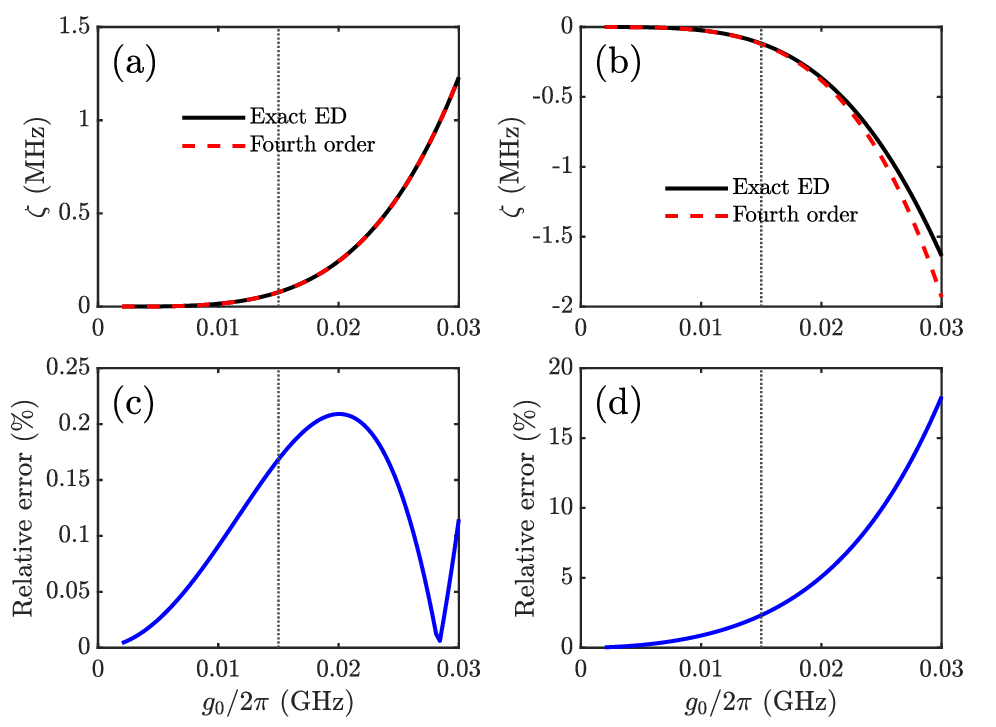}
\caption{
Comparison of the fourth-order perturbative \(ZZ\) interaction
with exact diagonalization (ED).
Panels (a) and (b) show the \(ZZ\) interaction strength
as a function of the qubit--waveguide coupling \(g_0\)
for the two-mode and three-mode configurations, respectively.
The black solid curves are obtained from ED,
and the red dashed curves represent the fourth-order
perturbative results.
Panels (c) and (d) show the corresponding relative errors,
defined as
\(\left|\zeta^{(4)}-\zeta_{\rm ED}\right|/
\left|\zeta_{\rm ED}\right|\times100\%\).
The vertical dotted lines indicate
\(g_0/2\pi=0.015~{\rm GHz}\).
The waveguide-mode frequencies are
\(\omega_{r,m}/2\pi=(4.8,5.2)~{\rm GHz}\)
for the two-mode configuration and
\(\omega_{r,m}/2\pi=(4.6,5.0,5.4)~{\rm GHz}\)
for the three-mode configuration.
The qubit frequencies are
\(\omega_1/2\pi=\omega_2/2\pi=5.0~{\rm GHz}\)
for the two-mode configuration and \(\omega_1/2\pi=5.176~{\rm GHz}\),
\(\omega_2/2\pi=4.866~{\rm GHz}\)
for the three-mode configuration.
The transmon anharmonicities are
\(\alpha_1/2\pi=\alpha_2/2\pi=-0.25~{\rm GHz}\).
}
\label{fig:zz_perturbation_validation}
\end{figure}

To verify the fourth-order perturbative expression,
we compare the analytical \(ZZ\) interaction
with that obtained by exact diagonalization
of the full Hamiltonian.
As shown in Fig.~\ref{fig:zz_perturbation_validation},
the fourth-order results agree well with
the exact results for both the two-mode
and three-mode configurations.
At \(g_0/2\pi=0.015~{\rm GHz}\),
the relative errors are approximately
\(0.17\%\) and \(2.32\%\), respectively.
The deviations become more pronounced
as the coupling strength increases,
indicating the growing importance of
higher-order corrections.
For the two-mode configuration,
the qubits are resonant, and the two
single-excitation eigenenergies are used
together to evaluate the conditional
energy shift.
The corresponding fourth-order result
is obtained from the resonant limit
of Eq.~\eqref{eq:app_zz_simplified}.
These comparisons confirm that the
fourth-order perturbative expression
provides an accurate description of
the effective \(ZZ\) interaction
in the weak-coupling regime considered here.

We now examine the virtual-channel contributions
for the three-mode configuration used in the
gate-dynamics simulation.
The purpose of this analysis is to identify
the virtual processes associated with the
fourth-order energy corrections and to examine
whether the selected operating point is close
to the corresponding resonances.
We consider the four path contributions
\(E_{20}^{(4)}\),
\(E_{02}^{(4)}\),
\(E_{\rm same}^{(4)}\), and
\(E_{\rm cross}^{(4)}\).
These quantities characterize the
individual classes of four-step virtual
transitions contributing to \(E_{11}^{(4)}\).
\begin{figure}[t]
\centering
\includegraphics[width=0.47\textwidth]{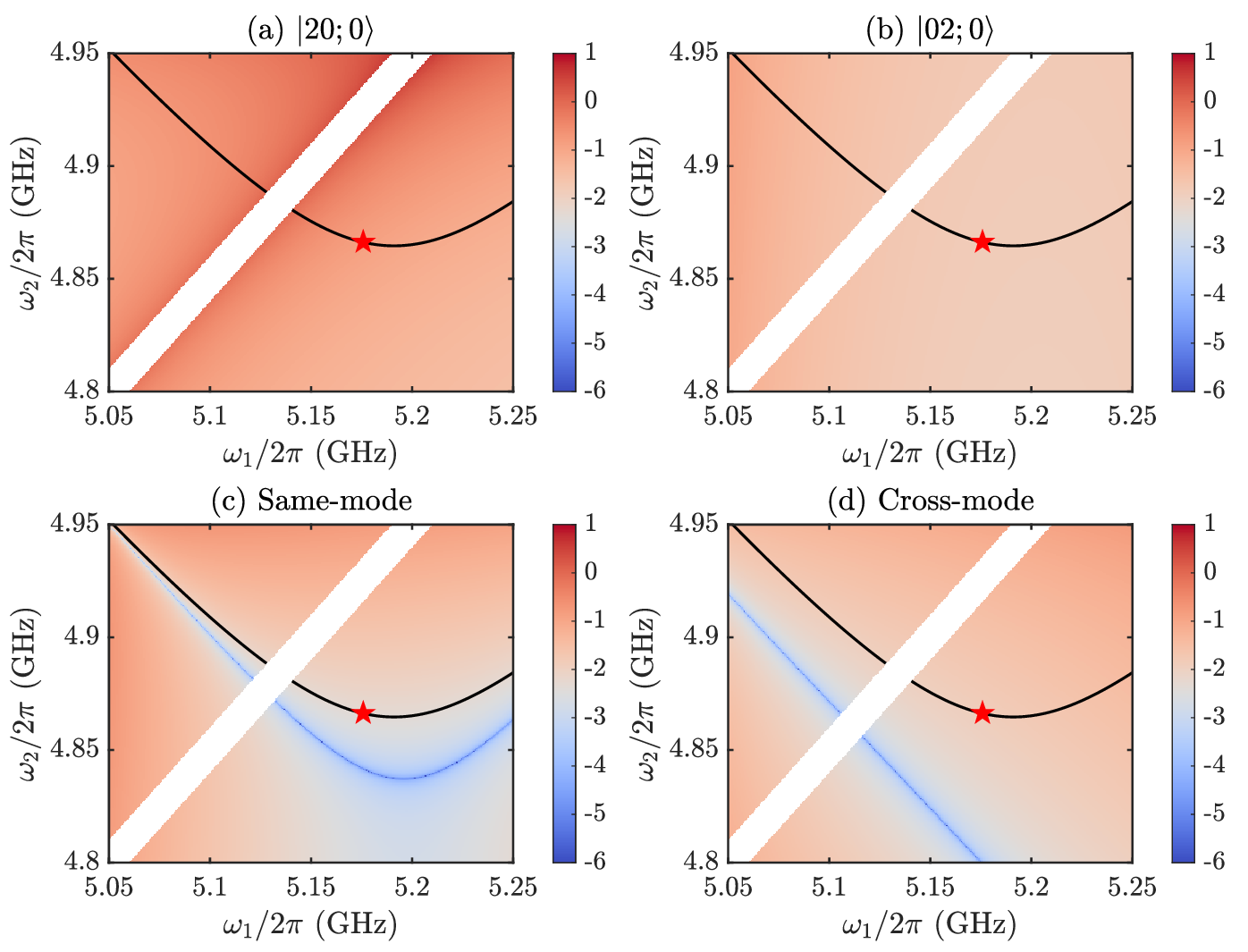}
\caption{
Fourth-order virtual-channel contributions
in the three-mode configuration.
The common parameters are
\(\omega_{r,0}/2\pi=5.0~{\rm GHz}\),
\(\Delta_{\rm FSR}=0.40~{\rm GHz}\),
\(g_\mu/2\pi=0.015~{\rm GHz}\), and
\(\alpha_1/2\pi=\alpha_2/2\pi=-0.25~{\rm GHz}\).
Panels (a)--(d) show the contributions
associated with \(|20;0\rangle\),
\(|02;0\rangle\), same-mode two-photon states,
and cross-mode two-photon states, respectively.
The black curve denotes the second-order
exchange-cancellation condition
\(J_{\rm eff}=0\), and the red star marks
the selected operating point
\(\omega_2/2\pi=4.866~{\rm GHz}\) and
\(\omega_1/2\pi=5.176~{\rm GHz}\).
}
\label{fig:zz_channels}
\end{figure}
Figure~\ref{fig:zz_channels} compares the
four classes of fourth-order virtual transitions.
Panels (a) and (b) correspond to the
doubly excited transmon states,
whereas panels (c) and (d) correspond
to the same- and cross-mode
two-photon states.
The white regions indicate parameter
regions excluded from the perturbative
analysis because the corresponding
energy differences become small.

The selected operating point lies in the region where the \(XY\) interaction is suppressed and
away from these bare degeneracy conditions.
Thus, the suppression of the second-order
\(XY\) interaction is not obtained by
tuning the qubits close to a particular
higher-excitation resonance.
The channel analysis illustrates how
doubly excited transmon states and
multimode two-photon states contribute
to the fourth-order energy corrections.
The complete \(ZZ\) interaction,
however, is determined by the
conditional energy shift of the
dressed computational states in
Eq.~\eqref{eq:app_zz_simplified}.

\section{$XY$ interaction Cancellation Outside the Central Frequency Window}
\label{app:boundary_effect}
\begin{figure}[t]
    \centering
    \includegraphics[width=0.45\textwidth]{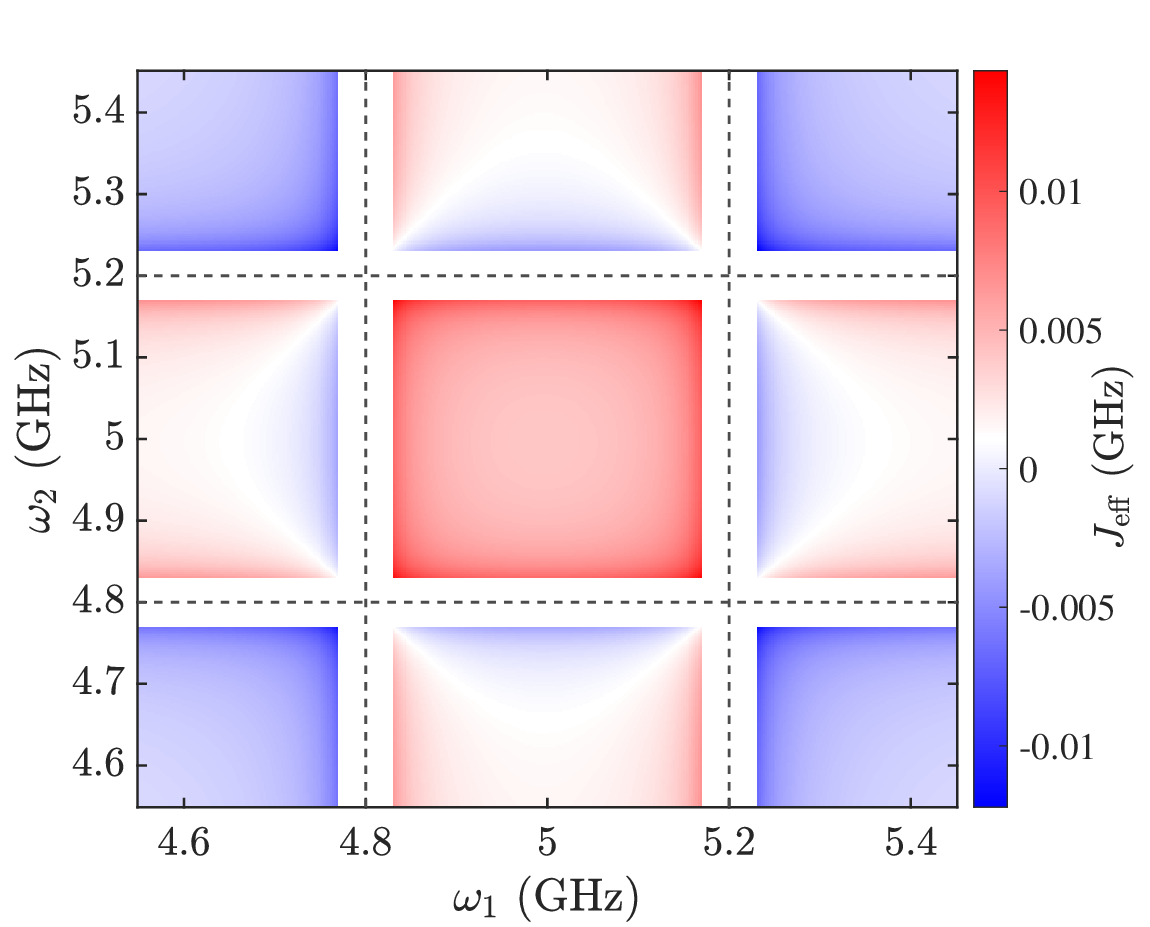}
    \caption{
 $XY$   interaction \(J_{\rm eff}\) as a function of the two
    qubit frequencies for the two waveguide modes. The calculation uses two neighboring waveguide modes
    with frequencies \(\omega_{r,1}=4.8~\mathrm{GHz}\) and
    \(\omega_{r,2}=5.2~\mathrm{GHz}\), with coupling strength
    \(g_0=20~\mathrm{MHz}\).
    The qubit frequencies are varied within
    \(4.55\leq\omega_{1,2}\leq5.45~\mathrm{GHz}\).
    The dashed lines indicate the retained waveguide mode frequencies.
    }
    \label{fig:boundary_effect}
\end{figure}
The odd--even distinction discussed in the main text applies to qubit
frequencies assigned within the corresponding central finite-mode window.
It does not imply a mathematical prohibition of $XY$ interaction cancellation for
an arbitrary two-mode truncation. Here we show that even-mode
configurations can exhibit destructive exchange interference once the
qubit frequencies are moved outside the central window. The relevant
criterion is therefore the spectral arrangement of the qubits relative
to the retained waveguide modes rather than the mode number alone.

We consider two neighboring waveguide modes with frequencies
\(\omega_{r,1}<\omega_{r,2}\) and opposite standing-wave parity factors
\(s_2=-s_1\). The mode-resolved exchange interaction can be written as
\begin{equation}
J_m
=
\frac{g_{1m}g_{2m}}{2}A_m ,
\end{equation}
where
\(A_m=1/\Delta_{1m}+1/\Delta_{2m}\) and
\(\Delta_{\mu m}=\omega_\mu-\omega_{r,m}\).
For symmetry-related qubit positions, the parity structure of neighboring
standing-wave modes gives
\(g_{12}g_{22}=-g_{11}g_{21}\),
so that the relative sign between the two exchange pathways is determined
by the competition between the parity factors and the detuning-dependent
terms \(A_m\).
Within the central two-mode window,
\(\omega_{r,1}<\omega_1,\omega_2<\omega_{r,2}\),
the lower mode lies below both qubits while the upper mode lies above both
qubits. Consequently,
\(A_1>0\) and \(A_2<0\).
The opposite signs of \(A_m\) compensate the parity reversal between
neighboring modes, \(s_2=-s_1\), resulting in
\(\operatorname{sgn}(J_1)=\operatorname{sgn}(J_2)\).
Therefore, the two mode-resolved exchange pathways interfere
constructively rather than destructively. The corresponding sign structure
is summarized in Table~\ref{table2}.

Outside the central frequency window, this sign constraint is removed.
The factors \(A_m\) are no longer restricted to fixed signs because the
qubit frequencies can be located on different sides of a retained mode.
Therefore, the parity-induced sign relation between neighboring modes does
not necessarily enforce constructive interference. The total $XY$
interaction is given by \(J_{\rm eff}=J_1+J_2\), and can consequently vanish
for suitable frequency arrangements.
To illustrate this point, Fig.~\ref{fig:boundary_effect} shows the
$XY$ interaction strength in the full qubit-frequency parameter space.
The white boundary corresponds to the cancellation condition
\(J_{\rm eff}=0\), demonstrating that destructive interference can indeed
occur outside the central two-mode window.

Such cancellation points, however, should not be interpreted as a
counterexample to the central-window odd--even behavior discussed in the
main text. Once a qubit frequency is moved outside the spectral interval
bounded by the retained modes, additional neighboring waveguide modes can
become spectrally comparable and contribute to the effective interaction.
Therefore, retaining only the two selected modes no longer follows a
uniform nearest-mode truncation criterion.
The three-mode configuration avoids this ambiguity. Both qubit frequencies
remain within the spectral interval, and the
neighboring modes responsible for the interference are explicitly included.
In this sense, the three-mode case represents the minimal central-window
configuration in which destructive exchange interference can be realized
while maintaining a self-consistent finite-mode description.
These observations clarify that the odd--even behavior discussed in the
main text is not a universal property determined solely by the number of
retained modes. Instead, it originates from the combined constraints of
the standing-wave parity structure and the spectral arrangement of the
operational qubit frequencies.

\bibliography{Multimode}

\begin{thebibliography}{74}%
\makeatletter
\providecommand \@ifxundefined [1]{%
 \@ifx{#1\undefined}
}%
\providecommand \@ifnum [1]{%
 \ifnum #1\expandafter \@firstoftwo
 \else \expandafter \@secondoftwo
 \fi
}%
\providecommand \@ifx [1]{%
 \ifx #1\expandafter \@firstoftwo
 \else \expandafter \@secondoftwo
 \fi
}%
\providecommand \natexlab [1]{#1}%
\providecommand \enquote  [1]{``#1''}%
\providecommand \bibnamefont  [1]{#1}%
\providecommand \bibfnamefont [1]{#1}%
\providecommand \citenamefont [1]{#1}%
\providecommand \href@noop [0]{\@secondoftwo}%
\providecommand \href [0]{\begingroup \@sanitize@url \@href}%
\providecommand \@href[1]{\@@startlink{#1}\@@href}%
\providecommand \@@href[1]{\endgroup#1\@@endlink}%
\providecommand \@sanitize@url [0]{\catcode `\\12\catcode `\$12\catcode
  `\&12\catcode `\#12\catcode `\^12\catcode `\_12\catcode `\%12\relax}%
\providecommand \@@startlink[1]{}%
\providecommand \@@endlink[0]{}%
\providecommand \url  [0]{\begingroup\@sanitize@url \@url }%
\providecommand \@url [1]{\endgroup\@href {#1}{\urlprefix }}%
\providecommand \urlprefix  [0]{URL }%
\providecommand \Eprint [0]{\href }%
\providecommand \doibase [0]{https://doi.org/}%
\providecommand \selectlanguage [0]{\@gobble}%
\providecommand \bibinfo  [0]{\@secondoftwo}%
\providecommand \bibfield  [0]{\@secondoftwo}%
\providecommand \translation [1]{[#1]}%
\providecommand \BibitemOpen [0]{}%
\providecommand \bibitemStop [0]{}%
\providecommand \bibitemNoStop [0]{.\EOS\space}%
\providecommand \EOS [0]{\spacefactor3000\relax}%
\providecommand \BibitemShut  [1]{\csname bibitem#1\endcsname}%
\let\auto@bib@innerbib\@empty
%</preamble>
\bibitem [{\citenamefont {Arute}\ \emph {et~al.}(2019)\citenamefont {Arute}
  \emph {et~al.}}]{arute2019}%
  \BibitemOpen
  \bibfield  {author} {\bibinfo {author} {\bibfnamefont {F.}~\bibnamefont
  {Arute}} \emph {et~al.},\ }\bibfield  {title} {\bibinfo {title} {Quantum
  supremacy using a programmable superconducting processor},\ }\href
  {https://doi.org/10.1038/s41586-019-1666-5} {\bibfield  {journal} {\bibinfo
  {journal} {Nature}\ }\textbf {\bibinfo {volume} {574}},\ \bibinfo {pages}
  {505} (\bibinfo {year} {2019})}\BibitemShut {NoStop}%
\bibitem [{\citenamefont {Kjaergaard}\ \emph {et~al.}(2020)\citenamefont
  {Kjaergaard}, \citenamefont {Schwartz}, \citenamefont {Braum"uller},
  \citenamefont {Krantz}, \citenamefont {Wang}, \citenamefont {Gustavsson},\
  and\ \citenamefont {Oliver}}]{kjaergaard2020}%
  \BibitemOpen
  \bibfield  {author} {\bibinfo {author} {\bibfnamefont {M.}~\bibnamefont
  {Kjaergaard}}, \bibinfo {author} {\bibfnamefont {M.~E.}\ \bibnamefont
  {Schwartz}}, \bibinfo {author} {\bibfnamefont {J.}~\bibnamefont
  {Braum"uller}}, \bibinfo {author} {\bibfnamefont {P.}~\bibnamefont {Krantz}},
  \bibinfo {author} {\bibfnamefont {J.~I.-J.}\ \bibnamefont {Wang}}, \bibinfo
  {author} {\bibfnamefont {S.}~\bibnamefont {Gustavsson}},\ and\ \bibinfo
  {author} {\bibfnamefont {W.~D.}\ \bibnamefont {Oliver}},\ }\bibfield  {title}
  {\bibinfo {title} {Superconducting qubits: Current state of play},\ }\href
  {https://doi.org/10.1146/annurev-conmatphys-031119-050605} {\bibfield
  {journal} {\bibinfo  {journal} {Annu. Rev. Condens. Matter Phys.}\ }\textbf
  {\bibinfo {volume} {11}},\ \bibinfo {pages} {369} (\bibinfo {year}
  {2020})}\BibitemShut {NoStop}%
\bibitem [{\citenamefont {Krantz}\ \emph {et~al.}(2019)\citenamefont {Krantz},
  \citenamefont {Kjaergaard}, \citenamefont {Yan}, \citenamefont {Orlando},
  \citenamefont {Gustavsson},\ and\ \citenamefont {Oliver}}]{krantz2019}%
  \BibitemOpen
  \bibfield  {author} {\bibinfo {author} {\bibfnamefont {P.}~\bibnamefont
  {Krantz}}, \bibinfo {author} {\bibfnamefont {M.}~\bibnamefont {Kjaergaard}},
  \bibinfo {author} {\bibfnamefont {F.}~\bibnamefont {Yan}}, \bibinfo {author}
  {\bibfnamefont {T.~P.}\ \bibnamefont {Orlando}}, \bibinfo {author}
  {\bibfnamefont {S.}~\bibnamefont {Gustavsson}},\ and\ \bibinfo {author}
  {\bibfnamefont {W.~D.}\ \bibnamefont {Oliver}},\ }\bibfield  {title}
  {\bibinfo {title} {A quantum engineer's guide to superconducting qubits},\
  }\href {https://doi.org/10.1063/1.5089550} {\bibfield  {journal} {\bibinfo
  {journal} {Appl. Phys. Rev.}\ }\textbf {\bibinfo {volume} {6}},\ \bibinfo
  {pages} {021318} (\bibinfo {year} {2019})}\BibitemShut {NoStop}%
\bibitem [{\citenamefont {Gu}\ \emph {et~al.}(2017)\citenamefont {Gu},
  \citenamefont {Kockum}, \citenamefont {Miranowicz}, \citenamefont {X.~Liu},\
  and\ \citenamefont {Nori}}]{Gu2017}%
  \BibitemOpen
  \bibfield  {author} {\bibinfo {author} {\bibfnamefont {X.}~\bibnamefont
  {Gu}}, \bibinfo {author} {\bibfnamefont {A.~F.}\ \bibnamefont {Kockum}},
  \bibinfo {author} {\bibfnamefont {A.}~\bibnamefont {Miranowicz}}, \bibinfo
  {author} {\bibfnamefont {Y.}~\bibnamefont {X.~Liu}},\ and\ \bibinfo {author}
  {\bibfnamefont {F.}~\bibnamefont {Nori}},\ }\bibfield  {title} {\bibinfo
  {title} {{Microwave photonics with superconducting quantum circuits}},\
  }\href {https://doi.org/10.1016/j.physrep.2017.10.002} {\bibfield  {journal}
  {\bibinfo  {journal} {Phys. Rep.}\ }\textbf {\bibinfo {volume} {718-719}},\
  \bibinfo {pages} {1} (\bibinfo {year} {2017})}\BibitemShut {NoStop}%
\bibitem [{\citenamefont {Monroe}\ \emph {et~al.}(2014)\citenamefont {Monroe},
  \citenamefont {Raussendorf}, \citenamefont {Ruthven}, \citenamefont {Brown},
  \citenamefont {Maunz}, \citenamefont {Duan},\ and\ \citenamefont
  {Kim}}]{monroe2014}%
  \BibitemOpen
  \bibfield  {author} {\bibinfo {author} {\bibfnamefont {C.}~\bibnamefont
  {Monroe}}, \bibinfo {author} {\bibfnamefont {R.}~\bibnamefont {Raussendorf}},
  \bibinfo {author} {\bibfnamefont {A.}~\bibnamefont {Ruthven}}, \bibinfo
  {author} {\bibfnamefont {K.~R.}\ \bibnamefont {Brown}}, \bibinfo {author}
  {\bibfnamefont {P.}~\bibnamefont {Maunz}}, \bibinfo {author} {\bibfnamefont
  {L.-M.}\ \bibnamefont {Duan}},\ and\ \bibinfo {author} {\bibfnamefont
  {J.}~\bibnamefont {Kim}},\ }\bibfield  {title} {\bibinfo {title} {Large-scale
  modular quantum-computer architecture with atomic memory and photonic
  interconnects},\ }\href {https://doi.org/10.1103/PhysRevA.89.022317}
  {\bibfield  {journal} {\bibinfo  {journal} {Phys. Rev. A}\ }\textbf {\bibinfo
  {volume} {89}},\ \bibinfo {pages} {022317} (\bibinfo {year}
  {2014})}\BibitemShut {NoStop}%
\bibitem [{\citenamefont {Wu}\ \emph {et~al.}(2024)\citenamefont {Wu},
  \citenamefont {Yan}, \citenamefont {Andersson}, \citenamefont {Anferov},
  \citenamefont {Chou}, \citenamefont {Conner}, \citenamefont {Grebel},
  \citenamefont {Joshi}, \citenamefont {Li}, \citenamefont {Miller},
  \citenamefont {Povey}, \citenamefont {Qiao},\ and\ \citenamefont
  {Cleland}}]{PhysRevX.14.041030}%
  \BibitemOpen
  \bibfield  {author} {\bibinfo {author} {\bibfnamefont {X.}~\bibnamefont
  {Wu}}, \bibinfo {author} {\bibfnamefont {H.}~\bibnamefont {Yan}}, \bibinfo
  {author} {\bibfnamefont {G.}~\bibnamefont {Andersson}}, \bibinfo {author}
  {\bibfnamefont {A.}~\bibnamefont {Anferov}}, \bibinfo {author} {\bibfnamefont
  {M.-H.}\ \bibnamefont {Chou}}, \bibinfo {author} {\bibfnamefont {C.~R.}\
  \bibnamefont {Conner}}, \bibinfo {author} {\bibfnamefont {J.}~\bibnamefont
  {Grebel}}, \bibinfo {author} {\bibfnamefont {Y.~J.}\ \bibnamefont {Joshi}},
  \bibinfo {author} {\bibfnamefont {S.}~\bibnamefont {Li}}, \bibinfo {author}
  {\bibfnamefont {J.~M.}\ \bibnamefont {Miller}}, \bibinfo {author}
  {\bibfnamefont {R.~G.}\ \bibnamefont {Povey}}, \bibinfo {author}
  {\bibfnamefont {H.}~\bibnamefont {Qiao}},\ and\ \bibinfo {author}
  {\bibfnamefont {A.~N.}\ \bibnamefont {Cleland}},\ }\bibfield  {title}
  {\bibinfo {title} {Modular quantum processor with an all-to-all
  reconfigurable router},\ }\href {https://doi.org/10.1103/PhysRevX.14.041030}
  {\bibfield  {journal} {\bibinfo  {journal} {Phys. Rev. X}\ }\textbf {\bibinfo
  {volume} {14}},\ \bibinfo {pages} {041030} (\bibinfo {year}
  {2024})}\BibitemShut {NoStop}%
\bibitem [{\citenamefont {Blais}\ \emph {et~al.}(2004)\citenamefont {Blais},
  \citenamefont {Huang}, \citenamefont {Wallraff}, \citenamefont {Girvin},\
  and\ \citenamefont {Schoelkopf}}]{blais2004}%
  \BibitemOpen
  \bibfield  {author} {\bibinfo {author} {\bibfnamefont {A.}~\bibnamefont
  {Blais}}, \bibinfo {author} {\bibfnamefont {R.-S.}\ \bibnamefont {Huang}},
  \bibinfo {author} {\bibfnamefont {A.}~\bibnamefont {Wallraff}}, \bibinfo
  {author} {\bibfnamefont {S.~M.}\ \bibnamefont {Girvin}},\ and\ \bibinfo
  {author} {\bibfnamefont {R.~J.}\ \bibnamefont {Schoelkopf}},\ }\bibfield
  {title} {\bibinfo {title} {Cavity quantum electrodynamics for superconducting
  electrical circuits: An architecture for quantum computation},\ }\href
  {https://doi.org/10.1103/PhysRevA.69.062320} {\bibfield  {journal} {\bibinfo
  {journal} {Phys. Rev. A}\ }\textbf {\bibinfo {volume} {69}},\ \bibinfo
  {pages} {062320} (\bibinfo {year} {2004})}\BibitemShut {NoStop}%
\bibitem [{\citenamefont {Blais}\ \emph {et~al.}(2021)\citenamefont {Blais},
  \citenamefont {Grimsmo}, \citenamefont {Girvin},\ and\ \citenamefont
  {Wallraff}}]{blais2021}%
  \BibitemOpen
  \bibfield  {author} {\bibinfo {author} {\bibfnamefont {A.}~\bibnamefont
  {Blais}}, \bibinfo {author} {\bibfnamefont {A.~L.}\ \bibnamefont {Grimsmo}},
  \bibinfo {author} {\bibfnamefont {S.~M.}\ \bibnamefont {Girvin}},\ and\
  \bibinfo {author} {\bibfnamefont {A.}~\bibnamefont {Wallraff}},\ }\bibfield
  {title} {\bibinfo {title} {Circuit quantum electrodynamics},\ }\href
  {https://doi.org/10.1103/RevModPhys.93.025005} {\bibfield  {journal}
  {\bibinfo  {journal} {Rev. Mod. Phys.}\ }\textbf {\bibinfo {volume} {93}},\
  \bibinfo {pages} {025005} (\bibinfo {year} {2021})}\BibitemShut {NoStop}%
\bibitem [{\citenamefont {Wallraff}\ \emph {et~al.}(2004)\citenamefont
  {Wallraff}, \citenamefont {Schuster}, \citenamefont {Blais}, \citenamefont
  {Frunzio}, \citenamefont {Huang}, \citenamefont {Majer}, \citenamefont
  {Kumar}, \citenamefont {Girvin},\ and\ \citenamefont
  {Schoelkopf}}]{wallraff2004}%
  \BibitemOpen
  \bibfield  {author} {\bibinfo {author} {\bibfnamefont {A.}~\bibnamefont
  {Wallraff}}, \bibinfo {author} {\bibfnamefont {D.~I.}\ \bibnamefont
  {Schuster}}, \bibinfo {author} {\bibfnamefont {A.}~\bibnamefont {Blais}},
  \bibinfo {author} {\bibfnamefont {L.}~\bibnamefont {Frunzio}}, \bibinfo
  {author} {\bibfnamefont {R.-S.}\ \bibnamefont {Huang}}, \bibinfo {author}
  {\bibfnamefont {J.}~\bibnamefont {Majer}}, \bibinfo {author} {\bibfnamefont
  {S.}~\bibnamefont {Kumar}}, \bibinfo {author} {\bibfnamefont {S.~M.}\
  \bibnamefont {Girvin}},\ and\ \bibinfo {author} {\bibfnamefont {R.~J.}\
  \bibnamefont {Schoelkopf}},\ }\bibfield  {title} {\bibinfo {title} {Strong
  coupling of a single photon to a superconducting qubit using circuit quantum
  electrodynamics},\ }\href {https://doi.org/10.1038/nature02851} {\bibfield
  {journal} {\bibinfo  {journal} {Nature}\ }\textbf {\bibinfo {volume} {431}},\
  \bibinfo {pages} {162} (\bibinfo {year} {2004})}\BibitemShut {NoStop}%
\bibitem [{\citenamefont {Majer}\ \emph {et~al.}(2007)\citenamefont {Majer},
  \citenamefont {Chow}, \citenamefont {Gambetta}, \citenamefont {Koch},
  \citenamefont {Johnson}, \citenamefont {Schreier}, \citenamefont {Frunzio},
  \citenamefont {Schuster}, \citenamefont {Houck}, \citenamefont {Wallraff}
  \emph {et~al.}}]{majer2007}%
  \BibitemOpen
  \bibfield  {author} {\bibinfo {author} {\bibfnamefont {J.}~\bibnamefont
  {Majer}}, \bibinfo {author} {\bibfnamefont {J.~M.}\ \bibnamefont {Chow}},
  \bibinfo {author} {\bibfnamefont {J.~M.}\ \bibnamefont {Gambetta}}, \bibinfo
  {author} {\bibfnamefont {J.}~\bibnamefont {Koch}}, \bibinfo {author}
  {\bibfnamefont {B.~R.}\ \bibnamefont {Johnson}}, \bibinfo {author}
  {\bibfnamefont {J.~A.}\ \bibnamefont {Schreier}}, \bibinfo {author}
  {\bibfnamefont {L.}~\bibnamefont {Frunzio}}, \bibinfo {author} {\bibfnamefont
  {D.~I.}\ \bibnamefont {Schuster}}, \bibinfo {author} {\bibfnamefont {A.~A.}\
  \bibnamefont {Houck}}, \bibinfo {author} {\bibfnamefont {A.}~\bibnamefont
  {Wallraff}}, \emph {et~al.},\ }\bibfield  {title} {\bibinfo {title} {Coupling
  superconducting qubits via a cavity bus},\ }\href
  {https://doi.org/10.1038/nature06184} {\bibfield  {journal} {\bibinfo
  {journal} {Nature}\ }\textbf {\bibinfo {volume} {449}},\ \bibinfo {pages}
  {443} (\bibinfo {year} {2007})}\BibitemShut {NoStop}%
\bibitem [{\citenamefont {Sundaresan}\ \emph {et~al.}(2019)\citenamefont
  {Sundaresan}, \citenamefont {Lundgren}, \citenamefont {Zhu}, \citenamefont
  {Gorshkov},\ and\ \citenamefont {Houck}}]{PhysRevX.9.011021}%
  \BibitemOpen
  \bibfield  {author} {\bibinfo {author} {\bibfnamefont {N.~M.}\ \bibnamefont
  {Sundaresan}}, \bibinfo {author} {\bibfnamefont {R.}~\bibnamefont
  {Lundgren}}, \bibinfo {author} {\bibfnamefont {G.}~\bibnamefont {Zhu}},
  \bibinfo {author} {\bibfnamefont {A.~V.}\ \bibnamefont {Gorshkov}},\ and\
  \bibinfo {author} {\bibfnamefont {A.~A.}\ \bibnamefont {Houck}},\ }\bibfield
  {title} {\bibinfo {title} {Interacting qubit-photon bound states with
  superconducting circuits},\ }\href
  {https://doi.org/10.1103/PhysRevX.9.011021} {\bibfield  {journal} {\bibinfo
  {journal} {Phys. Rev. X}\ }\textbf {\bibinfo {volume} {9}},\ \bibinfo {pages}
  {011021} (\bibinfo {year} {2019})}\BibitemShut {NoStop}%
\bibitem [{\citenamefont {Brehm}\ \emph {et~al.}(2021)\citenamefont {Brehm},
  \citenamefont {Poddubny}, \citenamefont {Stehli}, \citenamefont {Wolz},
  \citenamefont {Schneider}, \citenamefont {Rotzinger}, \citenamefont
  {Ustinov},\ and\ \citenamefont {Weides}}]{brehm2021}%
  \BibitemOpen
  \bibfield  {author} {\bibinfo {author} {\bibfnamefont {J.~D.}\ \bibnamefont
  {Brehm}}, \bibinfo {author} {\bibfnamefont {A.~N.}\ \bibnamefont {Poddubny}},
  \bibinfo {author} {\bibfnamefont {A.}~\bibnamefont {Stehli}}, \bibinfo
  {author} {\bibfnamefont {T.}~\bibnamefont {Wolz}}, \bibinfo {author}
  {\bibfnamefont {H.}~\bibnamefont {Schneider}}, \bibinfo {author}
  {\bibfnamefont {H.}~\bibnamefont {Rotzinger}}, \bibinfo {author}
  {\bibfnamefont {A.~V.}\ \bibnamefont {Ustinov}},\ and\ \bibinfo {author}
  {\bibfnamefont {M.}~\bibnamefont {Weides}},\ }\bibfield  {title} {\bibinfo
  {title} {Waveguide bandgap engineering with an array of superconducting
  qubits},\ }\href {https://doi.org/10.1038/s41535-021-00310-z} {\bibfield
  {journal} {\bibinfo  {journal} {npj Quantum Materials}\ }\textbf {\bibinfo
  {volume} {6}},\ \bibinfo {pages} {10} (\bibinfo {year} {2021})}\BibitemShut
  {NoStop}%
\bibitem [{\citenamefont {Sillanp{\"a}{\"a}}\ \emph {et~al.}(2007)\citenamefont
  {Sillanp{\"a}{\"a}}, \citenamefont {Park},\ and\ \citenamefont
  {Simmonds}}]{sillanpaa2007}%
  \BibitemOpen
  \bibfield  {author} {\bibinfo {author} {\bibfnamefont {M.~A.}\ \bibnamefont
  {Sillanp{\"a}{\"a}}}, \bibinfo {author} {\bibfnamefont {J.}~\bibnamefont
  {Park}},\ and\ \bibinfo {author} {\bibfnamefont {R.~W.}\ \bibnamefont
  {Simmonds}},\ }\bibfield  {title} {\bibinfo {title} {Coherent quantum state
  storage and transfer between two phase qubits via a resonant cavity},\ }\href
  {https://doi.org/10.1038/nature06124} {\bibfield  {journal} {\bibinfo
  {journal} {Nature}\ }\textbf {\bibinfo {volume} {449}},\ \bibinfo {pages}
  {438} (\bibinfo {year} {2007})}\BibitemShut {NoStop}%
\bibitem [{\citenamefont {Malekakhlagh}\ \emph {et~al.}(2016)\citenamefont
  {Malekakhlagh}, \citenamefont {Petrescu},\ and\ \citenamefont
  {T"ureci}}]{malekakhlagh2016}%
  \BibitemOpen
  \bibfield  {author} {\bibinfo {author} {\bibfnamefont {M.}~\bibnamefont
  {Malekakhlagh}}, \bibinfo {author} {\bibfnamefont {A.}~\bibnamefont
  {Petrescu}},\ and\ \bibinfo {author} {\bibfnamefont {H.~E.}\ \bibnamefont
  {T"ureci}},\ }\bibfield  {title} {\bibinfo {title} {Non-markovian dynamics of
  a superconducting qubit in an open multimode resonator},\ }\href
  {https://doi.org/10.1103/PhysRevA.94.063848} {\bibfield  {journal} {\bibinfo
  {journal} {Phys. Rev. A}\ }\textbf {\bibinfo {volume} {94}},\ \bibinfo
  {pages} {063848} (\bibinfo {year} {2016})}\BibitemShut {NoStop}%
\bibitem [{\citenamefont {Egger}\ and\ \citenamefont
  {Wilhelm}(2013)}]{egger2013}%
  \BibitemOpen
  \bibfield  {author} {\bibinfo {author} {\bibfnamefont {D.~J.}\ \bibnamefont
  {Egger}}\ and\ \bibinfo {author} {\bibfnamefont {F.~K.}\ \bibnamefont
  {Wilhelm}},\ }\bibfield  {title} {\bibinfo {title} {Multimode circuit quantum
  electrodynamics with hybrid metamaterial transmission lines},\ }\href
  {https://doi.org/10.1103/PhysRevLett.111.163601} {\bibfield  {journal}
  {\bibinfo  {journal} {Phys. Rev. Lett.}\ }\textbf {\bibinfo {volume} {111}},\
  \bibinfo {pages} {163601} (\bibinfo {year} {2013})}\BibitemShut {NoStop}%
\bibitem [{\citenamefont {Solgun}\ and\ \citenamefont
  {Srinivasan}(2022)}]{solgun2022}%
  \BibitemOpen
  \bibfield  {author} {\bibinfo {author} {\bibfnamefont {F.}~\bibnamefont
  {Solgun}}\ and\ \bibinfo {author} {\bibfnamefont {S.}~\bibnamefont
  {Srinivasan}},\ }\bibfield  {title} {\bibinfo {title} {Direct calculation of
  {$ZZ$} interaction rates in multimode circuit quantum electrodynamics},\
  }\href {https://doi.org/10.1103/PhysRevApplied.18.044025} {\bibfield
  {journal} {\bibinfo  {journal} {Phys. Rev. Applied}\ }\textbf {\bibinfo
  {volume} {18}},\ \bibinfo {pages} {044025} (\bibinfo {year}
  {2022})}\BibitemShut {NoStop}%
\bibitem [{\citenamefont {He}\ and\ \citenamefont {Zhang}(2025)}]{he2025}%
  \BibitemOpen
  \bibfield  {author} {\bibinfo {author} {\bibfnamefont {Y.}~\bibnamefont
  {He}}\ and\ \bibinfo {author} {\bibfnamefont {Y.-X.}\ \bibnamefont {Zhang}},\
  }\bibfield  {title} {\bibinfo {title} {Quantum state transfer via a multimode
  resonator},\ }\href {https://doi.org/10.1103/PhysRevLett.134.023602}
  {\bibfield  {journal} {\bibinfo  {journal} {Phys. Rev. Lett.}\ }\textbf
  {\bibinfo {volume} {134}},\ \bibinfo {pages} {023602} (\bibinfo {year}
  {2025})}\BibitemShut {NoStop}%
\bibitem [{\citenamefont {Wang}\ \emph {et~al.}(2026)\citenamefont {Wang},
  \citenamefont {Zhao}, \citenamefont {Hao-Ai}, \citenamefont {Cheng},\ and\
  \citenamefont {Liu}}]{7gr4-nk6m}%
  \BibitemOpen
  \bibfield  {author} {\bibinfo {author} {\bibfnamefont {D.-W.}\ \bibnamefont
  {Wang}}, \bibinfo {author} {\bibfnamefont {C.}~\bibnamefont {Zhao}}, \bibinfo
  {author} {\bibnamefont {Hao-Ai}}, \bibinfo {author} {\bibfnamefont
  {W.}~\bibnamefont {Cheng}},\ and\ \bibinfo {author} {\bibfnamefont {Y.-x.}\
  \bibnamefont {Liu}},\ }\bibfield  {title} {\bibinfo {title} {Thermal
  noise-immunized quantum state transfer through a microwave waveguide},\
  }\href {https://doi.org/10.1103/7gr4-nk6m} {\bibfield  {journal} {\bibinfo
  {journal} {Phys. Rev. A}\ }\textbf {\bibinfo {volume} {113}},\ \bibinfo
  {pages} {042437} (\bibinfo {year} {2026})}\BibitemShut {NoStop}%
\bibitem [{\citenamefont {Heya}\ \emph {et~al.}(2025)\citenamefont {Heya},
  \citenamefont {Phung}, \citenamefont {Malekakhlagh}, \citenamefont {Steiner},
  \citenamefont {Turchetti}, \citenamefont {Shanks}, \citenamefont {Mamin},
  \citenamefont {Lu}, \citenamefont {Kandel}, \citenamefont {Sundaresan},\ and\
  \citenamefont {Orcutt}}]{xx24-r7q6}%
  \BibitemOpen
  \bibfield  {author} {\bibinfo {author} {\bibfnamefont {K.}~\bibnamefont
  {Heya}}, \bibinfo {author} {\bibfnamefont {T.}~\bibnamefont {Phung}},
  \bibinfo {author} {\bibfnamefont {M.}~\bibnamefont {Malekakhlagh}}, \bibinfo
  {author} {\bibfnamefont {R.}~\bibnamefont {Steiner}}, \bibinfo {author}
  {\bibfnamefont {M.}~\bibnamefont {Turchetti}}, \bibinfo {author}
  {\bibfnamefont {W.}~\bibnamefont {Shanks}}, \bibinfo {author} {\bibfnamefont
  {J.}~\bibnamefont {Mamin}}, \bibinfo {author} {\bibfnamefont {W.-S.}\
  \bibnamefont {Lu}}, \bibinfo {author} {\bibfnamefont {Y.~P.}\ \bibnamefont
  {Kandel}}, \bibinfo {author} {\bibfnamefont {N.}~\bibnamefont {Sundaresan}},\
  and\ \bibinfo {author} {\bibfnamefont {J.}~\bibnamefont {Orcutt}},\
  }\bibfield  {title} {\bibinfo {title} {Randomized benchmarking of a remote
  cnot gate via a meter-scale microwave link},\ }\href
  {https://doi.org/10.1103/xx24-r7q6} {\bibfield  {journal} {\bibinfo
  {journal} {Phys. Rev. Lett.}\ }\textbf {\bibinfo {volume} {135}},\ \bibinfo
  {pages} {200801} (\bibinfo {year} {2025})}\BibitemShut {NoStop}%
\bibitem [{\citenamefont {Kurpiers}\ \emph {et~al.}(2018)\citenamefont
  {Kurpiers}, \citenamefont {Magnard}, \citenamefont {Walter}, \citenamefont
  {Poto\v{c}nik}, \citenamefont {Salath\'e}, \citenamefont {Pechal},
  \citenamefont {Mondal}, \citenamefont {Oppliger}, \citenamefont {Eichler},\
  and\ \citenamefont {Wallraff}}]{kurpiers2018}%
  \BibitemOpen
  \bibfield  {author} {\bibinfo {author} {\bibfnamefont {P.}~\bibnamefont
  {Kurpiers}}, \bibinfo {author} {\bibfnamefont {P.}~\bibnamefont {Magnard}},
  \bibinfo {author} {\bibfnamefont {T.}~\bibnamefont {Walter}}, \bibinfo
  {author} {\bibfnamefont {A.}~\bibnamefont {Poto\v{c}nik}}, \bibinfo {author}
  {\bibfnamefont {Y.}~\bibnamefont {Salath\'e}}, \bibinfo {author}
  {\bibfnamefont {M.}~\bibnamefont {Pechal}}, \bibinfo {author} {\bibfnamefont
  {M.}~\bibnamefont {Mondal}}, \bibinfo {author} {\bibfnamefont
  {M.}~\bibnamefont {Oppliger}}, \bibinfo {author} {\bibfnamefont
  {C.}~\bibnamefont {Eichler}},\ and\ \bibinfo {author} {\bibfnamefont
  {A.}~\bibnamefont {Wallraff}},\ }\bibfield  {title} {\bibinfo {title}
  {Deterministic quantum state transfer and generation of remote entanglement
  using microwave photons},\ }\href {https://doi.org/10.1038/s41586-018-0195-y}
  {\bibfield  {journal} {\bibinfo  {journal} {Nature}\ }\textbf {\bibinfo
  {volume} {558}},\ \bibinfo {pages} {264} (\bibinfo {year}
  {2018})}\BibitemShut {NoStop}%
\bibitem [{\citenamefont {Axline}\ \emph {et~al.}(2018)\citenamefont {Axline},
  \citenamefont {Burkhart}, \citenamefont {Pfaff}, \citenamefont {Zhang},
  \citenamefont {Chou}, \citenamefont {Campagne-Ibarcq}, \citenamefont
  {Reinhold}, \citenamefont {Pop}, \citenamefont {Frunzio}, \citenamefont
  {Devoret},\ and\ \citenamefont {Schoelkopf}}]{axline2018}%
  \BibitemOpen
  \bibfield  {author} {\bibinfo {author} {\bibfnamefont {C.~J.}\ \bibnamefont
  {Axline}}, \bibinfo {author} {\bibfnamefont {L.~D.}\ \bibnamefont
  {Burkhart}}, \bibinfo {author} {\bibfnamefont {W.}~\bibnamefont {Pfaff}},
  \bibinfo {author} {\bibfnamefont {M.}~\bibnamefont {Zhang}}, \bibinfo
  {author} {\bibfnamefont {K.~S.}\ \bibnamefont {Chou}}, \bibinfo {author}
  {\bibfnamefont {P.}~\bibnamefont {Campagne-Ibarcq}}, \bibinfo {author}
  {\bibfnamefont {P.}~\bibnamefont {Reinhold}}, \bibinfo {author}
  {\bibfnamefont {I.~M.}\ \bibnamefont {Pop}}, \bibinfo {author} {\bibfnamefont
  {L.}~\bibnamefont {Frunzio}}, \bibinfo {author} {\bibfnamefont {M.~H.}\
  \bibnamefont {Devoret}},\ and\ \bibinfo {author} {\bibfnamefont {R.~J.}\
  \bibnamefont {Schoelkopf}},\ }\bibfield  {title} {\bibinfo {title} {On-demand
  quantum state transfer and entanglement between remote microwave cavity
  memories},\ }\href {https://doi.org/10.1038/s41567-018-0115-y} {\bibfield
  {journal} {\bibinfo  {journal} {Nat. Phys.}\ }\textbf {\bibinfo {volume}
  {14}},\ \bibinfo {pages} {705} (\bibinfo {year} {2018})}\BibitemShut
  {NoStop}%
\bibitem [{\citenamefont {Campagne-Ibarcq}\ \emph {et~al.}(2018)\citenamefont
  {Campagne-Ibarcq}, \citenamefont {Eickbusch}, \citenamefont {Touzard},
  \citenamefont {Zalys-Geller}, \citenamefont {Frattini}, \citenamefont
  {Sivak}, \citenamefont {Reinhold}, \citenamefont {Puri}, \citenamefont
  {Shankar},\ and\ \citenamefont {Devoret}}]{campagne2018}%
  \BibitemOpen
  \bibfield  {author} {\bibinfo {author} {\bibfnamefont {P.}~\bibnamefont
  {Campagne-Ibarcq}}, \bibinfo {author} {\bibfnamefont {A.}~\bibnamefont
  {Eickbusch}}, \bibinfo {author} {\bibfnamefont {S.}~\bibnamefont {Touzard}},
  \bibinfo {author} {\bibfnamefont {E.}~\bibnamefont {Zalys-Geller}}, \bibinfo
  {author} {\bibfnamefont {N.~E.}\ \bibnamefont {Frattini}}, \bibinfo {author}
  {\bibfnamefont {V.~V.}\ \bibnamefont {Sivak}}, \bibinfo {author}
  {\bibfnamefont {P.}~\bibnamefont {Reinhold}}, \bibinfo {author}
  {\bibfnamefont {S.}~\bibnamefont {Puri}}, \bibinfo {author} {\bibfnamefont
  {S.}~\bibnamefont {Shankar}},\ and\ \bibinfo {author} {\bibfnamefont {M.~H.}\
  \bibnamefont {Devoret}},\ }\bibfield  {title} {\bibinfo {title}
  {Deterministic remote entanglement of superconducting circuits through
  microwave two-photon transitions},\ }\href
  {https://doi.org/10.1103/PhysRevLett.120.200501} {\bibfield  {journal}
  {\bibinfo  {journal} {Phys. Rev. Lett.}\ }\textbf {\bibinfo {volume} {120}},\
  \bibinfo {pages} {200501} (\bibinfo {year} {2018})}\BibitemShut {NoStop}%
\bibitem [{\citenamefont {Zhong}\ \emph {et~al.}(2019)\citenamefont {Zhong},
  \citenamefont {Chang}, \citenamefont {Bienfait} \emph {et~al.}}]{zhong2019}%
  \BibitemOpen
  \bibfield  {author} {\bibinfo {author} {\bibfnamefont {Y.~P.}\ \bibnamefont
  {Zhong}}, \bibinfo {author} {\bibfnamefont {H.~S.}\ \bibnamefont {Chang}},
  \bibinfo {author} {\bibfnamefont {A.}~\bibnamefont {Bienfait}}, \emph
  {et~al.},\ }\bibfield  {title} {\bibinfo {title} {Violating bell's inequality
  with remotely connected superconducting qubits},\ }\href
  {https://doi.org/10.1038/s41567-019-0507-7} {\bibfield  {journal} {\bibinfo
  {journal} {Nat. Phys.}\ }\textbf {\bibinfo {volume} {15}},\ \bibinfo {pages}
  {741} (\bibinfo {year} {2019})}\BibitemShut {NoStop}%
\bibitem [{\citenamefont {Leung}\ \emph {et~al.}(2019)\citenamefont {Leung},
  \citenamefont {Chakram}, \citenamefont {Hazard}, \citenamefont {Gyenis},\
  and\ \citenamefont {Houck}}]{leung2019}%
  \BibitemOpen
  \bibfield  {author} {\bibinfo {author} {\bibfnamefont {N.}~\bibnamefont
  {Leung}}, \bibinfo {author} {\bibfnamefont {S.}~\bibnamefont {Chakram}},
  \bibinfo {author} {\bibfnamefont {T.~M.}\ \bibnamefont {Hazard}}, \bibinfo
  {author} {\bibfnamefont {A.}~\bibnamefont {Gyenis}},\ and\ \bibinfo {author}
  {\bibfnamefont {A.~A.}\ \bibnamefont {Houck}},\ }\bibfield  {title} {\bibinfo
  {title} {Deterministic bidirectional communication and remote entanglement
  generation between superconducting quantum processors},\ }\href
  {https://doi.org/10.1038/s41534-019-0128-0} {\bibfield  {journal} {\bibinfo
  {journal} {npj Quantum Inf.}\ }\textbf {\bibinfo {volume} {5}},\ \bibinfo
  {pages} {18} (\bibinfo {year} {2019})}\BibitemShut {NoStop}%
\bibitem [{\citenamefont {Magnard}\ \emph {et~al.}(2020)\citenamefont
  {Magnard}, \citenamefont {Storz}, \citenamefont {Kurpiers}, \citenamefont
  {Sch\"ar}, \citenamefont {Marxer}, \citenamefont {L\"utolf}, \citenamefont
  {Walter}, \citenamefont {Besse}, \citenamefont {Gabureac}, \citenamefont
  {Reuer}, \citenamefont {Akin}, \citenamefont {Royer}, \citenamefont {Blais},\
  and\ \citenamefont {Wallraff}}]{magnard2020}%
  \BibitemOpen
  \bibfield  {author} {\bibinfo {author} {\bibfnamefont {P.}~\bibnamefont
  {Magnard}}, \bibinfo {author} {\bibfnamefont {S.}~\bibnamefont {Storz}},
  \bibinfo {author} {\bibfnamefont {P.}~\bibnamefont {Kurpiers}}, \bibinfo
  {author} {\bibfnamefont {J.}~\bibnamefont {Sch\"ar}}, \bibinfo {author}
  {\bibfnamefont {F.}~\bibnamefont {Marxer}}, \bibinfo {author} {\bibfnamefont
  {J.}~\bibnamefont {L\"utolf}}, \bibinfo {author} {\bibfnamefont
  {T.}~\bibnamefont {Walter}}, \bibinfo {author} {\bibfnamefont {J.-C.}\
  \bibnamefont {Besse}}, \bibinfo {author} {\bibfnamefont {M.}~\bibnamefont
  {Gabureac}}, \bibinfo {author} {\bibfnamefont {K.}~\bibnamefont {Reuer}},
  \bibinfo {author} {\bibfnamefont {A.}~\bibnamefont {Akin}}, \bibinfo {author}
  {\bibfnamefont {B.}~\bibnamefont {Royer}}, \bibinfo {author} {\bibfnamefont
  {A.}~\bibnamefont {Blais}},\ and\ \bibinfo {author} {\bibfnamefont
  {A.}~\bibnamefont {Wallraff}},\ }\bibfield  {title} {\bibinfo {title}
  {Microwave quantum link between superconducting circuits housed in spatially
  separated cryogenic systems},\ }\href
  {https://doi.org/10.1103/PhysRevLett.125.260502} {\bibfield  {journal}
  {\bibinfo  {journal} {Phys. Rev. Lett.}\ }\textbf {\bibinfo {volume} {125}},\
  \bibinfo {pages} {260502} (\bibinfo {year} {2020})}\BibitemShut {NoStop}%
\bibitem [{\citenamefont {Chang}\ \emph {et~al.}(2020)\citenamefont {Chang},
  \citenamefont {Zhong}, \citenamefont {Bienfait}, \citenamefont {Chou},
  \citenamefont {Conner}, \citenamefont {Dumur}, \citenamefont {Grebel},
  \citenamefont {Peairs}, \citenamefont {Povey}, \citenamefont {Satzinger},\
  and\ \citenamefont {Cleland}}]{chang2020}%
  \BibitemOpen
  \bibfield  {author} {\bibinfo {author} {\bibfnamefont {H.-S.}\ \bibnamefont
  {Chang}}, \bibinfo {author} {\bibfnamefont {Y.~P.}\ \bibnamefont {Zhong}},
  \bibinfo {author} {\bibfnamefont {A.}~\bibnamefont {Bienfait}}, \bibinfo
  {author} {\bibfnamefont {M.-H.}\ \bibnamefont {Chou}}, \bibinfo {author}
  {\bibfnamefont {C.~R.}\ \bibnamefont {Conner}}, \bibinfo {author}
  {\bibfnamefont {E.}~\bibnamefont {Dumur}}, \bibinfo {author} {\bibfnamefont
  {J.}~\bibnamefont {Grebel}}, \bibinfo {author} {\bibfnamefont {G.~A.}\
  \bibnamefont {Peairs}}, \bibinfo {author} {\bibfnamefont {R.~G.}\
  \bibnamefont {Povey}}, \bibinfo {author} {\bibfnamefont {K.~J.}\ \bibnamefont
  {Satzinger}},\ and\ \bibinfo {author} {\bibfnamefont {A.~N.}\ \bibnamefont
  {Cleland}},\ }\bibfield  {title} {\bibinfo {title} {Remote entanglement via
  adiabatic passage using a tunably dissipative quantum communication system},\
  }\href {https://doi.org/10.1103/PhysRevLett.124.240502} {\bibfield  {journal}
  {\bibinfo  {journal} {Phys. Rev. Lett.}\ }\textbf {\bibinfo {volume} {124}},\
  \bibinfo {pages} {240502} (\bibinfo {year} {2020})}\BibitemShut {NoStop}%
\bibitem [{\citenamefont {Grebel}\ \emph {et~al.}(2024)\citenamefont {Grebel},
  \citenamefont {Yan}, \citenamefont {Chou}, \citenamefont {Andersson},
  \citenamefont {Conner}, \citenamefont {Joshi}, \citenamefont {Miller},
  \citenamefont {Povey}, \citenamefont {Qiao}, \citenamefont {Wu},\ and\
  \citenamefont {Cleland}}]{grebel2024}%
  \BibitemOpen
  \bibfield  {author} {\bibinfo {author} {\bibfnamefont {J.}~\bibnamefont
  {Grebel}}, \bibinfo {author} {\bibfnamefont {H.}~\bibnamefont {Yan}},
  \bibinfo {author} {\bibfnamefont {M.-H.}\ \bibnamefont {Chou}}, \bibinfo
  {author} {\bibfnamefont {G.}~\bibnamefont {Andersson}}, \bibinfo {author}
  {\bibfnamefont {C.~R.}\ \bibnamefont {Conner}}, \bibinfo {author}
  {\bibfnamefont {Y.~J.}\ \bibnamefont {Joshi}}, \bibinfo {author}
  {\bibfnamefont {J.~M.}\ \bibnamefont {Miller}}, \bibinfo {author}
  {\bibfnamefont {R.~G.}\ \bibnamefont {Povey}}, \bibinfo {author}
  {\bibfnamefont {H.}~\bibnamefont {Qiao}}, \bibinfo {author} {\bibfnamefont
  {X.}~\bibnamefont {Wu}},\ and\ \bibinfo {author} {\bibfnamefont {A.~N.}\
  \bibnamefont {Cleland}},\ }\bibfield  {title} {\bibinfo {title}
  {Bidirectional multiphoton communication between remote superconducting
  nodes},\ }\href {https://doi.org/10.1103/PhysRevLett.132.047001} {\bibfield
  {journal} {\bibinfo  {journal} {Phys. Rev. Lett.}\ }\textbf {\bibinfo
  {volume} {132}},\ \bibinfo {pages} {047001} (\bibinfo {year}
  {2024})}\BibitemShut {NoStop}%
\bibitem [{\citenamefont {Song}\ \emph {et~al.}(2025)\citenamefont {Song},
  \citenamefont {Yang}, \citenamefont {Liu}, \citenamefont {Zhang},
  \citenamefont {Xue}, \citenamefont {Mi}, \citenamefont {Zhang}, \citenamefont
  {Yan}, \citenamefont {Jin},\ and\ \citenamefont {Yu}}]{song2025}%
  \BibitemOpen
  \bibfield  {author} {\bibinfo {author} {\bibfnamefont {J.}~\bibnamefont
  {Song}}, \bibinfo {author} {\bibfnamefont {S.}~\bibnamefont {Yang}}, \bibinfo
  {author} {\bibfnamefont {P.}~\bibnamefont {Liu}}, \bibinfo {author}
  {\bibfnamefont {H.-L.}\ \bibnamefont {Zhang}}, \bibinfo {author}
  {\bibfnamefont {G.-M.}\ \bibnamefont {Xue}}, \bibinfo {author} {\bibfnamefont
  {Z.-Y.}\ \bibnamefont {Mi}}, \bibinfo {author} {\bibfnamefont {W.-G.}\
  \bibnamefont {Zhang}}, \bibinfo {author} {\bibfnamefont {F.}~\bibnamefont
  {Yan}}, \bibinfo {author} {\bibfnamefont {Y.-R.}\ \bibnamefont {Jin}},\ and\
  \bibinfo {author} {\bibfnamefont {H.-F.}\ \bibnamefont {Yu}},\ }\bibfield
  {title} {\bibinfo {title} {Realization of high-fidelity perfect entanglers
  between remote superconducting quantum processors},\ }\href
  {https://doi.org/10.1103/npr7-b7kq} {\bibfield  {journal} {\bibinfo
  {journal} {Phys. Rev. Lett.}\ }\textbf {\bibinfo {volume} {135}},\ \bibinfo
  {pages} {050603} (\bibinfo {year} {2025})}\BibitemShut {NoStop}%
\bibitem [{\citenamefont {Renger}\ \emph {et~al.}(2026)\citenamefont {Renger},
  \citenamefont {Verjauw}, \citenamefont {Wurz}, \citenamefont {Hosseinkhani},
  \citenamefont {Ockeloen-Korppi}, \citenamefont {Liu}, \citenamefont {Rath},
  \citenamefont {Thapa}, \citenamefont {Vigneau}, \citenamefont {Wybo},
  \citenamefont {Bergholm}, \citenamefont {Chan}, \citenamefont {Csat\'ari},
  \citenamefont {Dahl}, \citenamefont {Davletkaliyev}, \citenamefont {Giri},
  \citenamefont {Gusenkova}, \citenamefont {Heimonen}, \citenamefont
  {Hiltunen}, \citenamefont {Hsu}, \citenamefont {Hyypp\"a}, \citenamefont
  {Ikonen}, \citenamefont {Jones}, \citenamefont {Khalid}, \citenamefont {Kim},
  \citenamefont {Koistinen}, \citenamefont {Komlev}, \citenamefont {Kotilahti},
  \citenamefont {Kukushkin}, \citenamefont {Lamprich}, \citenamefont {Landra},
  \citenamefont {Lee}, \citenamefont {Li}, \citenamefont {Liebermann},
  \citenamefont {Majumder}, \citenamefont {M\"antyl\"a}, \citenamefont
  {Marxer}, \citenamefont {van~de Griend}, \citenamefont {Milchakov},
  \citenamefont {Mro\ifmmode~\dot{z}\else \.{z}\fi{}ek}, \citenamefont {Nath},
  \citenamefont {Orell}, \citenamefont {Papi\ifmmode~\check{c}\else
  \v{c}\fi{}}, \citenamefont {Partanen}, \citenamefont {Plyushch},
  \citenamefont {Pogorzalek}, \citenamefont {Ritvas}, \citenamefont {Romero},
  \citenamefont {Sampo}, \citenamefont {Sepp\"al\"a}, \citenamefont {Selinmaa},
  \citenamefont {Sundstr\"om}, \citenamefont {Takmakov}, \citenamefont
  {Tarasinski}, \citenamefont {Tuorila}, \citenamefont {Tyrkk\"o},
  \citenamefont {V\"alimaa}, \citenamefont {Wesdorp}, \citenamefont {Yang},
  \citenamefont {Yu}, \citenamefont {Heinsoo}, \citenamefont {Veps\"al\"ainen},
  \citenamefont {Kindel}, \citenamefont {Ku},\ and\ \citenamefont
  {Deppe}}]{8x68-cx1w}%
  \BibitemOpen
  \bibfield  {author} {\bibinfo {author} {\bibfnamefont {M.}~\bibnamefont
  {Renger}}, \bibinfo {author} {\bibfnamefont {J.}~\bibnamefont {Verjauw}},
  \bibinfo {author} {\bibfnamefont {N.}~\bibnamefont {Wurz}}, \bibinfo {author}
  {\bibfnamefont {A.}~\bibnamefont {Hosseinkhani}}, \bibinfo {author}
  {\bibfnamefont {C.}~\bibnamefont {Ockeloen-Korppi}}, \bibinfo {author}
  {\bibfnamefont {W.}~\bibnamefont {Liu}}, \bibinfo {author} {\bibfnamefont
  {A.}~\bibnamefont {Rath}}, \bibinfo {author} {\bibfnamefont {M.~J.}\
  \bibnamefont {Thapa}}, \bibinfo {author} {\bibfnamefont {F.}~\bibnamefont
  {Vigneau}}, \bibinfo {author} {\bibfnamefont {E.}~\bibnamefont {Wybo}},
  \bibinfo {author} {\bibfnamefont {V.}~\bibnamefont {Bergholm}}, \bibinfo
  {author} {\bibfnamefont {C.~F.}\ \bibnamefont {Chan}}, \bibinfo {author}
  {\bibfnamefont {B.}~\bibnamefont {Csat\'ari}}, \bibinfo {author}
  {\bibfnamefont {S.}~\bibnamefont {Dahl}}, \bibinfo {author} {\bibfnamefont
  {R.}~\bibnamefont {Davletkaliyev}}, \bibinfo {author} {\bibfnamefont
  {R.}~\bibnamefont {Giri}}, \bibinfo {author} {\bibfnamefont {D.}~\bibnamefont
  {Gusenkova}}, \bibinfo {author} {\bibfnamefont {H.}~\bibnamefont {Heimonen}},
  \bibinfo {author} {\bibfnamefont {T.}~\bibnamefont {Hiltunen}}, \bibinfo
  {author} {\bibfnamefont {H.}~\bibnamefont {Hsu}}, \bibinfo {author}
  {\bibfnamefont {E.}~\bibnamefont {Hyypp\"a}}, \bibinfo {author}
  {\bibfnamefont {J.}~\bibnamefont {Ikonen}}, \bibinfo {author} {\bibfnamefont
  {T.}~\bibnamefont {Jones}}, \bibinfo {author} {\bibfnamefont
  {S.}~\bibnamefont {Khalid}}, \bibinfo {author} {\bibfnamefont {S.-G.}\
  \bibnamefont {Kim}}, \bibinfo {author} {\bibfnamefont {M.}~\bibnamefont
  {Koistinen}}, \bibinfo {author} {\bibfnamefont {A.}~\bibnamefont {Komlev}},
  \bibinfo {author} {\bibfnamefont {J.}~\bibnamefont {Kotilahti}}, \bibinfo
  {author} {\bibfnamefont {V.}~\bibnamefont {Kukushkin}}, \bibinfo {author}
  {\bibfnamefont {J.}~\bibnamefont {Lamprich}}, \bibinfo {author}
  {\bibfnamefont {A.}~\bibnamefont {Landra}}, \bibinfo {author} {\bibfnamefont
  {L.-H.}\ \bibnamefont {Lee}}, \bibinfo {author} {\bibfnamefont
  {T.}~\bibnamefont {Li}}, \bibinfo {author} {\bibfnamefont {P.}~\bibnamefont
  {Liebermann}}, \bibinfo {author} {\bibfnamefont {S.}~\bibnamefont
  {Majumder}}, \bibinfo {author} {\bibfnamefont {J.}~\bibnamefont
  {M\"antyl\"a}}, \bibinfo {author} {\bibfnamefont {F.}~\bibnamefont {Marxer}},
  \bibinfo {author} {\bibfnamefont {A.~M.}\ \bibnamefont {van~de Griend}},
  \bibinfo {author} {\bibfnamefont {V.}~\bibnamefont {Milchakov}}, \bibinfo
  {author} {\bibfnamefont {J.}~\bibnamefont {Mro\ifmmode~\dot{z}\else
  \.{z}\fi{}ek}}, \bibinfo {author} {\bibfnamefont {J.}~\bibnamefont {Nath}},
  \bibinfo {author} {\bibfnamefont {T.}~\bibnamefont {Orell}}, \bibinfo
  {author} {\bibfnamefont {M.}~\bibnamefont {Papi\ifmmode~\check{c}\else
  \v{c}\fi{}}}, \bibinfo {author} {\bibfnamefont {M.}~\bibnamefont {Partanen}},
  \bibinfo {author} {\bibfnamefont {A.}~\bibnamefont {Plyushch}}, \bibinfo
  {author} {\bibfnamefont {S.}~\bibnamefont {Pogorzalek}}, \bibinfo {author}
  {\bibfnamefont {J.}~\bibnamefont {Ritvas}}, \bibinfo {author} {\bibfnamefont
  {P.~F.}\ \bibnamefont {Romero}}, \bibinfo {author} {\bibfnamefont
  {V.}~\bibnamefont {Sampo}}, \bibinfo {author} {\bibfnamefont
  {M.}~\bibnamefont {Sepp\"al\"a}}, \bibinfo {author} {\bibfnamefont
  {V.}~\bibnamefont {Selinmaa}}, \bibinfo {author} {\bibfnamefont
  {L.}~\bibnamefont {Sundstr\"om}}, \bibinfo {author} {\bibfnamefont
  {I.}~\bibnamefont {Takmakov}}, \bibinfo {author} {\bibfnamefont
  {B.}~\bibnamefont {Tarasinski}}, \bibinfo {author} {\bibfnamefont
  {J.}~\bibnamefont {Tuorila}}, \bibinfo {author} {\bibfnamefont
  {O.}~\bibnamefont {Tyrkk\"o}}, \bibinfo {author} {\bibfnamefont
  {A.}~\bibnamefont {V\"alimaa}}, \bibinfo {author} {\bibfnamefont
  {J.}~\bibnamefont {Wesdorp}}, \bibinfo {author} {\bibfnamefont
  {P.}~\bibnamefont {Yang}}, \bibinfo {author} {\bibfnamefont {L.}~\bibnamefont
  {Yu}}, \bibinfo {author} {\bibfnamefont {J.}~\bibnamefont {Heinsoo}},
  \bibinfo {author} {\bibfnamefont {A.}~\bibnamefont {Veps\"al\"ainen}},
  \bibinfo {author} {\bibfnamefont {W.}~\bibnamefont {Kindel}}, \bibinfo
  {author} {\bibfnamefont {H.-S.}\ \bibnamefont {Ku}},\ and\ \bibinfo {author}
  {\bibfnamefont {F.}~\bibnamefont {Deppe}},\ }\bibfield  {title} {\bibinfo
  {title} {Superconducting qubit-resonator quantum processor with effective
  all-to-all connectivity},\ }\href {https://doi.org/10.1103/8x68-cx1w}
  {\bibfield  {journal} {\bibinfo  {journal} {Phys. Rev. Res.}\ }\textbf
  {\bibinfo {volume} {8}},\ \bibinfo {pages} {013148} (\bibinfo {year}
  {2026})}\BibitemShut {NoStop}%
\bibitem [{\citenamefont {Marxer}\ \emph {et~al.}(2023)\citenamefont {Marxer},
  \citenamefont {Veps\"al\"ainen}, \citenamefont {Jolin}, \citenamefont
  {Tuorila}, \citenamefont {Landra}, \citenamefont {Ockeloen-Korppi},
  \citenamefont {Liu}, \citenamefont {Ahonen}, \citenamefont {Auer},
  \citenamefont {Belzane}, \citenamefont {Bergholm}, \citenamefont {Chan},
  \citenamefont {Chan}, \citenamefont {Hiltunen}, \citenamefont {Hotari},
  \citenamefont {Hyypp\"a}, \citenamefont {Ikonen}, \citenamefont {Janzso},
  \citenamefont {Koistinen}, \citenamefont {Kotilahti}, \citenamefont {Li},
  \citenamefont {Luus}, \citenamefont {Papic}, \citenamefont {Partanen},
  \citenamefont {R\"abin\"a}, \citenamefont {Rosti}, \citenamefont {Savytskyi},
  \citenamefont {Sepp\"al\"a}, \citenamefont {Sevriuk}, \citenamefont {Takala},
  \citenamefont {Tarasinski}, \citenamefont {Thapa}, \citenamefont {Tosto},
  \citenamefont {Vorobeva}, \citenamefont {Yu}, \citenamefont {Tan},
  \citenamefont {Hassel}, \citenamefont {M\"ott\"onen},\ and\ \citenamefont
  {Heinsoo}}]{PRXQuantum.4.010314}%
  \BibitemOpen
  \bibfield  {author} {\bibinfo {author} {\bibfnamefont {F.}~\bibnamefont
  {Marxer}}, \bibinfo {author} {\bibfnamefont {A.}~\bibnamefont
  {Veps\"al\"ainen}}, \bibinfo {author} {\bibfnamefont {S.~W.}\ \bibnamefont
  {Jolin}}, \bibinfo {author} {\bibfnamefont {J.}~\bibnamefont {Tuorila}},
  \bibinfo {author} {\bibfnamefont {A.}~\bibnamefont {Landra}}, \bibinfo
  {author} {\bibfnamefont {C.}~\bibnamefont {Ockeloen-Korppi}}, \bibinfo
  {author} {\bibfnamefont {W.}~\bibnamefont {Liu}}, \bibinfo {author}
  {\bibfnamefont {O.}~\bibnamefont {Ahonen}}, \bibinfo {author} {\bibfnamefont
  {A.}~\bibnamefont {Auer}}, \bibinfo {author} {\bibfnamefont {L.}~\bibnamefont
  {Belzane}}, \bibinfo {author} {\bibfnamefont {V.}~\bibnamefont {Bergholm}},
  \bibinfo {author} {\bibfnamefont {C.~F.}\ \bibnamefont {Chan}}, \bibinfo
  {author} {\bibfnamefont {K.~W.}\ \bibnamefont {Chan}}, \bibinfo {author}
  {\bibfnamefont {T.}~\bibnamefont {Hiltunen}}, \bibinfo {author}
  {\bibfnamefont {J.}~\bibnamefont {Hotari}}, \bibinfo {author} {\bibfnamefont
  {E.}~\bibnamefont {Hyypp\"a}}, \bibinfo {author} {\bibfnamefont
  {J.}~\bibnamefont {Ikonen}}, \bibinfo {author} {\bibfnamefont
  {D.}~\bibnamefont {Janzso}}, \bibinfo {author} {\bibfnamefont
  {M.}~\bibnamefont {Koistinen}}, \bibinfo {author} {\bibfnamefont
  {J.}~\bibnamefont {Kotilahti}}, \bibinfo {author} {\bibfnamefont
  {T.}~\bibnamefont {Li}}, \bibinfo {author} {\bibfnamefont {J.}~\bibnamefont
  {Luus}}, \bibinfo {author} {\bibfnamefont {M.}~\bibnamefont {Papic}},
  \bibinfo {author} {\bibfnamefont {M.}~\bibnamefont {Partanen}}, \bibinfo
  {author} {\bibfnamefont {J.}~\bibnamefont {R\"abin\"a}}, \bibinfo {author}
  {\bibfnamefont {J.}~\bibnamefont {Rosti}}, \bibinfo {author} {\bibfnamefont
  {M.}~\bibnamefont {Savytskyi}}, \bibinfo {author} {\bibfnamefont
  {M.}~\bibnamefont {Sepp\"al\"a}}, \bibinfo {author} {\bibfnamefont
  {V.}~\bibnamefont {Sevriuk}}, \bibinfo {author} {\bibfnamefont
  {E.}~\bibnamefont {Takala}}, \bibinfo {author} {\bibfnamefont
  {B.}~\bibnamefont {Tarasinski}}, \bibinfo {author} {\bibfnamefont {M.~J.}\
  \bibnamefont {Thapa}}, \bibinfo {author} {\bibfnamefont {F.}~\bibnamefont
  {Tosto}}, \bibinfo {author} {\bibfnamefont {N.}~\bibnamefont {Vorobeva}},
  \bibinfo {author} {\bibfnamefont {L.}~\bibnamefont {Yu}}, \bibinfo {author}
  {\bibfnamefont {K.~Y.}\ \bibnamefont {Tan}}, \bibinfo {author} {\bibfnamefont
  {J.}~\bibnamefont {Hassel}}, \bibinfo {author} {\bibfnamefont
  {M.}~\bibnamefont {M\"ott\"onen}},\ and\ \bibinfo {author} {\bibfnamefont
  {J.}~\bibnamefont {Heinsoo}},\ }\bibfield  {title} {\bibinfo {title}
  {Long-distance transmon coupler with {CZ}-gate fidelity above 99.8\%},\
  }\href {https://doi.org/10.1103/PRXQuantum.4.010314} {\bibfield  {journal}
  {\bibinfo  {journal} {PRX Quantum}\ }\textbf {\bibinfo {volume} {4}},\
  \bibinfo {pages} {010314} (\bibinfo {year} {2023})}\BibitemShut {NoStop}%
\bibitem [{\citenamefont {Yuan}\ \emph {et~al.}()\citenamefont {Yuan},
  \citenamefont {Zhang}, \citenamefont {He}, \citenamefont {Fei}, \citenamefont
  {Han}, \citenamefont {Wang}, \citenamefont {Sun}, \citenamefont {Mu},
  \citenamefont {Zhao}, \citenamefont {Liu}, \citenamefont {Wang},\ and\
  \citenamefont {Shan}}]{yuan2026}%
  \BibitemOpen
  \bibfield  {author} {\bibinfo {author} {\bibfnamefont {B.}~\bibnamefont
  {Yuan}}, \bibinfo {author} {\bibfnamefont {C.}~\bibnamefont {Zhang}},
  \bibinfo {author} {\bibfnamefont {H.}~\bibnamefont {He}}, \bibinfo {author}
  {\bibfnamefont {Y.}~\bibnamefont {Fei}}, \bibinfo {author} {\bibfnamefont
  {C.}~\bibnamefont {Han}}, \bibinfo {author} {\bibfnamefont {S.}~\bibnamefont
  {Wang}}, \bibinfo {author} {\bibfnamefont {H.}~\bibnamefont {Sun}}, \bibinfo
  {author} {\bibfnamefont {Q.}~\bibnamefont {Mu}}, \bibinfo {author}
  {\bibfnamefont {B.}~\bibnamefont {Zhao}}, \bibinfo {author} {\bibfnamefont
  {F.}~\bibnamefont {Liu}}, \bibinfo {author} {\bibfnamefont {W.}~\bibnamefont
  {Wang}},\ and\ \bibinfo {author} {\bibfnamefont {Z.}~\bibnamefont {Shan}},\
  }\bibfield  {title} {\bibinfo {title} {Tunable nonlocal zz interaction for
  remote controlled-z gates between distributed fixed-frequency qubits},\
  }\href {https://arxiv.org/abs/2603.28526} {\ }\Eprint
  {https://arxiv.org/abs/2603.28526} {arXiv:2603.28526 [quant-ph]} \BibitemShut
  {NoStop}%
\bibitem [{\citenamefont {Filipp}\ \emph {et~al.}(2011)\citenamefont {Filipp},
  \citenamefont {Fink}, \citenamefont {Baur}, \citenamefont {Bianchetti},
  \citenamefont {Leek}, \citenamefont {Göppl}, \citenamefont {Steffen},
  \citenamefont {Gambetta}, \citenamefont {Blais},\ and\ \citenamefont
  {Wallraff}}]{filipp2011}%
  \BibitemOpen
  \bibfield  {author} {\bibinfo {author} {\bibfnamefont {S.}~\bibnamefont
  {Filipp}}, \bibinfo {author} {\bibfnamefont {J.~M.}\ \bibnamefont {Fink}},
  \bibinfo {author} {\bibfnamefont {M.}~\bibnamefont {Baur}}, \bibinfo {author}
  {\bibfnamefont {R.}~\bibnamefont {Bianchetti}}, \bibinfo {author}
  {\bibfnamefont {P.~J.}\ \bibnamefont {Leek}}, \bibinfo {author}
  {\bibfnamefont {M.}~\bibnamefont {Göppl}}, \bibinfo {author} {\bibfnamefont
  {L.}~\bibnamefont {Steffen}}, \bibinfo {author} {\bibfnamefont {J.~M.}\
  \bibnamefont {Gambetta}}, \bibinfo {author} {\bibfnamefont {A.}~\bibnamefont
  {Blais}},\ and\ \bibinfo {author} {\bibfnamefont {A.}~\bibnamefont
  {Wallraff}},\ }\bibfield  {title} {\bibinfo {title} {Multimode mediated
  qubit-qubit coupling and dark-state symmetries in circuit quantum
  electrodynamics},\ }\href {https://doi.org/10.1103/PhysRevA.83.063827}
  {\bibfield  {journal} {\bibinfo  {journal} {Phys. Rev. A}\ }\textbf {\bibinfo
  {volume} {83}},\ \bibinfo {pages} {063827} (\bibinfo {year}
  {2011})}\BibitemShut {NoStop}%
\bibitem [{\citenamefont {Deng}\ \emph {et~al.}(2025)\citenamefont {Deng},
  \citenamefont {Zheng}, \citenamefont {Liao}, \citenamefont {Zhou},
  \citenamefont {Ge}, \citenamefont {Zhao}, \citenamefont {Lan}, \citenamefont
  {Tan}, \citenamefont {Zhang}, \citenamefont {Li},\ and\ \citenamefont
  {Yu}}]{deng2025}%
  \BibitemOpen
  \bibfield  {author} {\bibinfo {author} {\bibfnamefont {X.}~\bibnamefont
  {Deng}}, \bibinfo {author} {\bibfnamefont {W.}~\bibnamefont {Zheng}},
  \bibinfo {author} {\bibfnamefont {X.}~\bibnamefont {Liao}}, \bibinfo {author}
  {\bibfnamefont {H.}~\bibnamefont {Zhou}}, \bibinfo {author} {\bibfnamefont
  {Y.}~\bibnamefont {Ge}}, \bibinfo {author} {\bibfnamefont {J.}~\bibnamefont
  {Zhao}}, \bibinfo {author} {\bibfnamefont {D.}~\bibnamefont {Lan}}, \bibinfo
  {author} {\bibfnamefont {X.}~\bibnamefont {Tan}}, \bibinfo {author}
  {\bibfnamefont {Y.}~\bibnamefont {Zhang}}, \bibinfo {author} {\bibfnamefont
  {S.}~\bibnamefont {Li}},\ and\ \bibinfo {author} {\bibfnamefont
  {Y.}~\bibnamefont {Yu}},\ }\bibfield  {title} {\bibinfo {title} {Long-range
  zz interaction via resonator-induced phase in superconducting qubits},\
  }\href {https://doi.org/10.1103/PhysRevLett.134.020801} {\bibfield  {journal}
  {\bibinfo  {journal} {Phys. Rev. Lett.}\ }\textbf {\bibinfo {volume} {134}},\
  \bibinfo {pages} {020801} (\bibinfo {year} {2025})}\BibitemShut {NoStop}%
\bibitem [{\citenamefont {Morvan}\ \emph {et~al.}(2022)\citenamefont {Morvan},
  \citenamefont {Chen}, \citenamefont {Larson}, \citenamefont {Santiago},\ and\
  \citenamefont {Siddiqi}}]{PhysRevResearch.4.023079}%
  \BibitemOpen
  \bibfield  {author} {\bibinfo {author} {\bibfnamefont {A.}~\bibnamefont
  {Morvan}}, \bibinfo {author} {\bibfnamefont {L.}~\bibnamefont {Chen}},
  \bibinfo {author} {\bibfnamefont {J.~M.}\ \bibnamefont {Larson}}, \bibinfo
  {author} {\bibfnamefont {D.~I.}\ \bibnamefont {Santiago}},\ and\ \bibinfo
  {author} {\bibfnamefont {I.}~\bibnamefont {Siddiqi}},\ }\bibfield  {title}
  {\bibinfo {title} {Optimizing frequency allocation for fixed-frequency
  superconducting quantum processors},\ }\href
  {https://doi.org/10.1103/PhysRevResearch.4.023079} {\bibfield  {journal}
  {\bibinfo  {journal} {Phys. Rev. Res.}\ }\textbf {\bibinfo {volume} {4}},\
  \bibinfo {pages} {023079} (\bibinfo {year} {2022})}\BibitemShut {NoStop}%
\bibitem [{\citenamefont {Klimov}\ \emph {et~al.}(2024)\citenamefont {Klimov},
  \citenamefont {Bengtsson}, \citenamefont {Quintana}, \citenamefont
  {Bourassa}, \citenamefont {Hong}, \citenamefont {Dunsworth}, \citenamefont
  {Satzinger}, \citenamefont {Livingston}, \citenamefont {Sivak}, \citenamefont
  {Niu}, \citenamefont {Andersen}, \citenamefont {Zhang}, \citenamefont {Chik},
  \citenamefont {Chen}, \citenamefont {Neill}, \citenamefont {Erickson},
  \citenamefont {Grajales~Dau}, \citenamefont {Megrant}, \citenamefont
  {Roushan}, \citenamefont {Korotkov}, \citenamefont {Kelly}, \citenamefont
  {Smelyanskiy}, \citenamefont {Chen},\ and\ \citenamefont
  {Neven}}]{Klimov2024}%
  \BibitemOpen
  \bibfield  {author} {\bibinfo {author} {\bibfnamefont {P.~V.}\ \bibnamefont
  {Klimov}}, \bibinfo {author} {\bibfnamefont {A.}~\bibnamefont {Bengtsson}},
  \bibinfo {author} {\bibfnamefont {C.}~\bibnamefont {Quintana}}, \bibinfo
  {author} {\bibfnamefont {A.}~\bibnamefont {Bourassa}}, \bibinfo {author}
  {\bibfnamefont {S.}~\bibnamefont {Hong}}, \bibinfo {author} {\bibfnamefont
  {A.}~\bibnamefont {Dunsworth}}, \bibinfo {author} {\bibfnamefont {K.~J.}\
  \bibnamefont {Satzinger}}, \bibinfo {author} {\bibfnamefont {W.~P.}\
  \bibnamefont {Livingston}}, \bibinfo {author} {\bibfnamefont
  {V.}~\bibnamefont {Sivak}}, \bibinfo {author} {\bibfnamefont {M.~Y.}\
  \bibnamefont {Niu}}, \bibinfo {author} {\bibfnamefont {T.~I.}\ \bibnamefont
  {Andersen}}, \bibinfo {author} {\bibfnamefont {Y.}~\bibnamefont {Zhang}},
  \bibinfo {author} {\bibfnamefont {D.}~\bibnamefont {Chik}}, \bibinfo {author}
  {\bibfnamefont {Z.}~\bibnamefont {Chen}}, \bibinfo {author} {\bibfnamefont
  {C.}~\bibnamefont {Neill}}, \bibinfo {author} {\bibfnamefont
  {C.}~\bibnamefont {Erickson}}, \bibinfo {author} {\bibfnamefont
  {A.}~\bibnamefont {Grajales~Dau}}, \bibinfo {author} {\bibfnamefont
  {A.}~\bibnamefont {Megrant}}, \bibinfo {author} {\bibfnamefont
  {P.}~\bibnamefont {Roushan}}, \bibinfo {author} {\bibfnamefont {A.~N.}\
  \bibnamefont {Korotkov}}, \bibinfo {author} {\bibfnamefont {J.}~\bibnamefont
  {Kelly}}, \bibinfo {author} {\bibfnamefont {V.}~\bibnamefont {Smelyanskiy}},
  \bibinfo {author} {\bibfnamefont {Y.}~\bibnamefont {Chen}},\ and\ \bibinfo
  {author} {\bibfnamefont {H.}~\bibnamefont {Neven}},\ }\bibfield  {title}
  {\bibinfo {title} {Optimizing quantum gates towards the scale of logical
  qubits},\ }\href {https://doi.org/10.1038/s41467-024-46623-y} {\bibfield
  {journal} {\bibinfo  {journal} {Nature Communications}\ }\textbf {\bibinfo
  {volume} {15}},\ \bibinfo {pages} {2442} (\bibinfo {year}
  {2024})}\BibitemShut {NoStop}%
\bibitem [{\citenamefont {Kueng}\ \emph {et~al.}(2016)\citenamefont {Kueng},
  \citenamefont {Long}, \citenamefont {Doherty},\ and\ \citenamefont
  {Flammia}}]{PhysRevLett.117.170502}%
  \BibitemOpen
  \bibfield  {author} {\bibinfo {author} {\bibfnamefont {R.}~\bibnamefont
  {Kueng}}, \bibinfo {author} {\bibfnamefont {D.~M.}\ \bibnamefont {Long}},
  \bibinfo {author} {\bibfnamefont {A.~C.}\ \bibnamefont {Doherty}},\ and\
  \bibinfo {author} {\bibfnamefont {S.~T.}\ \bibnamefont {Flammia}},\
  }\bibfield  {title} {\bibinfo {title} {Comparing experiments to the
  fault-tolerance threshold},\ }\href
  {https://doi.org/10.1103/PhysRevLett.117.170502} {\bibfield  {journal}
  {\bibinfo  {journal} {Phys. Rev. Lett.}\ }\textbf {\bibinfo {volume} {117}},\
  \bibinfo {pages} {170502} (\bibinfo {year} {2016})}\BibitemShut {NoStop}%
\bibitem [{\citenamefont {Ai}\ and\ \citenamefont {Liu}(2025)}]{yr9d-7z8k}%
  \BibitemOpen
  \bibfield  {author} {\bibinfo {author} {\bibfnamefont {H.}~\bibnamefont
  {Ai}}\ and\ \bibinfo {author} {\bibfnamefont {Y.-x.}\ \bibnamefont {Liu}},\
  }\bibfield  {title} {\bibinfo {title} {Scalable parameter design for
  superconducting quantum circuits with graph neural networks},\ }\href
  {https://doi.org/10.1103/yr9d-7z8k} {\bibfield  {journal} {\bibinfo
  {journal} {Phys. Rev. Lett.}\ }\textbf {\bibinfo {volume} {135}},\ \bibinfo
  {pages} {040601} (\bibinfo {year} {2025})}\BibitemShut {NoStop}%
\bibitem [{\citenamefont {Mundada}\ \emph {et~al.}(2019)\citenamefont
  {Mundada}, \citenamefont {Zhang}, \citenamefont {Hazard},\ and\ \citenamefont
  {Houck}}]{mundada2019}%
  \BibitemOpen
  \bibfield  {author} {\bibinfo {author} {\bibfnamefont {P.}~\bibnamefont
  {Mundada}}, \bibinfo {author} {\bibfnamefont {G.}~\bibnamefont {Zhang}},
  \bibinfo {author} {\bibfnamefont {T.}~\bibnamefont {Hazard}},\ and\ \bibinfo
  {author} {\bibfnamefont {A.~A.}\ \bibnamefont {Houck}},\ }\bibfield  {title}
  {\bibinfo {title} {Suppression of qubit crosstalk in a tunable coupling
  superconducting circuit},\ }\href
  {https://doi.org/10.1103/PhysRevApplied.12.054023} {\bibfield  {journal}
  {\bibinfo  {journal} {Phys. Rev. Applied}\ }\textbf {\bibinfo {volume}
  {12}},\ \bibinfo {pages} {054023} (\bibinfo {year} {2019})}\BibitemShut
  {NoStop}%
\bibitem [{\citenamefont {Sung}\ \emph {et~al.}(2021)\citenamefont {Sung},
  \citenamefont {Veps"al"ainen}, \citenamefont {Kannan} \emph
  {et~al.}}]{sung2021}%
  \BibitemOpen
  \bibfield  {author} {\bibinfo {author} {\bibfnamefont {Y.}~\bibnamefont
  {Sung}}, \bibinfo {author} {\bibfnamefont {A.}~\bibnamefont {Veps"al"ainen}},
  \bibinfo {author} {\bibfnamefont {B.}~\bibnamefont {Kannan}}, \emph
  {et~al.},\ }\bibfield  {title} {\bibinfo {title} {Realization of
  high-fidelity cz and zz-free iswap gates with a tunable coupler},\ }\href
  {https://doi.org/10.1103/PhysRevX.11.021058} {\bibfield  {journal} {\bibinfo
  {journal} {Phys. Rev. X}\ }\textbf {\bibinfo {volume} {11}},\ \bibinfo
  {pages} {021058} (\bibinfo {year} {2021})}\BibitemShut {NoStop}%
\bibitem [{\citenamefont {Zhao}\ \emph
  {et~al.}(2023{\natexlab{a}})\citenamefont {Zhao}, \citenamefont {Yang},
  \citenamefont {Wang}, \citenamefont {Yan}, \citenamefont {Li}, \citenamefont
  {Wang},\ and\ \citenamefont {Zhou}}]{PhysRevA.108.043703}%
  \BibitemOpen
  \bibfield  {author} {\bibinfo {author} {\bibfnamefont {C.}~\bibnamefont
  {Zhao}}, \bibinfo {author} {\bibfnamefont {Z.}~\bibnamefont {Yang}}, \bibinfo
  {author} {\bibfnamefont {D.}~\bibnamefont {Wang}}, \bibinfo {author}
  {\bibfnamefont {Y.}~\bibnamefont {Yan}}, \bibinfo {author} {\bibfnamefont
  {C.}~\bibnamefont {Li}}, \bibinfo {author} {\bibfnamefont {Z.}~\bibnamefont
  {Wang}},\ and\ \bibinfo {author} {\bibfnamefont {L.}~\bibnamefont {Zhou}},\
  }\bibfield  {title} {\bibinfo {title} {Quantum networks assisted by dark
  modes in optomagnonic systems},\ }\href
  {https://doi.org/10.1103/PhysRevA.108.043703} {\bibfield  {journal} {\bibinfo
   {journal} {Phys. Rev. A}\ }\textbf {\bibinfo {volume} {108}},\ \bibinfo
  {pages} {043703} (\bibinfo {year} {2023}{\natexlab{a}})}\BibitemShut
  {NoStop}%
\bibitem [{\citenamefont {Strauch}\ \emph {et~al.}(2003)\citenamefont
  {Strauch}, \citenamefont {Johnson}, \citenamefont {Dragt}, \citenamefont
  {Lobb}, \citenamefont {Anderson},\ and\ \citenamefont
  {Wellstood}}]{PhysRevLett.91.167005}%
  \BibitemOpen
  \bibfield  {author} {\bibinfo {author} {\bibfnamefont {F.~W.}\ \bibnamefont
  {Strauch}}, \bibinfo {author} {\bibfnamefont {P.~R.}\ \bibnamefont
  {Johnson}}, \bibinfo {author} {\bibfnamefont {A.~J.}\ \bibnamefont {Dragt}},
  \bibinfo {author} {\bibfnamefont {C.~J.}\ \bibnamefont {Lobb}}, \bibinfo
  {author} {\bibfnamefont {J.~R.}\ \bibnamefont {Anderson}},\ and\ \bibinfo
  {author} {\bibfnamefont {F.~C.}\ \bibnamefont {Wellstood}},\ }\bibfield
  {title} {\bibinfo {title} {Quantum logic gates for coupled superconducting
  phase qubits},\ }\href {https://doi.org/10.1103/PhysRevLett.91.167005}
  {\bibfield  {journal} {\bibinfo  {journal} {Phys. Rev. Lett.}\ }\textbf
  {\bibinfo {volume} {91}},\ \bibinfo {pages} {167005} (\bibinfo {year}
  {2003})}\BibitemShut {NoStop}%
\bibitem [{\citenamefont {Zhao}\ \emph {et~al.}(2026)\citenamefont {Zhao},
  \citenamefont {Xu},\ and\ \citenamefont {Xue}}]{xwny-vkft}%
  \BibitemOpen
  \bibfield  {author} {\bibinfo {author} {\bibfnamefont {P.}~\bibnamefont
  {Zhao}}, \bibinfo {author} {\bibfnamefont {P.}~\bibnamefont {Xu}},\ and\
  \bibinfo {author} {\bibfnamefont {Z.-Y.}\ \bibnamefont {Xue}},\ }\bibfield
  {title} {\bibinfo {title} {Scalable native multiqubit gates via engineered
  noncomputational-state interactions in superconducting fluxonium qubits},\
  }\href {https://doi.org/10.1103/xwny-vkft} {\bibfield  {journal} {\bibinfo
  {journal} {Phys. Rev. A}\ }\textbf {\bibinfo {volume} {113}},\ \bibinfo
  {pages} {022604} (\bibinfo {year} {2026})}\BibitemShut {NoStop}%
\bibitem [{\citenamefont {McKay}\ \emph {et~al.}(2016)\citenamefont {McKay},
  \citenamefont {Filipp}, \citenamefont {Mezzacapo}, \citenamefont {Magesan},
  \citenamefont {Chow},\ and\ \citenamefont {Gambetta}}]{mckay2016}%
  \BibitemOpen
  \bibfield  {author} {\bibinfo {author} {\bibfnamefont {D.~C.}\ \bibnamefont
  {McKay}}, \bibinfo {author} {\bibfnamefont {S.}~\bibnamefont {Filipp}},
  \bibinfo {author} {\bibfnamefont {A.}~\bibnamefont {Mezzacapo}}, \bibinfo
  {author} {\bibfnamefont {E.}~\bibnamefont {Magesan}}, \bibinfo {author}
  {\bibfnamefont {J.~M.}\ \bibnamefont {Chow}},\ and\ \bibinfo {author}
  {\bibfnamefont {J.~M.}\ \bibnamefont {Gambetta}},\ }\bibfield  {title}
  {\bibinfo {title} {Universal gate for fixed-frequency qubits via a tunable
  bus},\ }\href {https://doi.org/10.1103/PhysRevApplied.6.064007} {\bibfield
  {journal} {\bibinfo  {journal} {Phys. Rev. Appl.}\ }\textbf {\bibinfo
  {volume} {6}},\ \bibinfo {pages} {064007} (\bibinfo {year}
  {2016})}\BibitemShut {NoStop}%
\bibitem [{\citenamefont {Xu}\ \emph {et~al.}(2020)\citenamefont {Xu},
  \citenamefont {Chu}, \citenamefont {Yuan}, \citenamefont {Qiu}, \citenamefont
  {Zhou}, \citenamefont {Zhang}, \citenamefont {Tan}, \citenamefont {Yu},
  \citenamefont {Liu}, \citenamefont {Li}, \citenamefont {Yan},\ and\
  \citenamefont {Yu}}]{PhysRevLett.125.240503}%
  \BibitemOpen
  \bibfield  {author} {\bibinfo {author} {\bibfnamefont {Y.}~\bibnamefont
  {Xu}}, \bibinfo {author} {\bibfnamefont {J.}~\bibnamefont {Chu}}, \bibinfo
  {author} {\bibfnamefont {J.}~\bibnamefont {Yuan}}, \bibinfo {author}
  {\bibfnamefont {J.}~\bibnamefont {Qiu}}, \bibinfo {author} {\bibfnamefont
  {Y.}~\bibnamefont {Zhou}}, \bibinfo {author} {\bibfnamefont {L.}~\bibnamefont
  {Zhang}}, \bibinfo {author} {\bibfnamefont {X.}~\bibnamefont {Tan}}, \bibinfo
  {author} {\bibfnamefont {Y.}~\bibnamefont {Yu}}, \bibinfo {author}
  {\bibfnamefont {S.}~\bibnamefont {Liu}}, \bibinfo {author} {\bibfnamefont
  {J.}~\bibnamefont {Li}}, \bibinfo {author} {\bibfnamefont {F.}~\bibnamefont
  {Yan}},\ and\ \bibinfo {author} {\bibfnamefont {D.}~\bibnamefont {Yu}},\
  }\bibfield  {title} {\bibinfo {title} {High-fidelity, high-scalability
  two-qubit gate scheme for superconducting qubits},\ }\href
  {https://doi.org/10.1103/PhysRevLett.125.240503} {\bibfield  {journal}
  {\bibinfo  {journal} {Phys. Rev. Lett.}\ }\textbf {\bibinfo {volume} {125}},\
  \bibinfo {pages} {240503} (\bibinfo {year} {2020})}\BibitemShut {NoStop}%
\bibitem [{\citenamefont {Chen}\ \emph {et~al.}(2014)\citenamefont {Chen},
  \citenamefont {Neill}, \citenamefont {Roushan}, \citenamefont {Leung},
  \citenamefont {Fang}, \citenamefont {Barends}, \citenamefont {Kelly},
  \citenamefont {Campbell}, \citenamefont {Chen}, \citenamefont {Chiaro} \emph
  {et~al.}}]{chen2014}%
  \BibitemOpen
  \bibfield  {author} {\bibinfo {author} {\bibfnamefont {Y.}~\bibnamefont
  {Chen}}, \bibinfo {author} {\bibfnamefont {C.}~\bibnamefont {Neill}},
  \bibinfo {author} {\bibfnamefont {P.}~\bibnamefont {Roushan}}, \bibinfo
  {author} {\bibfnamefont {N.}~\bibnamefont {Leung}}, \bibinfo {author}
  {\bibfnamefont {M.}~\bibnamefont {Fang}}, \bibinfo {author} {\bibfnamefont
  {R.}~\bibnamefont {Barends}}, \bibinfo {author} {\bibfnamefont
  {J.}~\bibnamefont {Kelly}}, \bibinfo {author} {\bibfnamefont
  {B.}~\bibnamefont {Campbell}}, \bibinfo {author} {\bibfnamefont
  {Z.}~\bibnamefont {Chen}}, \bibinfo {author} {\bibfnamefont {B.}~\bibnamefont
  {Chiaro}}, \emph {et~al.},\ }\bibfield  {title} {\bibinfo {title} {Qubit
  architecture with high coherence and fast tunable coupling},\ }\href
  {https://doi.org/10.1103/PhysRevLett.113.220502} {\bibfield  {journal}
  {\bibinfo  {journal} {Phys. Rev. Lett.}\ }\textbf {\bibinfo {volume} {113}},\
  \bibinfo {pages} {220502} (\bibinfo {year} {2014})}\BibitemShut {NoStop}%
\bibitem [{\citenamefont {Yan}\ \emph {et~al.}(2018)\citenamefont {Yan},
  \citenamefont {Campbell}, \citenamefont {Krantz}, \citenamefont {Kjaergaard},
  \citenamefont {Kim}, \citenamefont {Yoder}, \citenamefont {Sung},
  \citenamefont {Orlando}, \citenamefont {Gustavsson},\ and\ \citenamefont
  {Oliver}}]{yan2018}%
  \BibitemOpen
  \bibfield  {author} {\bibinfo {author} {\bibfnamefont {F.}~\bibnamefont
  {Yan}}, \bibinfo {author} {\bibfnamefont {D.~L.}\ \bibnamefont {Campbell}},
  \bibinfo {author} {\bibfnamefont {P.}~\bibnamefont {Krantz}}, \bibinfo
  {author} {\bibfnamefont {M.}~\bibnamefont {Kjaergaard}}, \bibinfo {author}
  {\bibfnamefont {D.}~\bibnamefont {Kim}}, \bibinfo {author} {\bibfnamefont
  {J.~L.}\ \bibnamefont {Yoder}}, \bibinfo {author} {\bibfnamefont
  {Y.}~\bibnamefont {Sung}}, \bibinfo {author} {\bibfnamefont {T.~P.}\
  \bibnamefont {Orlando}}, \bibinfo {author} {\bibfnamefont {S.}~\bibnamefont
  {Gustavsson}},\ and\ \bibinfo {author} {\bibfnamefont {W.~D.}\ \bibnamefont
  {Oliver}},\ }\bibfield  {title} {\bibinfo {title} {Tunable coupling scheme
  for implementing high-fidelity two-qubit gates},\ }\href
  {https://doi.org/10.1103/PhysRevApplied.10.054062} {\bibfield  {journal}
  {\bibinfo  {journal} {Phys. Rev. Applied}\ }\textbf {\bibinfo {volume}
  {10}},\ \bibinfo {pages} {054062} (\bibinfo {year} {2018})}\BibitemShut
  {NoStop}%
\bibitem [{\citenamefont {Foxen}\ \emph {et~al.}(2020)\citenamefont {Foxen},
  \citenamefont {Neill}, \citenamefont {Dunsworth} \emph {et~al.}}]{foxen2020}%
  \BibitemOpen
  \bibfield  {author} {\bibinfo {author} {\bibfnamefont {B.}~\bibnamefont
  {Foxen}}, \bibinfo {author} {\bibfnamefont {C.}~\bibnamefont {Neill}},
  \bibinfo {author} {\bibfnamefont {A.}~\bibnamefont {Dunsworth}}, \emph
  {et~al.},\ }\bibfield  {title} {\bibinfo {title} {Demonstrating a continuous
  set of two-qubit gates for near-term quantum algorithms},\ }\href
  {https://doi.org/10.1103/PhysRevLett.125.120504} {\bibfield  {journal}
  {\bibinfo  {journal} {Phys. Rev. Lett.}\ }\textbf {\bibinfo {volume} {125}},\
  \bibinfo {pages} {120504} (\bibinfo {year} {2020})}\BibitemShut {NoStop}%
\bibitem [{\citenamefont {Ding}\ \emph {et~al.}(2023)\citenamefont {Ding},
  \citenamefont {Hays}, \citenamefont {Sung}, \citenamefont {Kannan},
  \citenamefont {An}, \citenamefont {Di~Paolo}, \citenamefont {Karamlou},
  \citenamefont {Hazard}, \citenamefont {Azar}, \citenamefont {Kim},
  \citenamefont {Niedzielski}, \citenamefont {Melville}, \citenamefont
  {Schwartz}, \citenamefont {Yoder}, \citenamefont {Orlando}, \citenamefont
  {Gustavsson}, \citenamefont {Grover}, \citenamefont {Serniak},\ and\
  \citenamefont {Oliver}}]{PhysRevX.13.031035}%
  \BibitemOpen
  \bibfield  {author} {\bibinfo {author} {\bibfnamefont {L.}~\bibnamefont
  {Ding}}, \bibinfo {author} {\bibfnamefont {M.}~\bibnamefont {Hays}}, \bibinfo
  {author} {\bibfnamefont {Y.}~\bibnamefont {Sung}}, \bibinfo {author}
  {\bibfnamefont {B.}~\bibnamefont {Kannan}}, \bibinfo {author} {\bibfnamefont
  {J.}~\bibnamefont {An}}, \bibinfo {author} {\bibfnamefont {A.}~\bibnamefont
  {Di~Paolo}}, \bibinfo {author} {\bibfnamefont {A.~H.}\ \bibnamefont
  {Karamlou}}, \bibinfo {author} {\bibfnamefont {T.~M.}\ \bibnamefont
  {Hazard}}, \bibinfo {author} {\bibfnamefont {K.}~\bibnamefont {Azar}},
  \bibinfo {author} {\bibfnamefont {D.~K.}\ \bibnamefont {Kim}}, \bibinfo
  {author} {\bibfnamefont {B.~M.}\ \bibnamefont {Niedzielski}}, \bibinfo
  {author} {\bibfnamefont {A.}~\bibnamefont {Melville}}, \bibinfo {author}
  {\bibfnamefont {M.~E.}\ \bibnamefont {Schwartz}}, \bibinfo {author}
  {\bibfnamefont {J.~L.}\ \bibnamefont {Yoder}}, \bibinfo {author}
  {\bibfnamefont {T.~P.}\ \bibnamefont {Orlando}}, \bibinfo {author}
  {\bibfnamefont {S.}~\bibnamefont {Gustavsson}}, \bibinfo {author}
  {\bibfnamefont {J.~A.}\ \bibnamefont {Grover}}, \bibinfo {author}
  {\bibfnamefont {K.}~\bibnamefont {Serniak}},\ and\ \bibinfo {author}
  {\bibfnamefont {W.~D.}\ \bibnamefont {Oliver}},\ }\bibfield  {title}
  {\bibinfo {title} {High-fidelity, frequency-flexible two-qubit fluxonium
  gates with a transmon coupler},\ }\href
  {https://doi.org/10.1103/PhysRevX.13.031035} {\bibfield  {journal} {\bibinfo
  {journal} {Phys. Rev. X}\ }\textbf {\bibinfo {volume} {13}},\ \bibinfo
  {pages} {031035} (\bibinfo {year} {2023})}\BibitemShut {NoStop}%
\bibitem [{\citenamefont {Kandala}\ \emph {et~al.}(2021)\citenamefont
  {Kandala}, \citenamefont {Wei}, \citenamefont {Srinivasan}, \citenamefont
  {Magesan}, \citenamefont {Carnevale}, \citenamefont {Keefe}, \citenamefont
  {Klaus}, \citenamefont {Dial},\ and\ \citenamefont
  {McKay}}]{PhysRevLett.127.130501}%
  \BibitemOpen
  \bibfield  {author} {\bibinfo {author} {\bibfnamefont {A.}~\bibnamefont
  {Kandala}}, \bibinfo {author} {\bibfnamefont {K.~X.}\ \bibnamefont {Wei}},
  \bibinfo {author} {\bibfnamefont {S.}~\bibnamefont {Srinivasan}}, \bibinfo
  {author} {\bibfnamefont {E.}~\bibnamefont {Magesan}}, \bibinfo {author}
  {\bibfnamefont {S.}~\bibnamefont {Carnevale}}, \bibinfo {author}
  {\bibfnamefont {G.~A.}\ \bibnamefont {Keefe}}, \bibinfo {author}
  {\bibfnamefont {D.}~\bibnamefont {Klaus}}, \bibinfo {author} {\bibfnamefont
  {O.}~\bibnamefont {Dial}},\ and\ \bibinfo {author} {\bibfnamefont {D.~C.}\
  \bibnamefont {McKay}},\ }\bibfield  {title} {\bibinfo {title} {Demonstration
  of a high-fidelity cnot gate for fixed-frequency transmons with engineered
  $zz$ suppression},\ }\href {https://doi.org/10.1103/PhysRevLett.127.130501}
  {\bibfield  {journal} {\bibinfo  {journal} {Phys. Rev. Lett.}\ }\textbf
  {\bibinfo {volume} {127}},\ \bibinfo {pages} {130501} (\bibinfo {year}
  {2021})}\BibitemShut {NoStop}%
\bibitem [{\citenamefont {Ku}\ \emph {et~al.}(2020)\citenamefont {Ku},
  \citenamefont {Xu}, \citenamefont {Brink}, \citenamefont {McKay},
  \citenamefont {Hertzberg}, \citenamefont {Ansari},\ and\ \citenamefont
  {Plourde}}]{PhysRevLett.125.200504}%
  \BibitemOpen
  \bibfield  {author} {\bibinfo {author} {\bibfnamefont {J.}~\bibnamefont
  {Ku}}, \bibinfo {author} {\bibfnamefont {X.}~\bibnamefont {Xu}}, \bibinfo
  {author} {\bibfnamefont {M.}~\bibnamefont {Brink}}, \bibinfo {author}
  {\bibfnamefont {D.~C.}\ \bibnamefont {McKay}}, \bibinfo {author}
  {\bibfnamefont {J.~B.}\ \bibnamefont {Hertzberg}}, \bibinfo {author}
  {\bibfnamefont {M.~H.}\ \bibnamefont {Ansari}},\ and\ \bibinfo {author}
  {\bibfnamefont {B.~L.~T.}\ \bibnamefont {Plourde}},\ }\bibfield  {title}
  {\bibinfo {title} {Suppression of unwanted $zz$ interactions in a hybrid
  two-qubit system},\ }\href {https://doi.org/10.1103/PhysRevLett.125.200504}
  {\bibfield  {journal} {\bibinfo  {journal} {Phys. Rev. Lett.}\ }\textbf
  {\bibinfo {volume} {125}},\ \bibinfo {pages} {200504} (\bibinfo {year}
  {2020})}\BibitemShut {NoStop}%
\bibitem [{\citenamefont {Li}\ \emph {et~al.}(2024)\citenamefont {Li},
  \citenamefont {Kubo}, \citenamefont {Ho}, \citenamefont {Yan}, \citenamefont
  {Nakamura},\ and\ \citenamefont {Goto}}]{PhysRevX.14.041050}%
  \BibitemOpen
  \bibfield  {author} {\bibinfo {author} {\bibfnamefont {R.}~\bibnamefont
  {Li}}, \bibinfo {author} {\bibfnamefont {K.}~\bibnamefont {Kubo}}, \bibinfo
  {author} {\bibfnamefont {Y.}~\bibnamefont {Ho}}, \bibinfo {author}
  {\bibfnamefont {Z.}~\bibnamefont {Yan}}, \bibinfo {author} {\bibfnamefont
  {Y.}~\bibnamefont {Nakamura}},\ and\ \bibinfo {author} {\bibfnamefont
  {H.}~\bibnamefont {Goto}},\ }\bibfield  {title} {\bibinfo {title}
  {Realization of high-fidelity cz gate based on a double-transmon coupler},\
  }\href {https://doi.org/10.1103/PhysRevX.14.041050} {\bibfield  {journal}
  {\bibinfo  {journal} {Phys. Rev. X}\ }\textbf {\bibinfo {volume} {14}},\
  \bibinfo {pages} {041050} (\bibinfo {year} {2024})}\BibitemShut {NoStop}%
\bibitem [{\citenamefont {Zhao}\ \emph {et~al.}(2021)\citenamefont {Zhao},
  \citenamefont {Lan}, \citenamefont {Xu}, \citenamefont {Xue}, \citenamefont
  {Blank}, \citenamefont {Tan}, \citenamefont {Yu},\ and\ \citenamefont
  {Yu}}]{PhysRevApplied.16.024037}%
  \BibitemOpen
  \bibfield  {author} {\bibinfo {author} {\bibfnamefont {P.}~\bibnamefont
  {Zhao}}, \bibinfo {author} {\bibfnamefont {D.}~\bibnamefont {Lan}}, \bibinfo
  {author} {\bibfnamefont {P.}~\bibnamefont {Xu}}, \bibinfo {author}
  {\bibfnamefont {G.}~\bibnamefont {Xue}}, \bibinfo {author} {\bibfnamefont
  {M.}~\bibnamefont {Blank}}, \bibinfo {author} {\bibfnamefont
  {X.}~\bibnamefont {Tan}}, \bibinfo {author} {\bibfnamefont {H.}~\bibnamefont
  {Yu}},\ and\ \bibinfo {author} {\bibfnamefont {Y.}~\bibnamefont {Yu}},\
  }\bibfield  {title} {\bibinfo {title} {Suppression of static $zz$ interaction
  in an all-transmon quantum processor},\ }\href
  {https://doi.org/10.1103/PhysRevApplied.16.024037} {\bibfield  {journal}
  {\bibinfo  {journal} {Phys. Rev. Appl.}\ }\textbf {\bibinfo {volume} {16}},\
  \bibinfo {pages} {024037} (\bibinfo {year} {2021})}\BibitemShut {NoStop}%
\bibitem [{\citenamefont {Zhang}\ \emph {et~al.}(2025)\citenamefont {Zhang},
  \citenamefont {Chu}, \citenamefont {Niu}, \citenamefont {Zhong},\ and\
  \citenamefont {Yu}}]{10.1063/5.0280529}%
  \BibitemOpen
  \bibfield  {author} {\bibinfo {author} {\bibfnamefont {J.}~\bibnamefont
  {Zhang}}, \bibinfo {author} {\bibfnamefont {J.}~\bibnamefont {Chu}}, \bibinfo
  {author} {\bibfnamefont {J.}~\bibnamefont {Niu}}, \bibinfo {author}
  {\bibfnamefont {Y.}~\bibnamefont {Zhong}},\ and\ \bibinfo {author}
  {\bibfnamefont {D.}~\bibnamefont {Yu}},\ }\bibfield  {title} {\bibinfo
  {title} {High-contrast interaction between remote superconducting qubits
  mediated by multimode cable coupling},\ }\href
  {https://doi.org/10.1063/5.0280529} {\bibfield  {journal} {\bibinfo
  {journal} {Appl. Phys. Lett.}\ }\textbf {\bibinfo {volume} {127}},\ \bibinfo
  {pages} {054001} (\bibinfo {year} {2025})}\BibitemShut {NoStop}%
\bibitem [{\citenamefont {Egorova}\ \emph {et~al.}(2025)\citenamefont
  {Egorova}, \citenamefont {Kazmina}, \citenamefont {Simakov}, \citenamefont
  {Moskalenko}, \citenamefont {Abramov}, \citenamefont {Kalacheva},
  \citenamefont {Lubsanov}, \citenamefont {Bolgar}, \citenamefont {Maleeva},\
  and\ \citenamefont {Besedin}}]{2h4m-mg2p}%
  \BibitemOpen
  \bibfield  {author} {\bibinfo {author} {\bibfnamefont {E.~Y.}\ \bibnamefont
  {Egorova}}, \bibinfo {author} {\bibfnamefont {A.~S.}\ \bibnamefont
  {Kazmina}}, \bibinfo {author} {\bibfnamefont {I.~A.}\ \bibnamefont
  {Simakov}}, \bibinfo {author} {\bibfnamefont {I.~N.}\ \bibnamefont
  {Moskalenko}}, \bibinfo {author} {\bibfnamefont {N.~N.}\ \bibnamefont
  {Abramov}}, \bibinfo {author} {\bibfnamefont {D.~A.}\ \bibnamefont
  {Kalacheva}}, \bibinfo {author} {\bibfnamefont {V.~B.}\ \bibnamefont
  {Lubsanov}}, \bibinfo {author} {\bibfnamefont {A.~N.}\ \bibnamefont
  {Bolgar}}, \bibinfo {author} {\bibfnamefont {N.}~\bibnamefont {Maleeva}},\
  and\ \bibinfo {author} {\bibfnamefont {I.~S.}\ \bibnamefont {Besedin}},\
  }\bibfield  {title} {\bibinfo {title} {Three-mode tunable coupler for
  superconducting two-qubit gates},\ }\href {https://doi.org/10.1103/2h4m-mg2p}
  {\bibfield  {journal} {\bibinfo  {journal} {Phys. Rev. Appl.}\ }\textbf
  {\bibinfo {volume} {23}},\ \bibinfo {pages} {064056} (\bibinfo {year}
  {2025})}\BibitemShut {NoStop}%
\bibitem [{\citenamefont {Houck}\ \emph {et~al.}(2008)\citenamefont {Houck},
  \citenamefont {Schreier}, \citenamefont {Johnson}, \citenamefont {Chow},
  \citenamefont {Koch}, \citenamefont {Gambetta}, \citenamefont {Schuster},
  \citenamefont {Frunzio}, \citenamefont {Devoret}, \citenamefont {Girvin},\
  and\ \citenamefont {Schoelkopf}}]{PhysRevLett.101.080502}%
  \BibitemOpen
  \bibfield  {author} {\bibinfo {author} {\bibfnamefont {A.~A.}\ \bibnamefont
  {Houck}}, \bibinfo {author} {\bibfnamefont {J.~A.}\ \bibnamefont {Schreier}},
  \bibinfo {author} {\bibfnamefont {B.~R.}\ \bibnamefont {Johnson}}, \bibinfo
  {author} {\bibfnamefont {J.~M.}\ \bibnamefont {Chow}}, \bibinfo {author}
  {\bibfnamefont {J.}~\bibnamefont {Koch}}, \bibinfo {author} {\bibfnamefont
  {J.~M.}\ \bibnamefont {Gambetta}}, \bibinfo {author} {\bibfnamefont {D.~I.}\
  \bibnamefont {Schuster}}, \bibinfo {author} {\bibfnamefont {L.}~\bibnamefont
  {Frunzio}}, \bibinfo {author} {\bibfnamefont {M.~H.}\ \bibnamefont
  {Devoret}}, \bibinfo {author} {\bibfnamefont {S.~M.}\ \bibnamefont
  {Girvin}},\ and\ \bibinfo {author} {\bibfnamefont {R.~J.}\ \bibnamefont
  {Schoelkopf}},\ }\bibfield  {title} {\bibinfo {title} {Controlling the
  spontaneous emission of a superconducting transmon qubit},\ }\href
  {https://doi.org/10.1103/PhysRevLett.101.080502} {\bibfield  {journal}
  {\bibinfo  {journal} {Phys. Rev. Lett.}\ }\textbf {\bibinfo {volume} {101}},\
  \bibinfo {pages} {080502} (\bibinfo {year} {2008})}\BibitemShut {NoStop}%
\bibitem [{\citenamefont {Zhong}\ \emph {et~al.}(2021)\citenamefont {Zhong},
  \citenamefont {Chang}, \citenamefont {Bienfait}, \citenamefont {Dumur},
  \citenamefont {Chou}, \citenamefont {Conner}, \citenamefont {Grebel},
  \citenamefont {Povey}, \citenamefont {Yan}, \citenamefont {Schuster},\ and\
  \citenamefont {Cleland}}]{Zhong2021}%
  \BibitemOpen
  \bibfield  {author} {\bibinfo {author} {\bibfnamefont {Y.}~\bibnamefont
  {Zhong}}, \bibinfo {author} {\bibfnamefont {H.-S.}\ \bibnamefont {Chang}},
  \bibinfo {author} {\bibfnamefont {A.}~\bibnamefont {Bienfait}}, \bibinfo
  {author} {\bibfnamefont {{\'E}.}~\bibnamefont {Dumur}}, \bibinfo {author}
  {\bibfnamefont {M.-H.}\ \bibnamefont {Chou}}, \bibinfo {author}
  {\bibfnamefont {C.~R.}\ \bibnamefont {Conner}}, \bibinfo {author}
  {\bibfnamefont {J.}~\bibnamefont {Grebel}}, \bibinfo {author} {\bibfnamefont
  {R.~G.}\ \bibnamefont {Povey}}, \bibinfo {author} {\bibfnamefont
  {H.}~\bibnamefont {Yan}}, \bibinfo {author} {\bibfnamefont {D.~I.}\
  \bibnamefont {Schuster}},\ and\ \bibinfo {author} {\bibfnamefont {A.~N.}\
  \bibnamefont {Cleland}},\ }\bibfield  {title} {\bibinfo {title}
  {Deterministic multi-qubit entanglement in a quantum network},\ }\href
  {https://doi.org/10.1038/s41586-021-03288-7} {\bibfield  {journal} {\bibinfo
  {journal} {Nature}\ }\textbf {\bibinfo {volume} {590}},\ \bibinfo {pages}
  {571} (\bibinfo {year} {2021})}\BibitemShut {NoStop}%
\bibitem [{\citenamefont {Pellizzari}(1997)}]{PhysRevLett.79.5242}%
  \BibitemOpen
  \bibfield  {author} {\bibinfo {author} {\bibfnamefont {T.}~\bibnamefont
  {Pellizzari}},\ }\bibfield  {title} {\bibinfo {title} {Quantum networking
  with optical fibres},\ }\href {https://doi.org/10.1103/PhysRevLett.79.5242}
  {\bibfield  {journal} {\bibinfo  {journal} {Phys. Rev. Lett.}\ }\textbf
  {\bibinfo {volume} {79}},\ \bibinfo {pages} {5242} (\bibinfo {year}
  {1997})}\BibitemShut {NoStop}%
\bibitem [{\citenamefont {Vogell}\ \emph {et~al.}(2017)\citenamefont {Vogell},
  \citenamefont {Vermersch}, \citenamefont {Northup}, \citenamefont {Lanyon},\
  and\ \citenamefont {Muschik}}]{Vogell2017}%
  \BibitemOpen
  \bibfield  {author} {\bibinfo {author} {\bibfnamefont {B.}~\bibnamefont
  {Vogell}}, \bibinfo {author} {\bibfnamefont {B.}~\bibnamefont {Vermersch}},
  \bibinfo {author} {\bibfnamefont {T.~E.}\ \bibnamefont {Northup}}, \bibinfo
  {author} {\bibfnamefont {B.~P.}\ \bibnamefont {Lanyon}},\ and\ \bibinfo
  {author} {\bibfnamefont {C.~A.}\ \bibnamefont {Muschik}},\ }\bibfield
  {title} {\bibinfo {title} {Deterministic quantum state transfer between
  remote qubits in cavities},\ }\href
  {https://doi.org/10.1088/2058-9565/aa868b} {\bibfield  {journal} {\bibinfo
  {journal} {Quantum Sci. Technol.}\ }\textbf {\bibinfo {volume} {2}},\
  \bibinfo {pages} {045003} (\bibinfo {year} {2017})}\BibitemShut {NoStop}%
\bibitem [{\citenamefont {Schrieffer}\ and\ \citenamefont
  {Wolff}(1966)}]{Schrieffer1966}%
  \BibitemOpen
  \bibfield  {author} {\bibinfo {author} {\bibfnamefont {J.~R.}\ \bibnamefont
  {Schrieffer}}\ and\ \bibinfo {author} {\bibfnamefont {P.~A.}\ \bibnamefont
  {Wolff}},\ }\bibfield  {title} {\bibinfo {title} {Relation between the
  anderson and kondo hamiltonians},\ }\href
  {https://doi.org/10.1103/PhysRev.149.491} {\bibfield  {journal} {\bibinfo
  {journal} {Phys. Rev.}\ }\textbf {\bibinfo {volume} {149}},\ \bibinfo {pages}
  {491} (\bibinfo {year} {1966})}\BibitemShut {NoStop}%
\bibitem [{\citenamefont {Schutjens}\ \emph {et~al.}(2013)\citenamefont
  {Schutjens}, \citenamefont {Dagga}, \citenamefont {Egger},\ and\
  \citenamefont {Wilhelm}}]{Wilhelm}%
  \BibitemOpen
  \bibfield  {author} {\bibinfo {author} {\bibfnamefont {R.}~\bibnamefont
  {Schutjens}}, \bibinfo {author} {\bibfnamefont {F.~A.}\ \bibnamefont
  {Dagga}}, \bibinfo {author} {\bibfnamefont {D.~J.}\ \bibnamefont {Egger}},\
  and\ \bibinfo {author} {\bibfnamefont {F.~K.}\ \bibnamefont {Wilhelm}},\
  }\bibfield  {title} {\bibinfo {title} {Single-qubit gates in
  frequency-crowded transmon systems},\ }\href
  {https://doi.org/10.1103/PhysRevA.88.052330} {\bibfield  {journal} {\bibinfo
  {journal} {Phys. Rev. A}\ }\textbf {\bibinfo {volume} {88}},\ \bibinfo
  {pages} {052330} (\bibinfo {year} {2013})}\BibitemShut {NoStop}%
\bibitem [{\citenamefont {Larssen}\ \emph {et~al.}()\citenamefont {Larssen},
  \citenamefont {Peng},\ and\ \citenamefont {Markidis}}]{larssen2026}%
  \BibitemOpen
  \bibfield  {author} {\bibinfo {author} {\bibfnamefont {F.}~\bibnamefont
  {Larssen}}, \bibinfo {author} {\bibfnamefont {I.}~\bibnamefont {Peng}},\ and\
  \bibinfo {author} {\bibfnamefont {S.}~\bibnamefont {Markidis}},\ }\bibfield
  {title} {\bibinfo {title} {Which superconducting qubit model is good enough?
  from effective two-level to circuit-based hamiltonians for pulse-level
  simulation},\ }\href {https://arxiv.org/abs/2605.23034} {\ }\Eprint
  {https://arxiv.org/abs/2605.23034} {arXiv:2605.23034} \BibitemShut {NoStop}%
\bibitem [{\citenamefont {Zhao}\ \emph
  {et~al.}(2023{\natexlab{b}})\citenamefont {Zhao}, \citenamefont {Wang},
  \citenamefont {Hu}, \citenamefont {Ma}, \citenamefont {Xu}, \citenamefont
  {Jin},\ and\ \citenamefont {Yu}}]{PhysRevApplied.19.054050}%
  \BibitemOpen
  \bibfield  {author} {\bibinfo {author} {\bibfnamefont {P.}~\bibnamefont
  {Zhao}}, \bibinfo {author} {\bibfnamefont {R.}~\bibnamefont {Wang}}, \bibinfo
  {author} {\bibfnamefont {M.-J.}\ \bibnamefont {Hu}}, \bibinfo {author}
  {\bibfnamefont {T.}~\bibnamefont {Ma}}, \bibinfo {author} {\bibfnamefont
  {P.}~\bibnamefont {Xu}}, \bibinfo {author} {\bibfnamefont {Y.}~\bibnamefont
  {Jin}},\ and\ \bibinfo {author} {\bibfnamefont {H.}~\bibnamefont {Yu}},\
  }\bibfield  {title} {\bibinfo {title} {Baseband control of superconducting
  qubits with shared microwave drives},\ }\href
  {https://doi.org/10.1103/PhysRevApplied.19.054050} {\bibfield  {journal}
  {\bibinfo  {journal} {Phys. Rev. Appl.}\ }\textbf {\bibinfo {volume} {19}},\
  \bibinfo {pages} {054050} (\bibinfo {year} {2023}{\natexlab{b}})}\BibitemShut
  {NoStop}%
\bibitem [{\citenamefont {Ghosh}\ \emph {et~al.}(2013)\citenamefont {Ghosh},
  \citenamefont {Galiautdinov}, \citenamefont {Zhou}, \citenamefont {Korotkov},
  \citenamefont {Martinis},\ and\ \citenamefont {Geller}}]{PhysRevA.87.022309}%
  \BibitemOpen
  \bibfield  {author} {\bibinfo {author} {\bibfnamefont {J.}~\bibnamefont
  {Ghosh}}, \bibinfo {author} {\bibfnamefont {A.}~\bibnamefont {Galiautdinov}},
  \bibinfo {author} {\bibfnamefont {Z.}~\bibnamefont {Zhou}}, \bibinfo {author}
  {\bibfnamefont {A.~N.}\ \bibnamefont {Korotkov}}, \bibinfo {author}
  {\bibfnamefont {J.~M.}\ \bibnamefont {Martinis}},\ and\ \bibinfo {author}
  {\bibfnamefont {M.~R.}\ \bibnamefont {Geller}},\ }\bibfield  {title}
  {\bibinfo {title} {High-fidelity controlled-${\ensuremath{\sigma}}^{Z}$ gate
  for resonator-based superconducting quantum computers},\ }\href
  {https://doi.org/10.1103/PhysRevA.87.022309} {\bibfield  {journal} {\bibinfo
  {journal} {Phys. Rev. A}\ }\textbf {\bibinfo {volume} {87}},\ \bibinfo
  {pages} {022309} (\bibinfo {year} {2013})}\BibitemShut {NoStop}%
\bibitem [{\citenamefont {Martinis}\ and\ \citenamefont
  {Geller}(2014)}]{PhysRevA.90.022307}%
  \BibitemOpen
  \bibfield  {author} {\bibinfo {author} {\bibfnamefont {J.~M.}\ \bibnamefont
  {Martinis}}\ and\ \bibinfo {author} {\bibfnamefont {M.~R.}\ \bibnamefont
  {Geller}},\ }\bibfield  {title} {\bibinfo {title} {Fast adiabatic qubit gates
  using only ${\ensuremath{\sigma}}_{z}$ control},\ }\href
  {https://doi.org/10.1103/PhysRevA.90.022307} {\bibfield  {journal} {\bibinfo
  {journal} {Phys. Rev. A}\ }\textbf {\bibinfo {volume} {90}},\ \bibinfo
  {pages} {022307} (\bibinfo {year} {2014})}\BibitemShut {NoStop}%
\bibitem [{\citenamefont {Barends}\ \emph {et~al.}(2019)\citenamefont
  {Barends}, \citenamefont {Quintana}, \citenamefont {Petukhov}, \citenamefont
  {Chen}, \citenamefont {Kafri}, \citenamefont {Kechedzhi}, \citenamefont
  {Collins}, \citenamefont {Naaman}, \citenamefont {Boixo}, \citenamefont
  {Arute}, \citenamefont {Arya}, \citenamefont {Buell}, \citenamefont
  {Burkett}, \citenamefont {Chen}, \citenamefont {Chiaro}, \citenamefont
  {Dunsworth}, \citenamefont {Foxen}, \citenamefont {Fowler}, \citenamefont
  {Gidney}, \citenamefont {Giustina}, \citenamefont {Graff}, \citenamefont
  {Huang}, \citenamefont {Jeffrey}, \citenamefont {Kelly}, \citenamefont
  {Klimov}, \citenamefont {Kostritsa}, \citenamefont {Landhuis}, \citenamefont
  {Lucero}, \citenamefont {McEwen}, \citenamefont {Megrant}, \citenamefont
  {Mi}, \citenamefont {Mutus}, \citenamefont {Neeley}, \citenamefont {Neill},
  \citenamefont {Ostby}, \citenamefont {Roushan}, \citenamefont {Sank},
  \citenamefont {Satzinger}, \citenamefont {Vainsencher}, \citenamefont
  {White}, \citenamefont {Yao}, \citenamefont {Yeh}, \citenamefont {Zalcman},
  \citenamefont {Neven}, \citenamefont {Smelyanskiy},\ and\ \citenamefont
  {Martinis}}]{PhysRevLett.123.210501}%
  \BibitemOpen
  \bibfield  {author} {\bibinfo {author} {\bibfnamefont {R.}~\bibnamefont
  {Barends}}, \bibinfo {author} {\bibfnamefont {C.~M.}\ \bibnamefont
  {Quintana}}, \bibinfo {author} {\bibfnamefont {A.~G.}\ \bibnamefont
  {Petukhov}}, \bibinfo {author} {\bibfnamefont {Y.}~\bibnamefont {Chen}},
  \bibinfo {author} {\bibfnamefont {D.}~\bibnamefont {Kafri}}, \bibinfo
  {author} {\bibfnamefont {K.}~\bibnamefont {Kechedzhi}}, \bibinfo {author}
  {\bibfnamefont {R.}~\bibnamefont {Collins}}, \bibinfo {author} {\bibfnamefont
  {O.}~\bibnamefont {Naaman}}, \bibinfo {author} {\bibfnamefont
  {S.}~\bibnamefont {Boixo}}, \bibinfo {author} {\bibfnamefont
  {F.}~\bibnamefont {Arute}}, \bibinfo {author} {\bibfnamefont
  {K.}~\bibnamefont {Arya}}, \bibinfo {author} {\bibfnamefont {D.}~\bibnamefont
  {Buell}}, \bibinfo {author} {\bibfnamefont {B.}~\bibnamefont {Burkett}},
  \bibinfo {author} {\bibfnamefont {Z.}~\bibnamefont {Chen}}, \bibinfo {author}
  {\bibfnamefont {B.}~\bibnamefont {Chiaro}}, \bibinfo {author} {\bibfnamefont
  {A.}~\bibnamefont {Dunsworth}}, \bibinfo {author} {\bibfnamefont
  {B.}~\bibnamefont {Foxen}}, \bibinfo {author} {\bibfnamefont
  {A.}~\bibnamefont {Fowler}}, \bibinfo {author} {\bibfnamefont
  {C.}~\bibnamefont {Gidney}}, \bibinfo {author} {\bibfnamefont
  {M.}~\bibnamefont {Giustina}}, \bibinfo {author} {\bibfnamefont
  {R.}~\bibnamefont {Graff}}, \bibinfo {author} {\bibfnamefont
  {T.}~\bibnamefont {Huang}}, \bibinfo {author} {\bibfnamefont
  {E.}~\bibnamefont {Jeffrey}}, \bibinfo {author} {\bibfnamefont
  {J.}~\bibnamefont {Kelly}}, \bibinfo {author} {\bibfnamefont {P.~V.}\
  \bibnamefont {Klimov}}, \bibinfo {author} {\bibfnamefont {F.}~\bibnamefont
  {Kostritsa}}, \bibinfo {author} {\bibfnamefont {D.}~\bibnamefont {Landhuis}},
  \bibinfo {author} {\bibfnamefont {E.}~\bibnamefont {Lucero}}, \bibinfo
  {author} {\bibfnamefont {M.}~\bibnamefont {McEwen}}, \bibinfo {author}
  {\bibfnamefont {A.}~\bibnamefont {Megrant}}, \bibinfo {author} {\bibfnamefont
  {X.}~\bibnamefont {Mi}}, \bibinfo {author} {\bibfnamefont {J.}~\bibnamefont
  {Mutus}}, \bibinfo {author} {\bibfnamefont {M.}~\bibnamefont {Neeley}},
  \bibinfo {author} {\bibfnamefont {C.}~\bibnamefont {Neill}}, \bibinfo
  {author} {\bibfnamefont {E.}~\bibnamefont {Ostby}}, \bibinfo {author}
  {\bibfnamefont {P.}~\bibnamefont {Roushan}}, \bibinfo {author} {\bibfnamefont
  {D.}~\bibnamefont {Sank}}, \bibinfo {author} {\bibfnamefont {K.~J.}\
  \bibnamefont {Satzinger}}, \bibinfo {author} {\bibfnamefont {A.}~\bibnamefont
  {Vainsencher}}, \bibinfo {author} {\bibfnamefont {T.}~\bibnamefont {White}},
  \bibinfo {author} {\bibfnamefont {J.}~\bibnamefont {Yao}}, \bibinfo {author}
  {\bibfnamefont {P.}~\bibnamefont {Yeh}}, \bibinfo {author} {\bibfnamefont
  {A.}~\bibnamefont {Zalcman}}, \bibinfo {author} {\bibfnamefont
  {H.}~\bibnamefont {Neven}}, \bibinfo {author} {\bibfnamefont {V.~N.}\
  \bibnamefont {Smelyanskiy}},\ and\ \bibinfo {author} {\bibfnamefont {J.~M.}\
  \bibnamefont {Martinis}},\ }\bibfield  {title} {\bibinfo {title} {Diabatic
  gates for frequency-tunable superconducting qubits},\ }\href
  {https://doi.org/10.1103/PhysRevLett.123.210501} {\bibfield  {journal}
  {\bibinfo  {journal} {Phys. Rev. Lett.}\ }\textbf {\bibinfo {volume} {123}},\
  \bibinfo {pages} {210501} (\bibinfo {year} {2019})}\BibitemShut {NoStop}%
\bibitem [{\citenamefont {Chu}\ and\ \citenamefont
  {Yan}(2021)}]{PhysRevApplied.16.054020}%
  \BibitemOpen
  \bibfield  {author} {\bibinfo {author} {\bibfnamefont {J.}~\bibnamefont
  {Chu}}\ and\ \bibinfo {author} {\bibfnamefont {F.}~\bibnamefont {Yan}},\
  }\bibfield  {title} {\bibinfo {title} {Coupler-assisted controlled-phase gate
  with enhanced adiabaticity},\ }\href
  {https://doi.org/10.1103/PhysRevApplied.16.054020} {\bibfield  {journal}
  {\bibinfo  {journal} {Phys. Rev. Appl.}\ }\textbf {\bibinfo {volume} {16}},\
  \bibinfo {pages} {054020} (\bibinfo {year} {2021})}\BibitemShut {NoStop}%
\bibitem [{\citenamefont {Xu}\ \emph {et~al.}(2026)\citenamefont {Xu},
  \citenamefont {Deng}, \citenamefont {Zheng}, \citenamefont {Yan},
  \citenamefont {Zhang}, \citenamefont {Zhang}, \citenamefont {Huang},
  \citenamefont {Xia}, \citenamefont {Liao}, \citenamefont {Zhang},
  \citenamefont {Zhao}, \citenamefont {Li}, \citenamefont {Tan}, \citenamefont
  {Lan},\ and\ \citenamefont {Yu}}]{ls5b-279m}%
  \BibitemOpen
  \bibfield  {author} {\bibinfo {author} {\bibfnamefont {J.}~\bibnamefont
  {Xu}}, \bibinfo {author} {\bibfnamefont {X.}~\bibnamefont {Deng}}, \bibinfo
  {author} {\bibfnamefont {W.}~\bibnamefont {Zheng}}, \bibinfo {author}
  {\bibfnamefont {W.}~\bibnamefont {Yan}}, \bibinfo {author} {\bibfnamefont
  {T.}~\bibnamefont {Zhang}}, \bibinfo {author} {\bibfnamefont
  {Z.}~\bibnamefont {Zhang}}, \bibinfo {author} {\bibfnamefont
  {W.}~\bibnamefont {Huang}}, \bibinfo {author} {\bibfnamefont
  {X.}~\bibnamefont {Xia}}, \bibinfo {author} {\bibfnamefont {X.}~\bibnamefont
  {Liao}}, \bibinfo {author} {\bibfnamefont {Y.}~\bibnamefont {Zhang}},
  \bibinfo {author} {\bibfnamefont {J.}~\bibnamefont {Zhao}}, \bibinfo {author}
  {\bibfnamefont {S.}~\bibnamefont {Li}}, \bibinfo {author} {\bibfnamefont
  {X.}~\bibnamefont {Tan}}, \bibinfo {author} {\bibfnamefont {D.}~\bibnamefont
  {Lan}},\ and\ \bibinfo {author} {\bibfnamefont {Y.}~\bibnamefont {Yu}},\
  }\bibfield  {title} {\bibinfo {title} {Tunable hybrid-mode coupler enabling
  strong interactions between transmons at centimeter-scale distance},\ }\href
  {https://doi.org/10.1103/ls5b-279m} {\bibfield  {journal} {\bibinfo
  {journal} {Phys. Rev. Appl.}\ }\textbf {\bibinfo {volume} {25}},\ \bibinfo
  {pages} {014016} (\bibinfo {year} {2026})}\BibitemShut {NoStop}%
\bibitem [{\citenamefont {Levy-Yeyati}\ \emph {et~al.}(2025)\citenamefont
  {Levy-Yeyati}, \citenamefont {Vega}, \citenamefont {Ramos},\ and\
  \citenamefont {Gonz\'alez-Tudela}}]{PRXQuantum.6.010342}%
  \BibitemOpen
  \bibfield  {author} {\bibinfo {author} {\bibfnamefont {T.}~\bibnamefont
  {Levy-Yeyati}}, \bibinfo {author} {\bibfnamefont {C.}~\bibnamefont {Vega}},
  \bibinfo {author} {\bibfnamefont {T.}~\bibnamefont {Ramos}},\ and\ \bibinfo
  {author} {\bibfnamefont {A.}~\bibnamefont {Gonz\'alez-Tudela}},\ }\bibfield
  {title} {\bibinfo {title} {Passive photonic cz gate with two-level emitters
  in chiral multimode waveguide qed},\ }\href
  {https://doi.org/10.1103/PRXQuantum.6.010342} {\bibfield  {journal} {\bibinfo
   {journal} {PRX Quantum}\ }\textbf {\bibinfo {volume} {6}},\ \bibinfo {pages}
  {010342} (\bibinfo {year} {2025})}\BibitemShut {NoStop}%
\bibitem [{\citenamefont {Rieck}\ \emph {et~al.}()\citenamefont {Rieck},
  \citenamefont {Soro}, \citenamefont {Kockum},\ and\ \citenamefont
  {Chen}}]{rieck2026}%
  \BibitemOpen
  \bibfield  {author} {\bibinfo {author} {\bibfnamefont {W.}~\bibnamefont
  {Rieck}}, \bibinfo {author} {\bibfnamefont {A.}~\bibnamefont {Soro}},
  \bibinfo {author} {\bibfnamefont {A.~F.}\ \bibnamefont {Kockum}},\ and\
  \bibinfo {author} {\bibfnamefont {G.}~\bibnamefont {Chen}},\ }\bibfield
  {title} {\bibinfo {title} {Controlled-z gates with giant atoms in structured
  waveguides},\ }\href {https://arxiv.org/abs/2603.26345} {\ }\Eprint
  {https://arxiv.org/abs/2603.26345} {arXiv:2603.26345 [quant-ph]} \BibitemShut
  {NoStop}%
\bibitem [{\citenamefont {Levy-Yeyati}\ \emph {et~al.}(2026)\citenamefont
  {Levy-Yeyati}, \citenamefont {Ramos},\ and\ \citenamefont
  {Gonz\'alez-Tudela}}]{kv28-r6w4}%
  \BibitemOpen
  \bibfield  {author} {\bibinfo {author} {\bibfnamefont {T.}~\bibnamefont
  {Levy-Yeyati}}, \bibinfo {author} {\bibfnamefont {T.}~\bibnamefont {Ramos}},\
  and\ \bibinfo {author} {\bibfnamefont {A.}~\bibnamefont
  {Gonz\'alez-Tudela}},\ }\bibfield  {title} {\bibinfo {title} {Engineering
  giant transmon molecules as mediators of conditional two-photon gates},\
  }\href {https://doi.org/10.1103/kv28-r6w4} {\bibfield  {journal} {\bibinfo
  {journal} {Phys. Rev. Appl.}\ }\textbf {\bibinfo {volume} {25}},\ \bibinfo
  {pages} {054060} (\bibinfo {year} {2026})}\BibitemShut {NoStop}%
\bibitem [{\citenamefont {Sakurai}\ and\ \citenamefont
  {Napolitano}(2017)}]{Sakurai2017}%
  \BibitemOpen
  \bibfield  {author} {\bibinfo {author} {\bibfnamefont {J.~J.}\ \bibnamefont
  {Sakurai}}\ and\ \bibinfo {author} {\bibfnamefont {J.}~\bibnamefont
  {Napolitano}},\ }\href@noop {} {\emph {\bibinfo {title} {Modern Quantum
  Mechanics}}},\ \bibinfo {edition} {2nd}\ ed.\ (\bibinfo  {publisher}
  {Cambridge University Press},\ \bibinfo {year} {2017})\BibitemShut {NoStop}%
\bibitem [{\citenamefont {DiCarlo}\ \emph {et~al.}(2009)\citenamefont
  {DiCarlo}, \citenamefont {Chow}, \citenamefont {Gambetta} \emph
  {et~al.}}]{DiCarlo2009}%
  \BibitemOpen
  \bibfield  {author} {\bibinfo {author} {\bibfnamefont {L.}~\bibnamefont
  {DiCarlo}}, \bibinfo {author} {\bibfnamefont {J.~M.}\ \bibnamefont {Chow}},
  \bibinfo {author} {\bibfnamefont {J.~M.}\ \bibnamefont {Gambetta}}, \emph
  {et~al.},\ }\bibfield  {title} {\bibinfo {title} {Demonstration of two-qubit
  algorithms with a superconducting quantum processor},\ }\href
  {https://doi.org/10.1038/nature08121} {\bibfield  {journal} {\bibinfo
  {journal} {Nature}\ }\textbf {\bibinfo {volume} {460}},\ \bibinfo {pages}
  {240} (\bibinfo {year} {2009})}\BibitemShut {NoStop}%
\bibitem [{\citenamefont {Zhao}\ \emph
  {et~al.}(2023{\natexlab{c}})\citenamefont {Zhao}, \citenamefont {He},
  \citenamefont {Geng}, \citenamefont {He}, \citenamefont {Dai}, \citenamefont
  {Liu},\ and\ \citenamefont {Chen}}]{Zhao2023}%
  \BibitemOpen
  \bibfield  {author} {\bibinfo {author} {\bibfnamefont {C.}~\bibnamefont
  {Zhao}}, \bibinfo {author} {\bibfnamefont {Y.}~\bibnamefont {He}}, \bibinfo
  {author} {\bibfnamefont {X.}~\bibnamefont {Geng}}, \bibinfo {author}
  {\bibfnamefont {K.}~\bibnamefont {He}}, \bibinfo {author} {\bibfnamefont
  {G.}~\bibnamefont {Dai}}, \bibinfo {author} {\bibfnamefont {J.}~\bibnamefont
  {Liu}},\ and\ \bibinfo {author} {\bibfnamefont {W.}~\bibnamefont {Chen}},\
  }\bibfield  {title} {\bibinfo {title} {Multi-mode bus coupling architecture
  of superconducting quantum processor},\ }\href
  {https://doi.org/10.1088/0256-307X/40/1/010301} {\bibfield  {journal}
  {\bibinfo  {journal} {Chinese Physics Letters}\ }\textbf {\bibinfo {volume}
  {40}},\ \bibinfo {pages} {010301} (\bibinfo {year}
  {2023}{\natexlab{c}})}\BibitemShut {NoStop}%
\bibitem [{\citenamefont {Zhu}\ \emph {et~al.}(2013)\citenamefont {Zhu},
  \citenamefont {Ferguson}, \citenamefont {Manucharyan},\ and\ \citenamefont
  {Koch}}]{PhysRevB.87.024510}%
  \BibitemOpen
  \bibfield  {author} {\bibinfo {author} {\bibfnamefont {G.}~\bibnamefont
  {Zhu}}, \bibinfo {author} {\bibfnamefont {D.~G.}\ \bibnamefont {Ferguson}},
  \bibinfo {author} {\bibfnamefont {V.~E.}\ \bibnamefont {Manucharyan}},\ and\
  \bibinfo {author} {\bibfnamefont {J.}~\bibnamefont {Koch}},\ }\bibfield
  {title} {\bibinfo {title} {Circuit qed with fluxonium qubits: Theory of the
  dispersive regime},\ }\href {https://doi.org/10.1103/PhysRevB.87.024510}
  {\bibfield  {journal} {\bibinfo  {journal} {Phys. Rev. B}\ }\textbf {\bibinfo
  {volume} {87}},\ \bibinfo {pages} {024510} (\bibinfo {year}
  {2013})}\BibitemShut {NoStop}%
\end{thebibliography}%
\end{document}